\documentclass[aps,notitlepage,twocolumn, nofootinbib,pra,10pt]{revtex4-1}
\usepackage{amsmath,amsfonts,amssymb,amsthm,bbm,graphicx,enumerate,times}
\usepackage{mathtools}
\usepackage[usenames,dvipsnames]{color}

\usepackage{hyperref}
\usepackage{comment}
\usepackage{tikz}
\usepackage{graphicx}
\usepackage{float}
\usetikzlibrary{shapes,positioning}
\usepackage{etoolbox}
\usepackage{comment}
\usepackage{todonotes}
\usepackage{accents}
\usepackage{stmaryrd}

\usepackage{tabularx}
\makeatletter
\def\hlinewd#1{%
\noalign{\ifnum0=`}\fi\hrule \@height #1 %
\futurelet\reserved@a\@xhline}
\makeatother

\definecolor{darkgreen}{rgb}{.15,.6,.15}
\definecolor{darkcyan}{rgb}{.15,.5,.5}
\definecolor{darkred}{rgb}{.6,.15,.15}

  \definecolor{marek}{rgb}{.5,.5,.1}
  \definecolor{alex}{rgb}{.5,.1,.5}

\newcommand{\widerule}{\hlinewd{1.0pt}}

\newcommand{\comma}{, }

\usepackage[normalem]{ulem}

\usepackage{dsfont}
\newcommand{\id}{\mathds{1}}

 \renewcommand{\i}{\,\ensuremath\mathrm{i}}

\newcommand{\sto}{\,{\to}\,}

\DeclareMathOperator{\arcosh}{arcosh}

\newcommand{\ket}[1]{\left.\left|{#1}\right.\right\rangle}

\DeclareMathAlphabet{\mathpzcc}{OT1}{pzc}{m}{it}
\DeclareMathAlphabet{\mathpzc}{T1}{pzc}{m}{it}{\huge}

\newcommand{\m}{\operatorname{\gamma}}

\def\and{\quad\text{and}\quad}

\begin{document}

\title{Central charges of aperiodic holographic tensor network models}

\author{Alexander Jahn,$^1$ Zolt{\'a}n Zimbor{\'a}s,$^{2,3,4}$ and Jens Eisert$^{1,5}$}
\affiliation{$^1$Dahlem Center for Complex Quantum Systems\comma Freie Universit{\"a}t Berlin\comma 14195 Berlin\comma Germany \\
$^2$Institute for Particle and Nuclear Physics\comma Wigner Research Centre for Physics\comma 1121 Budapest\comma Hungary \\
$^3$BME-MTA Lend\"ulet Quantum Information Theory Research Group\comma 1111 Budapest\comma Hungary\\
$^4$Institute for Mathematics\comma Budapest University of Technology and Economics\comma 1111 Budapest\comma Hungary \\
$^5$Department of Mathematics and Computer Science\comma Freie Universit{\"a}t Berlin\comma 14195 Berlin\comma Germany
}

\begin{abstract}
Central to the AdS/CFT correspondence is a precise relationship between the curvature of an anti-de Sitter (AdS) spacetime and the central charge of the dual conformal field theory (CFT) on its boundary.
Our work shows that such a relationship can also be established for tensor network models of AdS/CFT based on regular bulk geometries, leading to an analytical form of the maximal central charges exhibited by the boundary states. 
We identify a class of tensors based on Majorana dimer states that saturate these bounds in the large curvature limit, while also realizing perfect and block-perfect holographic quantum error correcting codes.
Furthermore, the renormalization group description of the resulting model is shown to be analogous to the strong disorder renormalization group, thus giving the first example of an exact quantum error correcting code that gives rise to a well-understood critical system.
These systems exhibit a large range of fractional central charges, tunable by the choice of bulk tiling. Our approach thus provides a precise physical interpretation of tensor network models on regular hyperbolic geometries and establishes quantitative connections to a wide range of existing models. 
\end{abstract}

\maketitle
\date{\today}
\section{Introduction}
Years before the formulation of the holographic principle, J.~D.~Brown and M.~Henneaux noticed a peculiar property of anti-de Sitter (AdS) spacetime, a solution to Einstein's equation with constant negative curvature: At its asymptotic boundary, the generators of the symmetry group $SO(2,2)$ of $2{+}1$-dimensional AdS$_3$ spacetime form a \emph{Virasoro algebra} describing a $2$-dimensional \emph{conformal field theory} (CFT) with an effective \emph{central charge} depending on the curvature of the AdS bulk. 
Rather than a mathematical coincidence, the AdS/CFT correspondence \cite{Maldacena98} propelled this observation to the cornerstone of a holographic duality between gravity in $d{+}2$-dimensional AdS$_{d+2}$ spacetime and a conformal field theory (CFT) on its $d{+}1$-dimensional boundary, with an equivalent action describing both sides of the duality \cite{Witten:1998qj}.
A key motivation for the holographic principle was the discovery that a black hole's entropy scales with its horizon area rather than its volume \cite{Bekenstein:1973ur,Hawking:1974sw}. The Bekenstein-Hawking entropy formula 
\begin{equation}
\label{EQ_BEK_HAW}
S_\text{BH} = \frac{A_\text{hor}}{4 G} \ ,
\end{equation}
where $A_\text{hor}$ is the horizon area and $G$ the gravitational constant, has a surprising generalization in the context of AdS/CFT: The entanglement entropy $S_A$ \cite{AreaReview}
of a boundary region $A$ follows the Ryu-Takayanagi (RT) 
formula
\cite{PhysRevLett.96.181602}
\begin{equation}
\label{EQ_RT}
S_\text{A} = \frac{|\gamma_A|}{4 G} \ ,
\end{equation}
where $|\gamma_A|$ is the area of an extremal surface $\gamma_A$ in the bulk whose boundary $\partial\gamma_A$ matches the boundary $\partial A$. In $2{+}1$ dimensions, $\gamma_A$ is simply a geodesic curve and $|\gamma_A|$ its length. Both formulae \eqref{EQ_BEK_HAW} and \eqref{EQ_RT} suggest an encoding of information in Planckian pieces of area of size 
${\sim}\, G = l_p^2$ (in $3{+}1$ bulk dimensions). 

While the AdS/CFT correspondence is formulated in the continuum, \emph{tensor networks} \cite{Orus-AnnPhys-2014,VerstraeteBig,AreaReview,SchuchReview,Handwaving} have become a popular approach for models built on a discretized AdS spacetime, as they naturally incorporate the RT formula in the form of an upper bound on entanglement and yield boundary quantum states that can be efficiently computed.
The \emph{multi-scale entanglement renormalization ansatz} (MERA) \cite{PhysRevLett.101.110501}, a tensor network that well approximates critical boundary states, was identified as a possible realization of discrete holography \cite{PhysRevD.86.065007,PhysRevD.97.026012}, but the bulk geometry of the MERA cannot be directly related to an AdS time-slice \cite{Beny:2011vh,Bao:2015uaa,Milsted:2018san}.
Instead, regular hyperbolic tilings have recently been used as the basis of numerous discrete holographic models \cite{Pastawski2015,PhysRevLett.119.141602,Jahn:2017tls,Osborne:2017woa,PhysRevA.98.052301,Kohler:2018kqk,Jahn:2019nmz}, elucidating many aspects of AdS/CFT, particularly its deep connection to \emph{quantum error correction} \cite{Almheiri15,Pastawski2015}, 
However, a clear interpretation of the resulting boundary states in terms of a critical system, as is possible for the MERA, remained elusive.

Resolving this question, we show that tensor networks on regular tilings lead to boundary quantum states whose symmetries naturally discretize conformal symmetries on time-slices, allowing their maximal central charges to be analytically computed for any tiling. Relating this central charge to the scalar curvature of the tiling then  results in a discrete generalization of the Brown-Henneaux formula \cite{Brown:1986nw}. We demonstrate these properties using a class of tensor networks based on \emph{Majorana dimer states}, whose exact central charges are computed and are shown to saturate the upper bound in the strong-curvature limit. This class of states includes the widely studied \emph{hyperbolic pentagon code} (HyPeC), an instance of the \emph{HaPPY} codes \cite{Pastawski2015}, a toy model for quantum error correction in AdS/CFT.
In this paper, we argue that these dimer models are a discrete approximation of a CFT with an aperiodic structure, the inflation rules of the tiling providing a local renormalization group transformation identified with the strong-disorder renormalization group (SDRG). The discrete boundary thus exhibits quasi-regular symmetries, describing a CFT discretization that breaks translation invariance and possesses disorder on all length scales. Such critical systems have been extensively studied in the condensed matter literature, but no connection to holographic models had been known until now.

\section{Central charges and curvature}
\begin{figure}[htb]
\includegraphics[width=0.5\textwidth]{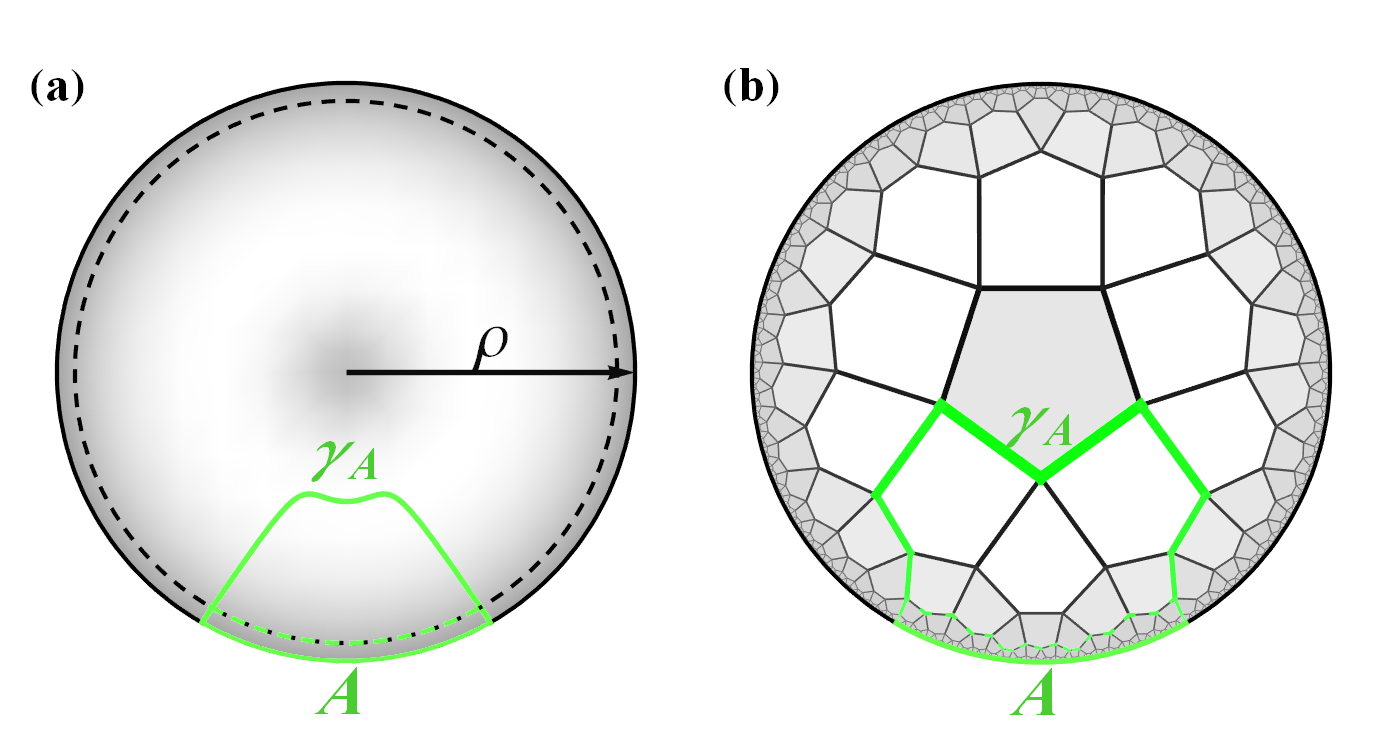}
\caption{(a) Continuous and (b) discretized geodesic $\gamma_A$ in the Poincar\'e disk with a deformation in the center and a boundary cutoff shown as a dashed curve. In the asymptotic region towards the boundary, the shape of $\gamma_A$ is independent of bulk deformations.}
\label{FIG_POINCARE_DISK}
\end{figure}

In \emph{global} AdS coordinates, AdS$_3$ spacetime takes the form
\begin{equation}
\label{EQ_ADS_GLOBAL}
\text{d}s^2 = -(1 + r^2/\alpha^2)\text{d}t^2 + \frac{\alpha^2 \text{d}r^2}{\alpha^2 + r^2} + r^2 \text{d}\phi \ ,
\end{equation}
where $\alpha$ is the \emph{AdS radius}. The scalar curvature or \emph{Ricci scalar} $R$ of AdS$_d$ spacetime with $d=2{+}1$ dimensions is given by
\begin{equation}
R = -\frac{d (d-1)}{\alpha^2} = - \frac{6}{\alpha^2}\ ,
\end{equation}
corresponding to a negative cosmological constant $\Lambda = -1/\alpha^2$. An AdS$_3$ time-slice can be more conveniently mapped to the \emph{Poincar\'e disk} with
\begin{equation}
\label{EQ_ADS_PDISK}
\mathrm{d}s^2 = 4 \alpha^2 \frac{\mathrm{d}\rho^2 + \rho^2 \mathrm{d}\phi^2}{(1-\rho^2)^2} \ .
\end{equation}
Global and Poincar\'e disk coordinates are related by a radial transformation $r = 2\alpha\rho/(1-\rho^2)$ and the time-slice constraint $\text{d}t=0$. The global radius is defined in $r \in [0,\infty[$, so the AdS boundary is mapped from $r=\infty$ to $\rho=1$. 
Consider an \emph{asymptotically} AdS spacetime, i.e., one described by Eq.~\eqref{EQ_ADS_PDISK} near the AdS boundary.
In this asymptotic region $\rho \to 1$, a bulk geodesic $\gamma_A$ corresponding to a boundary region $A$ will be unaffected by massive deformations further in the bulk, simply following a radial direction (see Fig.\ \ref{FIG_POINCARE_DISK}).
At two different cutoff radii $\rho_1<\rho_2$ close to unity, the subsystem 
length $\ell=|A|$ at each cutoff is given by
\begin{equation}
\ell^{(k)} = \frac{2\alpha \rho_k}{1-\rho_k^2} \Delta\phi\ \approx \frac{\alpha }{1-\rho_k} \Delta\phi\ ,
\end{equation}
where $\Delta\phi$ is the Poincar\'e disk angle subtended by $A$. The difference in geodesic length $|\gamma_A|$ between both cutoffs is given by the lengths of two radial segments:
\begin{align}
|\gamma_A^{(2)}| - |\gamma_A^{(1)}| = 2 \int_{\rho_1}^{\rho_2} \frac{2\alpha}{1-\rho^2} \mathrm{d}\rho &\approx 2\alpha\ln\frac{\ell^{(2)}}{\ell^{(1)}} \ .
\end{align}
Compare this with the entanglement entropy of a conformal field theory for a small subsystem ($\Delta\phi \ll 2\pi$), given by \cite{CalabreseReview}
\begin{equation}
\label{EQ_CALABRESE_CARDY}
S_A = \frac{c}{3} \ln\left( \frac{2\ell}{\Delta\phi\, \epsilon} \sin\frac{\Delta\phi}{2} \right) \approx \frac{c}{3} \ln\frac{\ell}{\epsilon} \ ,
\end{equation}
where $\epsilon$ denotes the lattice spacing and $c$ is the central charge of the CFT. Assuming that the RT prescription holds, 
we recover the Brown-Henneaux formula \cite{Brown:1986nw}
\begin{equation}
\label{EQ_BH}
c = \frac{3\alpha}{2 G} \ .
\end{equation}

\section{Discrete tensor network models}

A natural discretization of an AdS time-slice, or equivalently, the Poincar\'e disk, is given by a \emph{regular} hyperbolic tiling (Fig.\ \ref{FIG_POINCARE_DISK}). A regular $\{n,k\}$ tiling, with $k$ $n$-gon tiles at each vertex, is hyperbolic if the sum of inner angles of each $n$-gon is smaller than $(n-2)\pi$, i.e., when $1/n + 1/k < 1/2$. 
We can relate such a discretized bulk geometry to a quantum state by identifying each $n$-gon tile with a rank $n$ tensor and contracting them over all edges, forming a \emph{tensor network} \cite{Orus-AnnPhys-2014,VerstraeteBig,AreaReview,SchuchReview,Handwaving}. The uncontracted edges on the boundary are then identified as the physical sites of a boundary state.
As the Poincar\'e disk \eqref{FIG_POINCARE_DISK} can contain infinitely many tiles, a prescription for constructing the tiling towards the asymptotic boundary is needed. Starting with a given tile/tensor, we iteratively ``grow'' our geometry by contracting layers of tensors in \emph{inflation steps}, each step corresponding to a UV cutoff.
A discretized boundary region $A$ at such a discrete cutoff does not follow a constant radius $\rho$ in the Poincar\'e disk, with its length $\ell$ being larger than expected for a radial cutoff in the continuum. 
Similarly, as shown in Fig.\ \ref{FIG_POINCARE_DISK}, discretized geodesic cuts $\gamma_A$ no longer follow a smooth curve, with their lengths $|\gamma_A|$ also being larger than in the continuum. 
However, we can still define a discrete analogue to the RT formula \eqref{EQ_RT} as a bound on the entanglement entropy of a boundary region $A$, given by
\begin{equation}
\label{EQ_RT_BOUND}
S_A \leq \frac{|\gamma_A|}{s} \ln \chi \ ,
\end{equation}
with $\partial A=\partial \gamma_A$ and $s$ being the length of each individual edge. 
The \emph{bond dimension} $\chi$ of each tensor index is assumed to be constant throughout the network.

\begin{figure}[tb]
\includegraphics[width=0.5\textwidth]{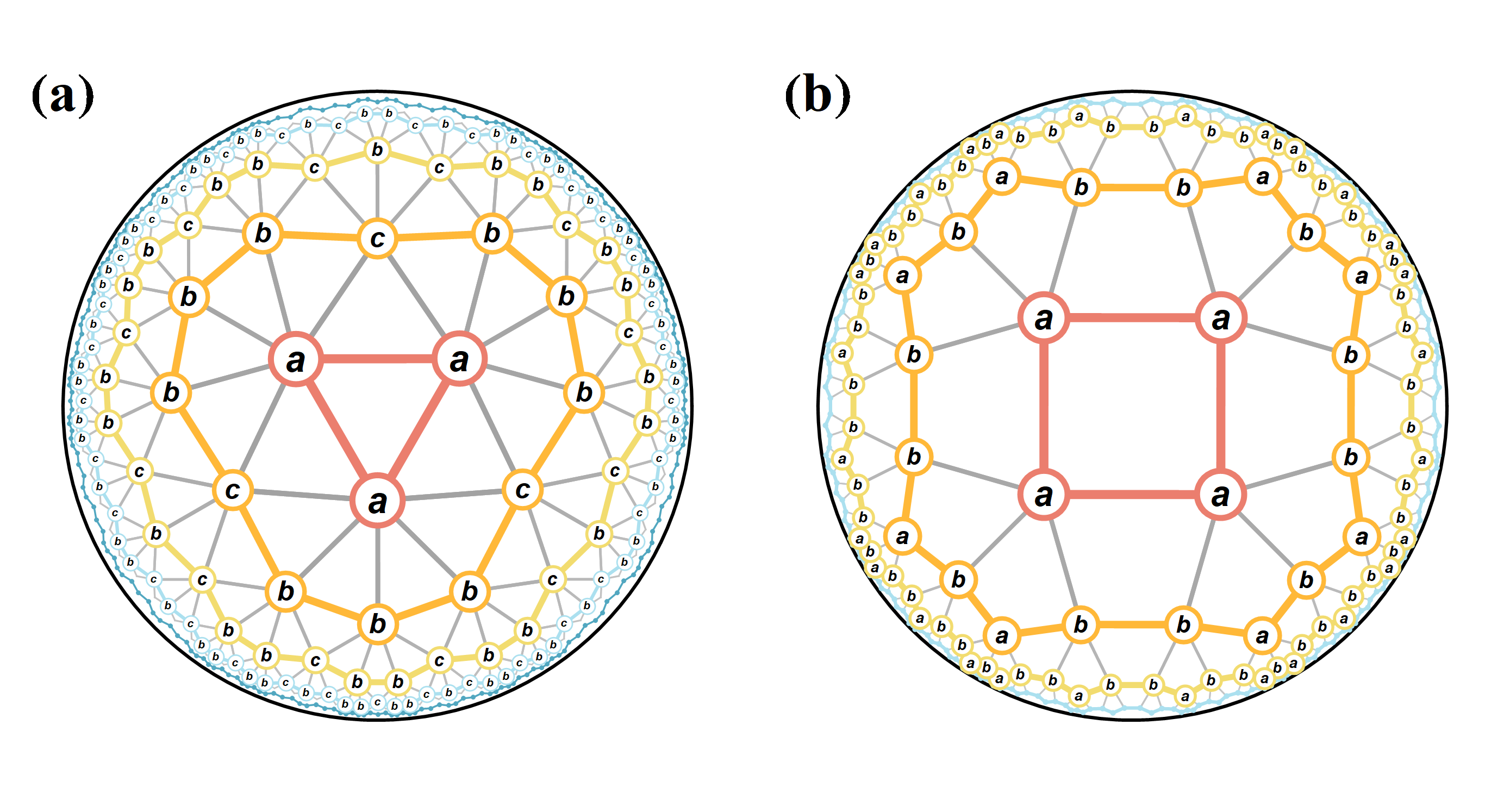}
\caption{Vertex inflation of (a) the $\{3,7\}$ and (b) the $\{4,5\}$ tiling, with vertices labeled by type and each inflation layer colour-coded.}
\label{FIG_VINFLATION}
\end{figure}

In analogy to the previous section, we now derive a discretized form of the Brown-Henneaux formula \eqref{EQ_BH} from the relative growth of boundary and geodesic lengths $\ell=|A|$ and $|\gamma_A|$ under inflation of the tiling.
We specifically consider \emph{vertex inflation}, whereby an inflation step consists of filling each open vertex with tiles. Vertices are labeled by their number of neighbors up to the given inflation level.
First consider the $n=3$ case, the triangular hyperbolic tiling, whose vertex inflation is shown in Fig.~\ref{FIG_VINFLATION} (left). We start with a single triangle with three vertices, each of which has two neighbours. The first inflation step gives each vertex $k{-}2$ additional neighbouring vertices, two of which are shared with its previous neighbours. Thus, the inflation step adds $k{-}3$ new vertices for each old one. After the first inflation step, all boundary vertices have either three or four 
neighbours, two of which are other boundary vertices. 
Denoting vertices with two, three, and four neighbours with the letters $a$, $b$, and $c$, respectively, this pattern is summarized in the \emph{inflation rule}
\begin{align}
a &\mapsto b^{k-4}c \ , &
b &\mapsto b^{k-5}c \ , &
c &\mapsto b^{k-6}c \ ,
\end{align}
where we encode the boundary vertices as a string of $a$, $b$ and $c$, $a^k$ denoting $k$ repetitions of $a$. 
The inflation rule for any hyperbolic $\{n,k\}$ tiling produces a \emph{quasi-regular} sequence \cite{Boyle:2018uiv} exhibiting self-similarity: After sufficiently many inflation steps, any starting sequence will lead to a sequence with the same distribution of letters. In this \emph{steady state} the relative frequency of letters is given by the largest eigenvalue of the \emph{substitution matrix} $M$, where $M_{i,j}$ is the number of $j$ vertices resulting from applying the inflation rule on an $i$ vertex.
For the $\{3,k\}$ tiling, it is given by
\begin{equation}
M =
\begin{pmatrix}
0 & k-4 & 1 \\
0 & k-5 & 1 \\
0 & k-6 & 1
\end{pmatrix} \ .
\end{equation}
Here the rows and columns correspod to $(a,b,c)$ vertices. The largest eigenvalue of $M$,
\begin{equation}
\lambda=\frac{1}{2} \left(\sqrt{k^2-8k+12}+k-4\right) \ ,
\end{equation}
is the scaling factor of the sequence (and sufficiently large subsystems thereof) in the steady state, i.e., after many inflation steps.
The scaling of discrete geodesics can also be computed: Coarse-graining a subsystem $A$ of the sequence by a \emph{deflation} step maps the two vertices that bound $A$ (and a few of its neighbours) onto two vertices at a lower inflation layer. For the $\{3,7\}$ tiling, this corresponds to removing two edges from the geodesic $\gamma_A$, one on either end. 
Thus, the average difference in entanglement entropy between both layers, denoted as $\overline{\Delta S_A}$, is bounded by $2\ln\chi$. Relating this to \eqref{EQ_CALABRESE_CARDY} leads to the central charge bound
\begin{equation}
c_{\{3,k\}} = \frac{3\, \overline{\Delta S_A}}{\ln\lambda} \leq \frac{6\ln\chi}{\ln\frac{\sqrt{k^2-8k+12}+k-4}{2}} =: c^\text{max}_{\{3,k\}} \ . 
\end{equation}
Generalizing this result to arbitrary hyperbolic $\{n,k\}$ tilings leads to further complications. For the $\{4,k\}$ tiling (Fig.\ \ref{FIG_VINFLATION}, right), the vertex inflation rule is
\begin{align}
a &\mapsto b(ab)^{k-3} \ , &
b &\mapsto b(ab)^{k-4} \ .
\end{align}
Again $a$ and $b$ denote  vertices with two and three neighbours up to a given inflation layer. The substitution matrix and its largest eigenvalues are found to be 
\begin{align}
M &=
\begin{pmatrix}
k-3 & k-2 \\
k-4 & k-3 
\end{pmatrix} \ , &
\lambda &= \sqrt{k^2-6 k+8}+k-3 \ .
\end{align}
Unfortunately, the change of geodesic length under deflation now depends on the vertices involved: As we can see in Fig.\ \ref{FIG_VINFLATION} (right), the deflation $a \mapsfrom b$ still only involves moving along one edge, but the deflation $a \mapsfrom a$ involves two. To determine the average change in geodesic length per deflation step, we first compute the left and right eigenvectors of $M$ for the eigenvalue $\lambda$, given by
\begin{align}
\vec{l} &= 
\begin{pmatrix}
\sqrt{8-6k+k^2} \\
k-2
\end{pmatrix} \ ,  &
\vec{r} &= 
\begin{pmatrix}
\sqrt{8-6k+k^2} \\
k-4 \ 
\end{pmatrix}.
\end{align}
When divided by their total sum, the components of $\vec{l}$ give the relative 
frequencies $P(a)$ and $P(b)$ of $a$ and $b$ vertices in the steady state.
This is not a probabilistic process; however, the relative frequencies can be captured on the formal 
level by a \emph{discrete Markov chain}. In this sense, 
we now wish to compute the probability of a deflation step $i \mapsfrom j$. 
Each vertex type corresponds to a state with transition probabilities to other states under a deflation step. After sufficiently many steps, the probability of reaching any given state becomes independent of the starting point.
While $M_{i,j} \propto P(i \mapsto j |i)$ is the (relative) transition probability of reaching a $j$ vertex from an $i$ one, we can construct the \emph{deflation matrix} $D$ giving the probability of the reverse process,
\begin{align}
D_{i,j} = P(i \mapsfrom j |j) &= \frac{P(i \mapsto j|i)P(i)}{\sum_k P(k \mapsto j|k)P(k)} \nonumber\\
&=
\frac{M_{i,j} l_i}{\sum_k M_{k,j} l_k} = \frac{M_{i,j} l_i}{\lambda\, l_j} \ .
\end{align}
The eigenvector $\vec{p}$ of $D$ with eigenvalue $1$ now encodes the average probability of reaching each vertex type through deflation. We find $p_i = l_i r_i$, as
\begin{equation}
\sum_j D_{i,j} p_j = \sum_j \frac{M_{i,j} l_i r_j}{\lambda} = l_i r_i = p_i\ .
\end{equation}
We normalize $\vec{p}$ so that $\sum_i p_i = 1$. 
If an inflation step $i \mapsto j$ adds $E_{i,j}$ edges to a geodesic ending at an $i$ vertex, i.e., adding $E_{i,j}\ln\chi$ to the entanglement bounded by the cut, then the average entanglement entropy loss per deflation step is given by
\begin{equation}
\overline{\Delta S_A} \leq \sum_{i,j} D_{i,j} E_{i,j} p_j \ln\chi = \frac{1}{\lambda} \sum_{i,j} M_{i,j} E_{i,j} l_i r_j \ln\chi \ .
\end{equation}
We thus call $E$ the \emph{entanglement matrix}. The central charge bound for the hyperbolic $\{n,k\}$ tiling thus becomes
\begin{equation}
\label{EQ_C_NK}
c_{\{n,k\}} \leq c^\text{max}_{\{n,k\}} = \frac{6 \sum_{i,j} M_{i,j} E_{i,j} l_i r_j \ln\chi}{\lambda \ln\lambda} \ .
\end{equation}
For the $\{4,k\}$ case, the entanglement matrix is simply
\begin{equation}
E = 
\begin{pmatrix}
1 & 2 \\
1 & 2 
\end{pmatrix} \ ,
\end{equation}
which yields a central charge bound
\begin{equation}
c^\text{max}_{\{4,k\}} = \frac{9 \ln\chi}{\ln \left(\sqrt{k^2-6 k+8}+k-3\right)} \ .
\end{equation}
Eq.\ \eqref{EQ_C_NK} can be used to derive central charge bounds for arbitrary $\{n,k\}$ tilings. For $k>3$, 
the inflation rules are as follows:
\begin{equation}
\label{EQ_REGINF_NK}
\begin{array}{c @{\hskip 1.2cm} c}
n=3: & n>3: \\
a \mapsto b^{k-4}\, c \ , & a \mapsto a^{n-4}\,  b\,  \left( a^{n-3}\, b \right)^{k-3} \ , \\
b \mapsto b^{k-5}\, c \ , & b \mapsto a^{n-4}\,  b\,  \left( a^{n-3}\, b \right)^{k-4} \ . \\
c \mapsto b^{k-6}\, c \ . 
\end{array}
\end{equation}
As before, the letters $a,b,c$ correspond to vertices with two, three, and four neighbors. 
In the $k=3$ case we also require three letters $a,b,c$, where $c$ now denotes a vertex to the right of a $b$-type vertex, leading to
\begin{align}
\label{EQ_REGINF_K3}
a &\mapsto c\, a^{n-5}\, b \ , &
b &\mapsto c\, a^{n-6}\, b \ , &
c &\mapsto \emptyset  \ .
\end{align}
Here $\emptyset$ is the \emph{empty set}, i.e., the letter disappears. 
While \eqref{EQ_REGINF_NK} and \eqref{EQ_REGINF_K3} reproduce the quasi-regular sequences resulting from vertex inflation, these forms are not sufficient to describe the propagation of geodesics for $n>4$. This requires distinguishing vertices by the graph distance of their neighbouring vertices to the center, which determines which paths from one inflation layer to the next correspond to discretized radial geodesics. 
As in the continuous case, where we considered radial geodesics in an asymptotically AdS geometry, our tiling can be non-regular in the center; only the tiling structure near the boundary of the Poincar\'e disk is relevant to the central charge of the boundary state.
The maximum central charges resulting from the full calculation for an arbitrary $\{n,k\}$ tiling are summarized in Tab.\ \ref{TAB_C}. The corresponding inflation rules and matrices $M$ and $E$ are given in the Appendix.

\begin{table*}
\centering
    \renewcommand{\arraystretch}{2.0}
\begin{tabular}{@{} l @{\hspace{0.8cm}}  c  c  c  c  c  c c c @{}}
\widerule
 & \multicolumn{6}{c}{Maximal central charge $c^\text{max}$} & \hspace{0.8cm} & Slope $c^\text{max}/\alpha$ \\
 \cline{2-7} \cline{9-9}
 & $k=3$ & $k=4$ & $k=5$ & $k=6$ & General $k$ & $k \sto \infty$ & & $k \sto\infty$ \\
 \cline{2-7} \cline{9-9}
$n=3$   &  -  &  -  &  -  &  -  & $\frac{6 \ln\chi}{\ln\frac{\sqrt{k^2-8k+12}+k-4}{2}}$ & $\frac{6 \ln\chi}{\ln (k-4)}$ & & $\frac{12 \ln\chi}{s}$ \\
$n=4$   &  -  &  -  &  $\frac{9 \ln\chi}{\ln (\sqrt{3}+2)}$  &  $\frac{9 \ln\chi}{\ln (2 \sqrt{2}+3)}$  &  $\frac{9 \ln\chi}{\ln (\sqrt{k^2-6 k+8}+k-3)}$  & $\frac{9 \ln\chi}{\ln (2k-6)}$  & & $\frac{18 \ln\chi}{s}$    \\
$n=5$   &  -  &  $\frac{10 \ln\chi}{\ln (\sqrt{3} + 2)}$  &  $\frac{10 \ln\chi}{\ln \frac{3 \sqrt{5}+7}{2}}$  &  $\frac{10 \ln\chi}{\ln \frac{4 \sqrt{6}+10}{2}}$   &  $\frac{10 \ln\chi}{\ln\frac{\sqrt{9 k^2-48 k+60}+3 k-8}{2}}$  &    $\frac{10 \ln\chi}{\ln (3k-8)}$  & & $\frac{20 \ln\chi}{s}$ \\
$n=6$   &  -  &  $\frac{12 \ln\chi}{\ln (2 \sqrt{2}+3)}$  &  $\frac{12 \ln\chi}{\ln (2 \sqrt{6}+5)}$  &  $\frac{2 \ln\chi}{\ln (4 \sqrt{3}+7)}$  &  $\frac{12 \ln\chi}{\ln (2 \sqrt{k^2-5 k+6}+2 k-5)}$  &  $\frac{12 \ln\chi}{\ln (4k-10)}$ & & $\frac{24 \ln\chi}{s}$ \\
$n=7$   &  $\frac{66 \ln\chi}{5 \ln\frac{3+\sqrt{5}}{2}}$  &  $\frac{66 \ln\chi}{5 \ln (\sqrt{15}+4)}$  &  $\frac{66 \ln\chi}{5 \ln \frac{\sqrt{165}+13}{2}}$  & $\frac{66 \ln\chi}{5 \ln (\sqrt{15}+4)}$  & $\frac{66 \ln\chi}{5 \ln \frac{5 k-12+\sqrt{(5k-10) (5 k-14)}}{2}}$ & $\frac{66 \ln\chi}{5 \ln (5k-12)}$ &  & $\frac{132 \ln\chi}{5 s}$ \\
$n=8$   &  $\frac{15 \ln\chi}{\ln (\sqrt{3}+2)}$ & $\frac{15 \ln\chi}{\ln (2 \sqrt{6}+5)}$ & $\frac{15 \ln\chi}{\ln (3 \sqrt{7}+8)}$ & $\frac{15 \ln\chi}{\ln (2 \sqrt{30}+11)}$ & $\frac{15 \ln\chi}{\ln (\sqrt{9 k^2-42 k+48}+3 k-7)}$ & $\frac{15 \ln\chi}{\ln (6k-14)}$ &  & $\frac{30 \ln\chi}{s}$ \\
$n=9$   &  $\frac{114 \ln\chi}{7\ln\frac{5+\sqrt{21}}{2}}$ & $\frac{114 \ln\chi}{7 \ln (\sqrt{35}+6)}$ & $\frac{114 \ln\chi}{119\ln\frac{7+\sqrt{357}}{2}}$ & $\frac{114 \ln\chi}{7\ln (2 \sqrt{42}+13)}$ & $\frac{114 \ln\chi}{7 \ln\frac{7 k+\sqrt{(16-7 k)^2-4}-16}{2}}$ & $\frac{114 \ln\chi}{7 \ln (7k-16)}$  & & $\frac{228 \ln\chi}{7 s}$ \\
$n \sto \infty$   &  $\frac{3 (n+2) \ln\chi}{2 \ln (n-4)}$ & $\frac{3 (n+2) \ln\chi}{2 \ln (2 (n-3))}$ & $\frac{3 (n+2) \ln\chi}{2 \ln (3 n-8)}$ & $\frac{3 (n+2) \ln\chi}{2 \ln (4 n-10)}$ & $\frac{3 (n+2) \ln\chi}{2 \ln ((n-2)(k-2) -2)}$ & $\frac{3 n \ln\chi}{2 \ln (n k)}$  & & $\frac{3 n \ln\chi}{s}$ \\
\widerule
\end{tabular}
\caption{Maximal central charges $c^\text{max}$ for the boundary state of a bond dimension $\chi$ tensor network embedded into a vertex-inflated regular $\{n,k\}$ tiling. 
The last column contains the slope of $c^\text{max}$ with respect to the AdS radius $\alpha$, given in terms of the geodesic edge length $d$.
Full derivations are given in Appendix \ref{APP_GEODESICS}.}
\label{TAB_C}
\end{table*}

\section{Curvature of regular tilings}

\begin{figure}[tb]
\includegraphics[width=0.3\textwidth]{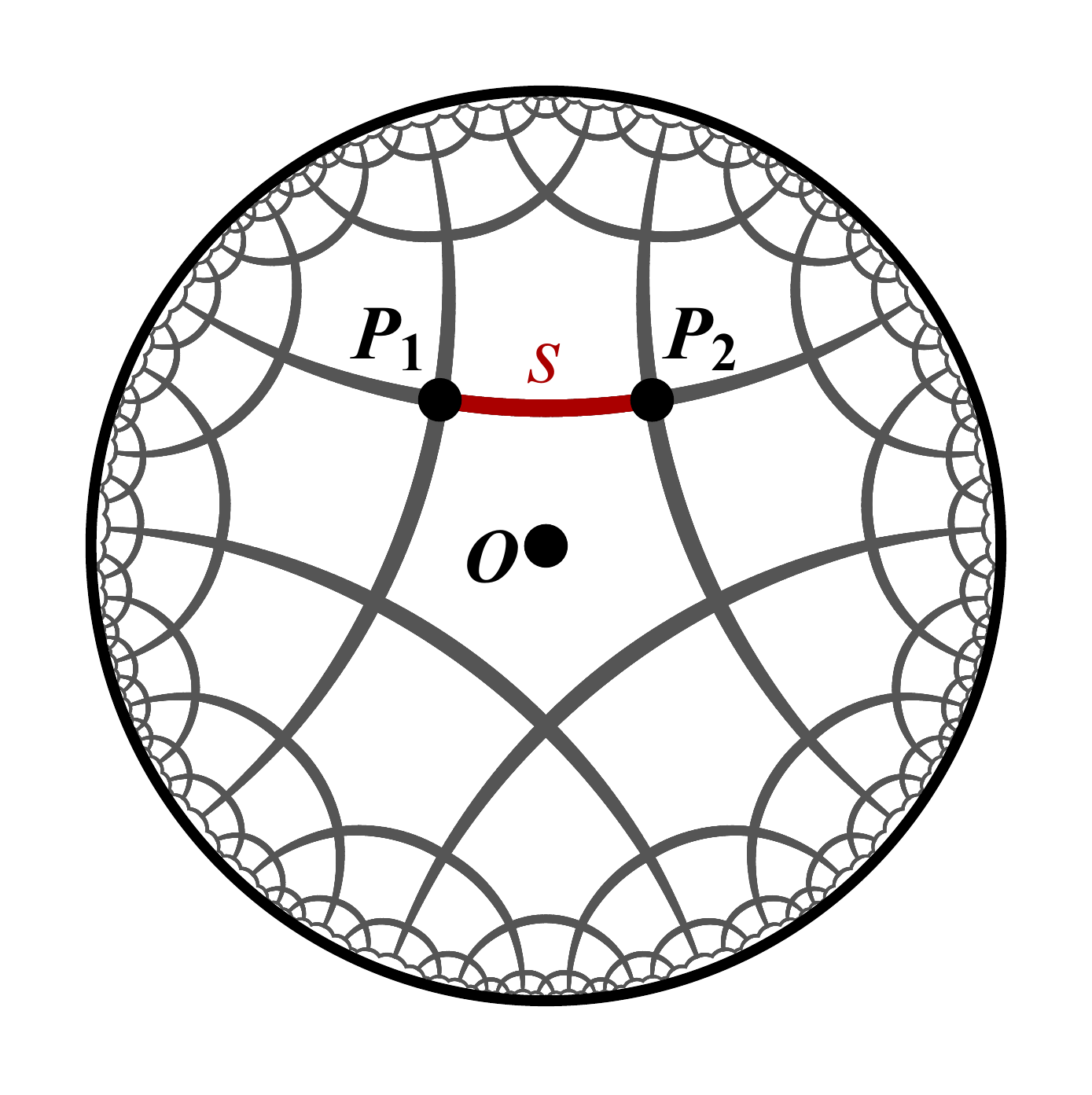}
\caption{Sketch of a $\{5,4\}$ tiling in the Poincar\'e disk with three reference points and one edge marked.}
\label{FIG_SKETCH}
\end{figure}

An $\{n,k\}$ tiling embedded into the Poincar\'e disk is constructed of identical $n$-gons with an angle of $2\pi/k$ at each corner (see Fig.\ \ref{FIG_SKETCH}).
The geodesic length $P_1 P_2 = s$ between two points $P_1$ and $P_2$ of the tiling determines the length between all other points in the tiling. The parameters $n$ and $k$ further fix the angles $\beta = \measuredangle(O P_1,O P_2)= 2\pi/n$ and $\gamma = \measuredangle(P_1 P_2, P_1 O)= \measuredangle(P_2 O, P_2 P_1)= \pi/k$. The hyperbolic law of cosines then states that
\begin{equation}
\label{EQ_HYP_LAW_COS}
\cos\beta = - \cos^2\gamma + \sin^2\gamma\, \cosh\frac{s}{\alpha}\ .
\end{equation}
Note that this form of the law of cosines holds for a \emph{Gaussian curvature} $K=R/2=-1/\alpha^2$ of the time-slice metric. Using this relation we can now express the AdS$_3$ radius in terms of the tiling parameters as
\begin{equation}
\frac{s}{\alpha} = 2 \arcosh\left( \frac{\cos\frac{\pi}{n}}{\sin\frac{\pi}{k}} \right) = 2\ln\left(\frac{2k}{\pi} \cos\frac{\pi}{n} \right) + O(k^{-2}) \ .
\end{equation}
Thus, $s/\alpha$ diverges logarithmically in the large $k$ limit. Note that the hyperbolic area $A = \alpha^2 (n - 2n/k - 2)$ is finite in this limit.

\begin{figure}[tb]
\includegraphics[width=0.45\textwidth]{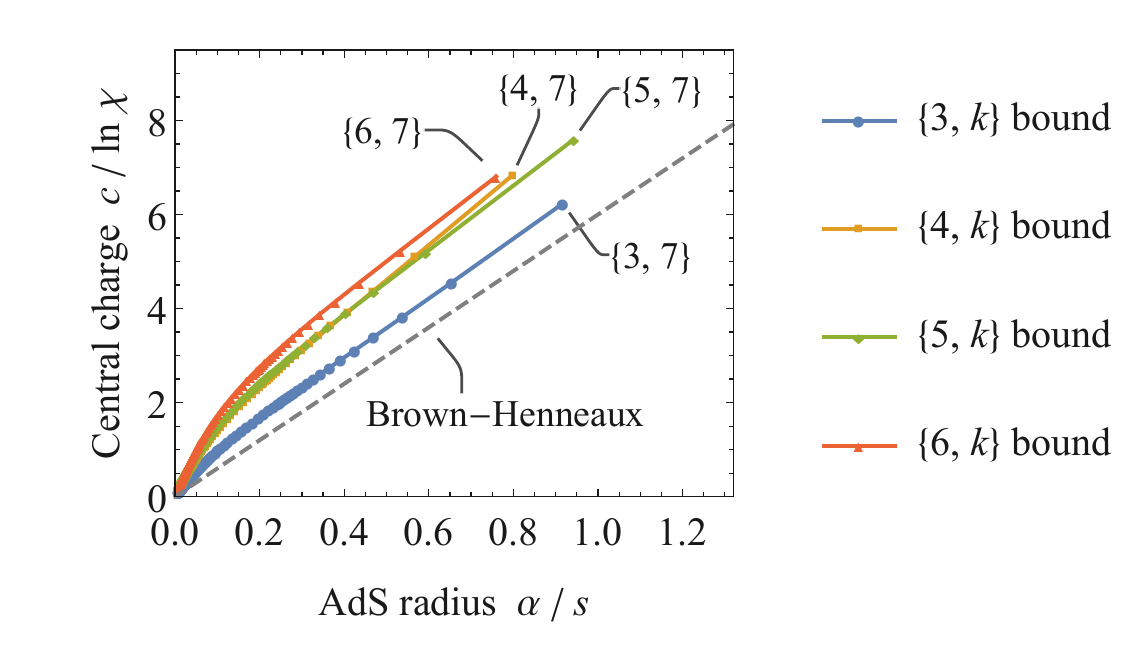}
\caption{Central charge bounds and AdS radii for $\{n,k\}$ tilings, with the continuum Brown-Hennaux formula for $G=s/4 \ln\chi$ shown as a dashed line. The data series start at $k=7$ for $n=3$, $k=5$ for $n=4$, and $k=4$ for both $n=5$ and $n=6$ (first data point of each series in the upper-right corner).
}
\label{FIG_CBOUNDS}
\end{figure}

We can now directly relate the previously derived bounds on central charges $c$ to the AdS radius $\alpha$ of the corresponding AdS geometry, with the results for various choices of $n$ shown in Fig.\ \ref{FIG_CBOUNDS}.
These bounds can be compared to the continuum Brown-Henneaux prescription \eqref{EQ_BH}, with the gravitational constant $G$ fixed through the RT formula: The length of a discretized minimal geodesic $\gamma_A$ corresponding to a boundary region $A$ can be written as $|\gamma_A| = N s$, where $N$ is the number of edges that $\gamma_A$ consists of (note that $N \to \infty$ in the asymptotic limit). As each edge contributes $\ln\chi$ to $S_A$, we find
\begin{equation}
S_A = \frac{|\gamma_A|}{4 G} = \frac{N s}{4 G} \overset{!}{=} N \ln\chi\ .
\end{equation}
We can thus rewrite \eqref{EQ_BH} as
\begin{equation}
\label{EQ_BH2}
c^\text{max} = \frac{6 \alpha \ln\chi}{s}  \ .
\end{equation}
Comparing this to the behaviour of boundary states of $\{n,k\}$ tilings in Fig.\ \ref{FIG_CBOUNDS}, we find that these bounds are always \emph{above} \eqref{EQ_BH2}. This implies that tensor networks with the same bulk curvature and entanglement entropy growth as a continuum model can always be constructed by choosing appropriate tensors.
Furthermore, we find a linear regime at large $k$ in all tilings with the slope depending on $n$. For example,
\begin{align}
\lim_{k \to \infty} \frac{c_{\{3,k\}}^\text{max}\, s}{\alpha_{\{3,k\}} \ln\chi} &= 12 \ , &
\lim_{k \to \infty} \frac{c_{\{4,k\}}^\text{max}\, s}{\alpha_{\{4,k\}} \ln\chi} &= 18 \ .
\end{align}
The general coefficients are given in Tab.\ \ref{TAB_C}. Note that they are significantly larger than the continuum value at small curvature, and increase monotonically with $n$.
At small $k$, a second linear regime appears, with a slope much closer to the Brown-Henneaux form, e.g.
\begin{equation}
\frac{c_{\{3,7\}}^\text{max}-c_{\{3,8\}}^\text{max}}{\alpha_{\{3,7\}}-\alpha_{\{3,8\}}} \approx 6.38 \frac{\ln\chi}{s} \ .
\end{equation}
As a tiling of lower curvature is a better approximation of a continuous geometry, a result closer to the BH formula is not unexpected; however, fixing $n$ while varying $k$ appears to produce a central charge shift relative to the BH result that remains constant for a large range of $k$, even as the curvature increases significantly.

\section{Strong disorder renormalization}

Having established the previous bounds on entanglement entropy asymptotics, we will consider cases when the central 
charge can be calculated exactly. Interestingly, the method that allows for such an exact calculation is deeply related to a very 
early approach to real-space renormalization group transformations that were originally introduced in Ref.\
 \cite{PhysRevLett.43.1434}  and later extended in Ref.\ \cite{PhysRevLett.69.534} to study the ground states, 
low-energy excitations and spatio-temporal correlations of random quantum spin chains. This technique, called the 
\emph{strong disorder renormalization group (SDRG)} \cite{StrongDisorder} 
has recently again gained considerable attention due to its role in studying 
\emph{many-body localization} \cite{MBL14}, 
quantum critical \emph{Floquet dynamics} \cite{Berdanier_2018} and models with highly 
area-law breaking ground states (rainbow states) \cite{Alba:2018ime}, see Ref.\ \cite{IgloiSDRGreview} 
and reference therein for recent development.

We now describe the basic results of SDRG on some aperiodic singlet models that share the quasi-regular symmetries of 
the boundary states described previously. One example is given by the \emph{Fibonacci XXZ chain} that is defined by the Hamiltonian
\begin{align}
H&=\sum_i J_i (S_i^xS_{i+1}^x+S_i^yS_{i+1}^y+\Delta S_i^zS_{i+1}^z),
\end{align}
where $S^{\alpha}_i$ (with $\alpha=x,y,z$) refers to spin-${1\over 2}$ operators.
The site-dependent couplings  $J_b > J_a>0$ are modulated according to 
the aperiodic Fibonacci sequence obtained from the inflation rule
\begin{align}
a &\mapsto ababa \ , &
b &\mapsto aba \ .
\end{align}
The SDRG procedure predicts that for this aperiodic Hamiltonian the ground state (in the large system size limit) 
is characterized by fully entangled pairs of sites \cite{JuhaszZimboras2007,Igloi_2007}.   
For example, inflating 
the letter $b$ twice leads to a Hamiltonian 
with the ground state given by
\begin{equation}
\begin{gathered}
\includegraphics[width=0.22\textwidth]{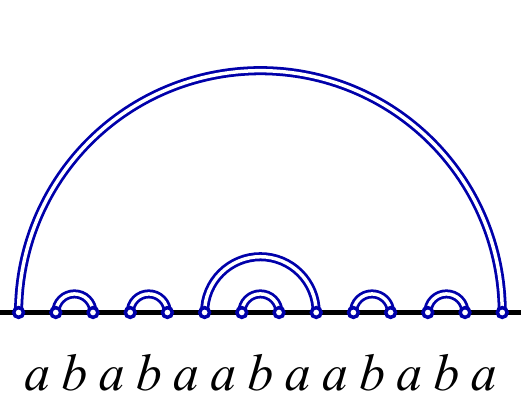}
\end{gathered}
\end{equation}
where each double-line denotes a singlet bond.
The entanglement entropy of a subsystem $A$ of such a singlet state is simply computed by counting the number of singlets connecting it to its complement $A^\text{C}$. For example, in the state
\begin{equation}
\begin{gathered}
\includegraphics[width=0.27\textwidth]{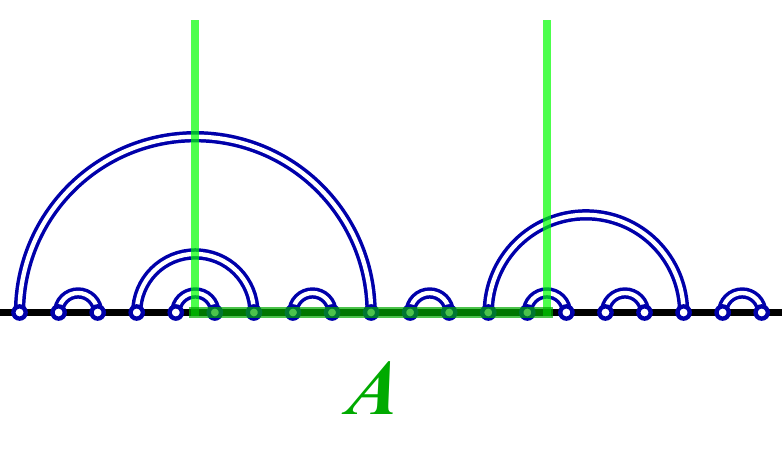}
\end{gathered}
\end{equation}
we find five singlets passing through the cuts between $A$ and $A^\text{C}$, resulting in an entanglement entropy $S_A = 5 \ln 2$.

\begin{figure}[tb]
\includegraphics[width=0.5\textwidth]{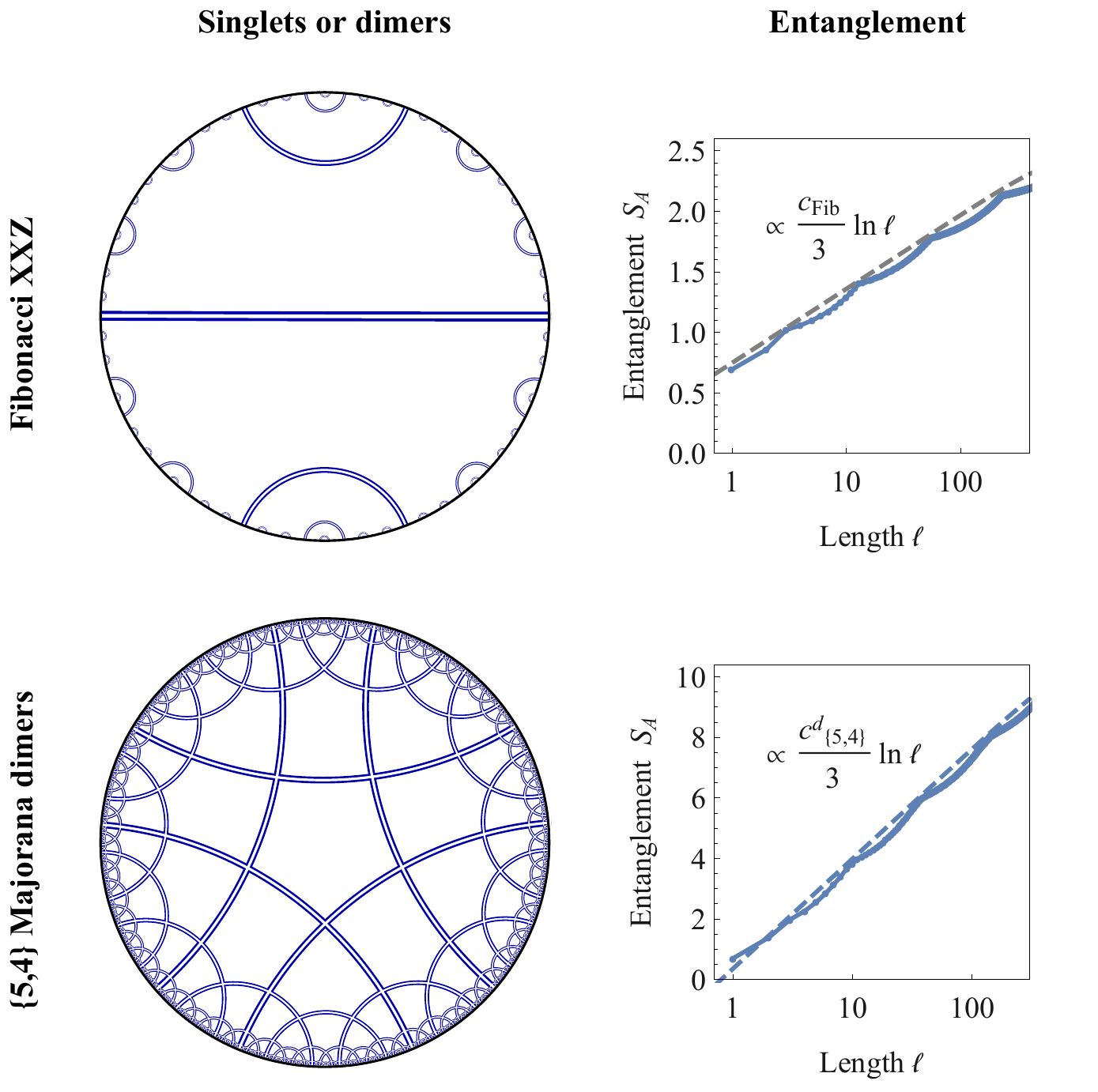}
\caption{Fibonacci singlets (top) and $\{5,4\}$ HyPeC Majorana dimers (bottom) shown in a disk projection along with their corresponding scaling of entanglement entropy $S_A$ with subsystem size $\ell=|A|$. 
The translation-invariant form \cite{CalabreseReview} with effective central charge $c=c_\text{Fib}$ from \eqref{EQ_C_FIB} and $c=c_{\{5,4\}}^\text{d}$ from \eqref{EQ_C_54_VERTEX} are shown as dashed curves.
}
\label{FIG_SINGLETS_VS_DIMERS}
\end{figure}

Applying the SDRG procedure to this model \cite{JuhaszZimboras2007}, it follows that one can systematically obtain the ground state 
corresponding to the Hamiltonian after inflating the letter $b$ for $n$ times by iterating the inverse of 
the renormalization steps, giving rise to the inflation rules

\begin{align}
\label{EQ_FIB_INFL_1}
\begin{gathered}
\includegraphics[height=0.1\textwidth]{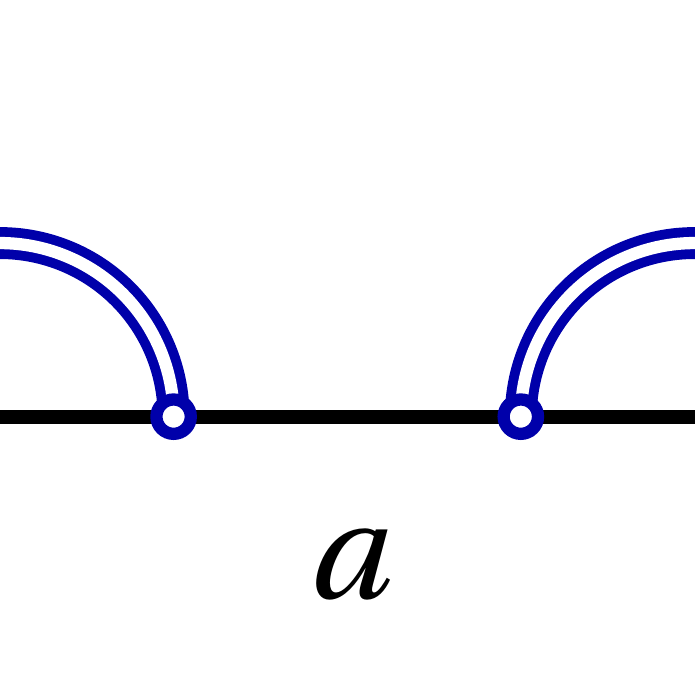}
\end{gathered}
&\scalebox{1.25}{$\quad\to\quad\;$}
\begin{gathered}
\includegraphics[height=0.1\textwidth]{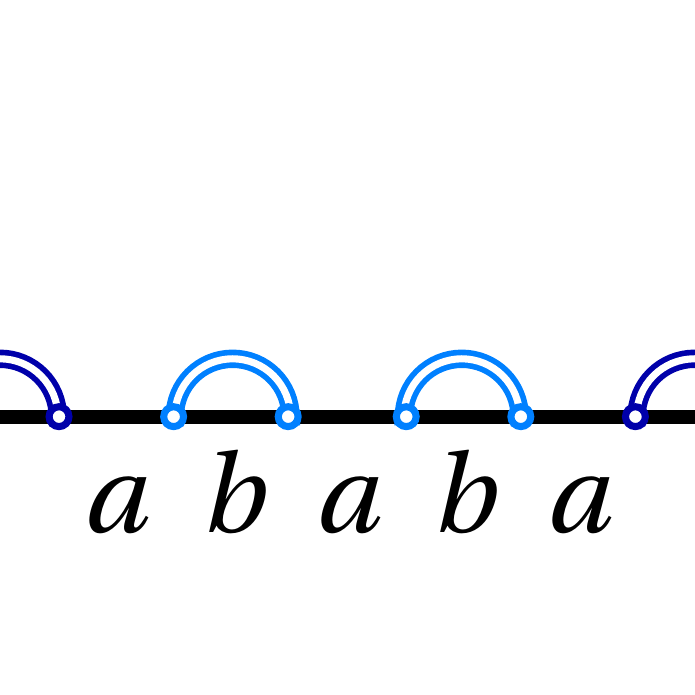}
\end{gathered} \ ,\\
\label{EQ_FIB_INFL_2}
\begin{gathered}
\includegraphics[height=0.1\textwidth]{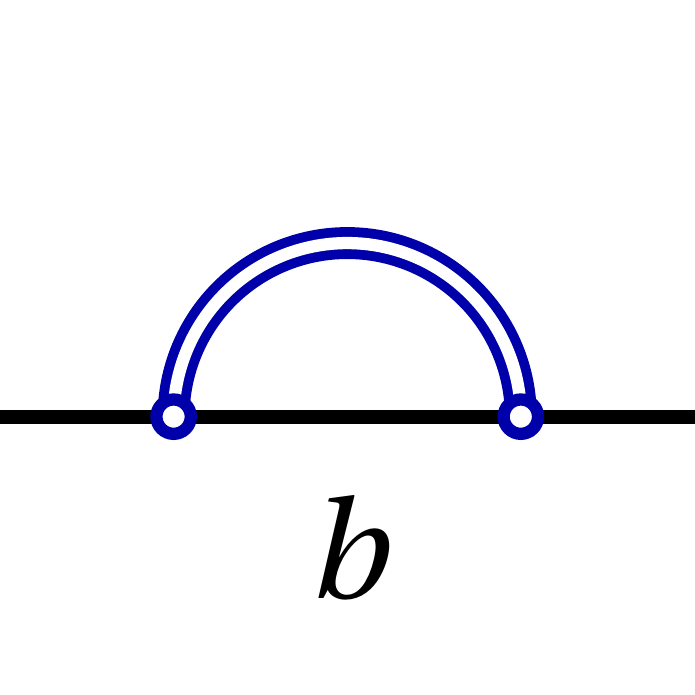}
\end{gathered}
&\scalebox{1.25}{$\quad\to\quad\;$}
\begin{gathered}
\includegraphics[height=0.1\textwidth]{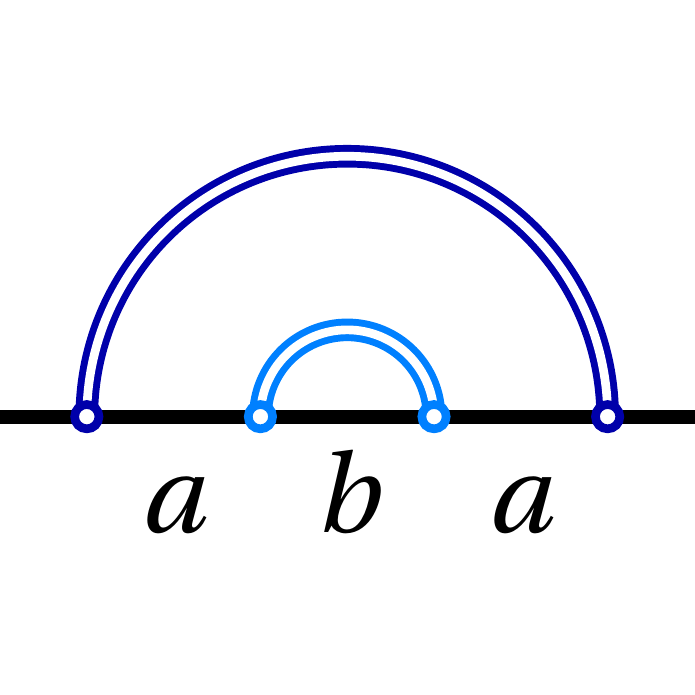}
\end{gathered} \ .
\end{align}
Newly added singlets are highlighted.
Intuitively, the fine-graining effected by the inflation rule places two sites within a singlet 
closer to each other than uncoupled neighbors.
Iteratively applying the inflation rule creates states with nonlocal entanglement. 
Each inflation step thus adds short-range entanglement on successively fine-grained scales, 
similar to the \emph{entanglement renormalization} produced by the MERA \cite{Vidal:2007hda}. 
Due to its \emph{strong disorder}, i.e., strong aperiodicity on all scales, this singlet model can be inflated \emph{locally} without changing the quasi-regular symmetry of the state.
For sufficiently large regions $A$, the dependency of the entanglement entropy on the region size $|A|$ 
can be computed from the aperiodic symmetries themselves and the central charge can be
obtained analytically \cite{JuhaszZimboras2007}.
For example, the Fibonacci XXZ chain has an effective central charge 
\begin{equation}
\label{EQ_C_FIB}
c_\text{Fib} = \frac{\left(3 - \frac{3}{\sqrt{5}} \right) \ln 2}{\ln(2 + \sqrt{5})} \approx 0.7962 \ . 
\end{equation}
The corresponding entanglement scaling is shown in Fig.\ \ref{FIG_SINGLETS_VS_DIMERS} (top), showing the result of seven successive application of the inflation rules \eqref{EQ_FIB_INFL_1} and  \eqref{EQ_FIB_INFL_2} onto one singlet.
Rather than a smooth logarithmic entanglement scaling with $\ell=|A|$ that we find in translation-invariant critical systems, a characteristic feature of multiscale aperiodicity is a linear growth of $S_A$ in fixed intervals of $\ln\ell$, with $S_A$ at the endpoints of these intervals growing logarithmically. 

In the next section, we give concrete examples of such multiscale aperiodic models as the boundary states of regular hyperbolic tilings and calculate their exact central charges.
Distinct from usual singlet models, we consider \emph{fractionalized} fermionic modes with large effective central charges.
The entangled pairs in these models exhibit crossing, requiring a new approach to computing their entanglement entropies.

\section{Majorana dimer models}

An efficiently contractible class of tensor networks with a holographic interpretation is given by \emph{Majorana dimer} states \cite{Jahn:2019nmz}: This versatile class of states corresponds to the intersection of 
stabilizer and free fermionic states; as part of the latter, they can also be efficiently described by \emph{matchgate} 
tensor networks \cite{Jahn:2017tls}.
In particular, the \emph{hyperbolic pentagon code} (HyPeC), a toy model of holographic quantum error correction \cite{Pastawski2015}, can be expressed in this form. 
This model is based on the $[[5,1,3]]$ stabilizer code, which encodes one logical qubit in five physical spins with a code distance of three, i.e., correcting one Pauli error \cite{Bennett:1996gf,Laflamme1996}. 
The tensors corresponding to its encoding isometry between logical and physical states are now contracted along a $\{5,4\}$ tiling, with each pentagon edge corresponding to a physical tensor leg. After contraction of this bulk tensor network, both the physical legs on the boundary of the contracted geometry as well as one logical leg per pentagon remain uncontracted. The whole tensor network thus serves as an isometry between bulk and boundary sites, with the bulk information fault-tolerantly encoded into the boundary as expected from continuum AdS/CFT \cite{Almheiri15}. 
While this model was originally formulated in terms of spin degrees of freedom, it can be mapped to fermions under a Jordan-Wigner transformation mapping Pauli operators $(\sigma^x,\sigma^y,\sigma^z)$ to Majorana operators $\m_k$, obeying $\{\m_j, \m_k\} = 2 \delta_{j,k}$, via
\begin{align}
\label{EQ_JW_TRANS}
{\m}_{2k-1} &=
   ({\sigma^z})^{{\otimes}(k-1)}{\otimes}\,\sigma^y\,{\otimes}\left( \id_2 \right)^{\otimes(r-k)} \text{ ,}\\
{\m}_{2k}  &=
   ({\sigma^z})^{{\otimes}(k-1)}{\otimes}\,\sigma^x\,{\otimes}\left( \id_2 \right)^{\otimes(r-k)} \text{ .}
\end{align}
Note that we have swapped the definition of even and odd operators with respect to Ref.\ \cite{Jahn:2019nmz} to simplify the following visualizations.
In this effective fermionic language, the two logical basis states $\bar{0}$ and $\bar{1}$ that span the logical qubit space of the $[[5,1,3]]$ code become Gaussian, as the stabilizer Hamiltonian is quadratic in Majorana operators when the parity is fixed.
Furthermore, these basis states are composed of paired Majorana modes -- Majorana dimers -- and can be represented graphically as
\begin{align}
\label{EQ_HYPEC_STATES}
\ket{\bar{0}}_5\; &= \,
\begin{gathered}
\includegraphics[height=0.1\textheight]{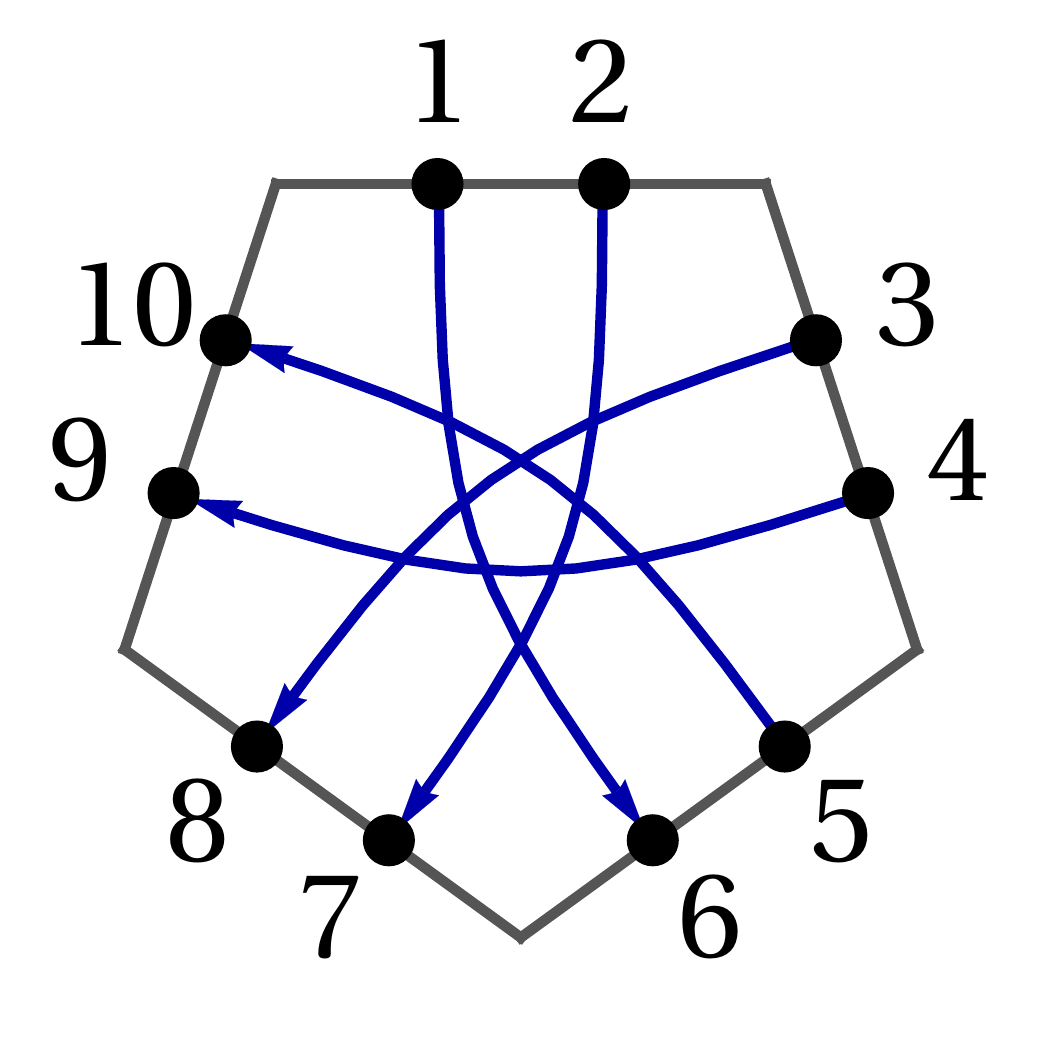}
\end{gathered} \ , &
\ket{\bar{1}}_5\; &= \,
\begin{gathered}
\includegraphics[height=0.1\textheight]{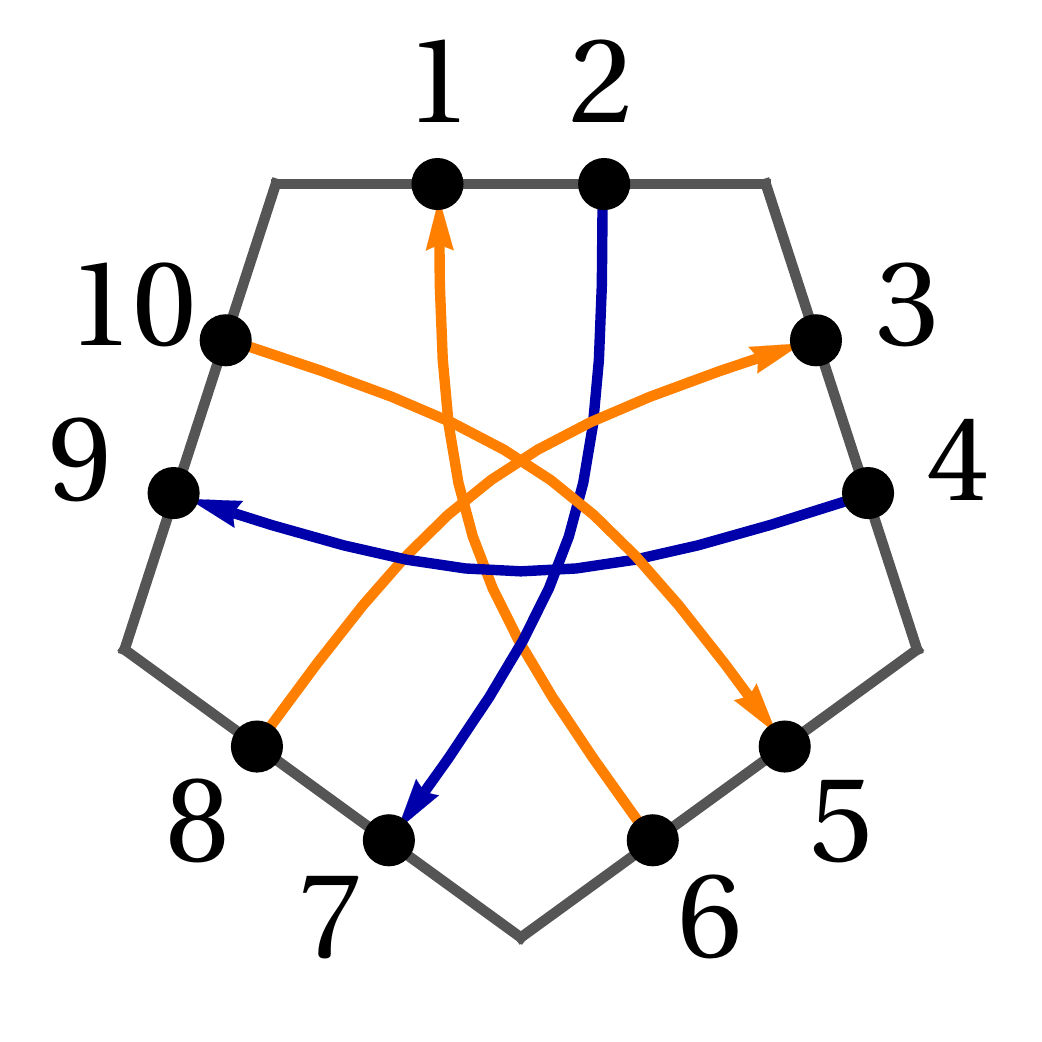}
\end{gathered} \ .
\end{align}
In this visualization, each edge of a pentagon tile is identified with two Majorana modes, with each arrow $j \to k$ between two modes $j$ and $k$ corresponding to a term $\i\m_j\m_k$ in the stabilizer Hamiltonian. The orientation of each arrow relative to the mode ordering gives it an associated \emph{dimer parity} $p_{j,k}$, with $p_{j,k}={+}1$ for $j<k$ (blue) and $p_{j,k}={-}1$ for $j>k$ (orange).
The dimer pattern becomes clearer when exchanging the ordering of odd and even Majorana modes, leading to
\begin{align}
\label{EQ_HYPEC_STATES2}
\ket{\bar{0}^\prime}_5\; &= \,
\begin{gathered}
\includegraphics[height=0.1\textheight]{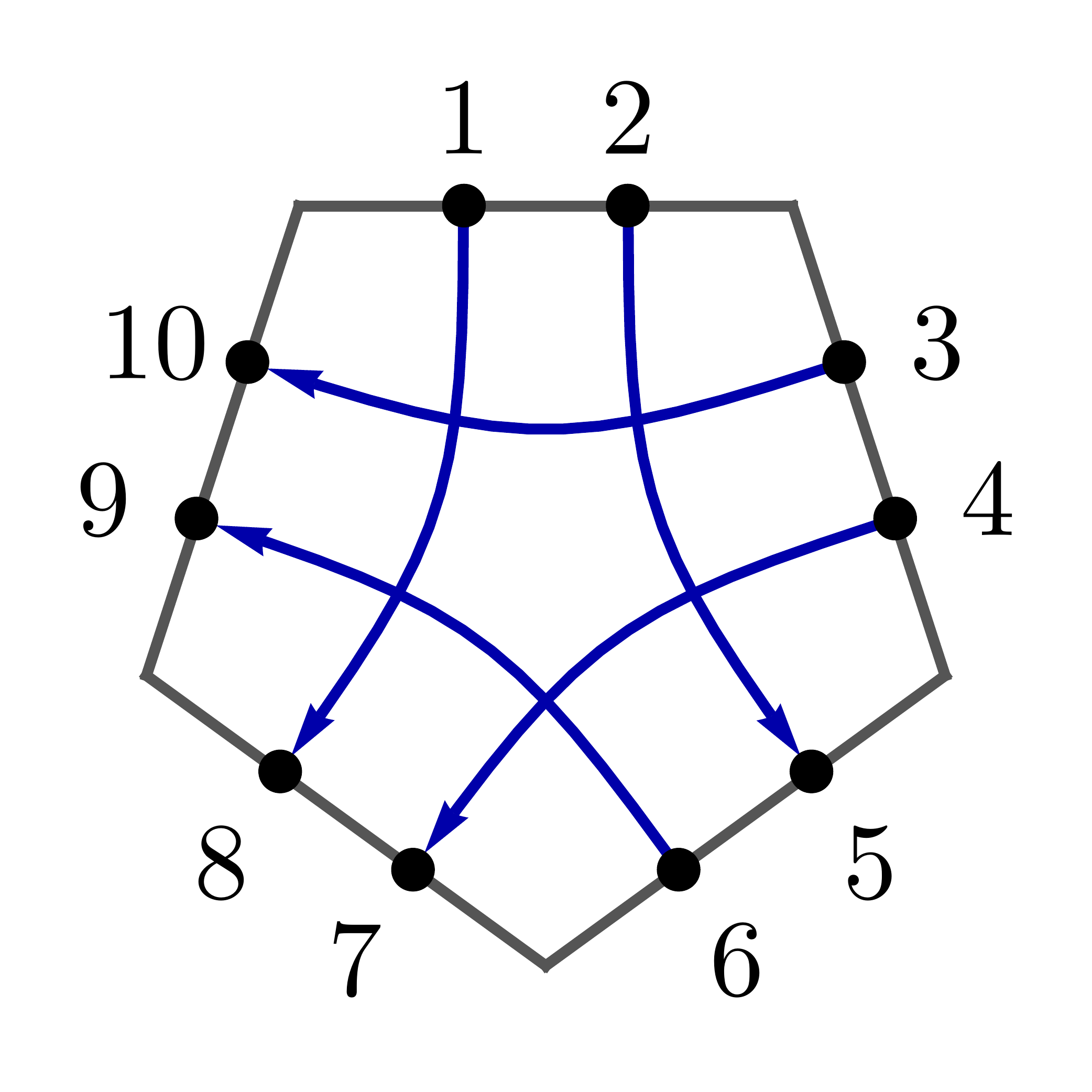}
\end{gathered} \ , &
\ket{\bar{1}^\prime}_5\; &= \,
\begin{gathered}
\includegraphics[height=0.1\textheight]{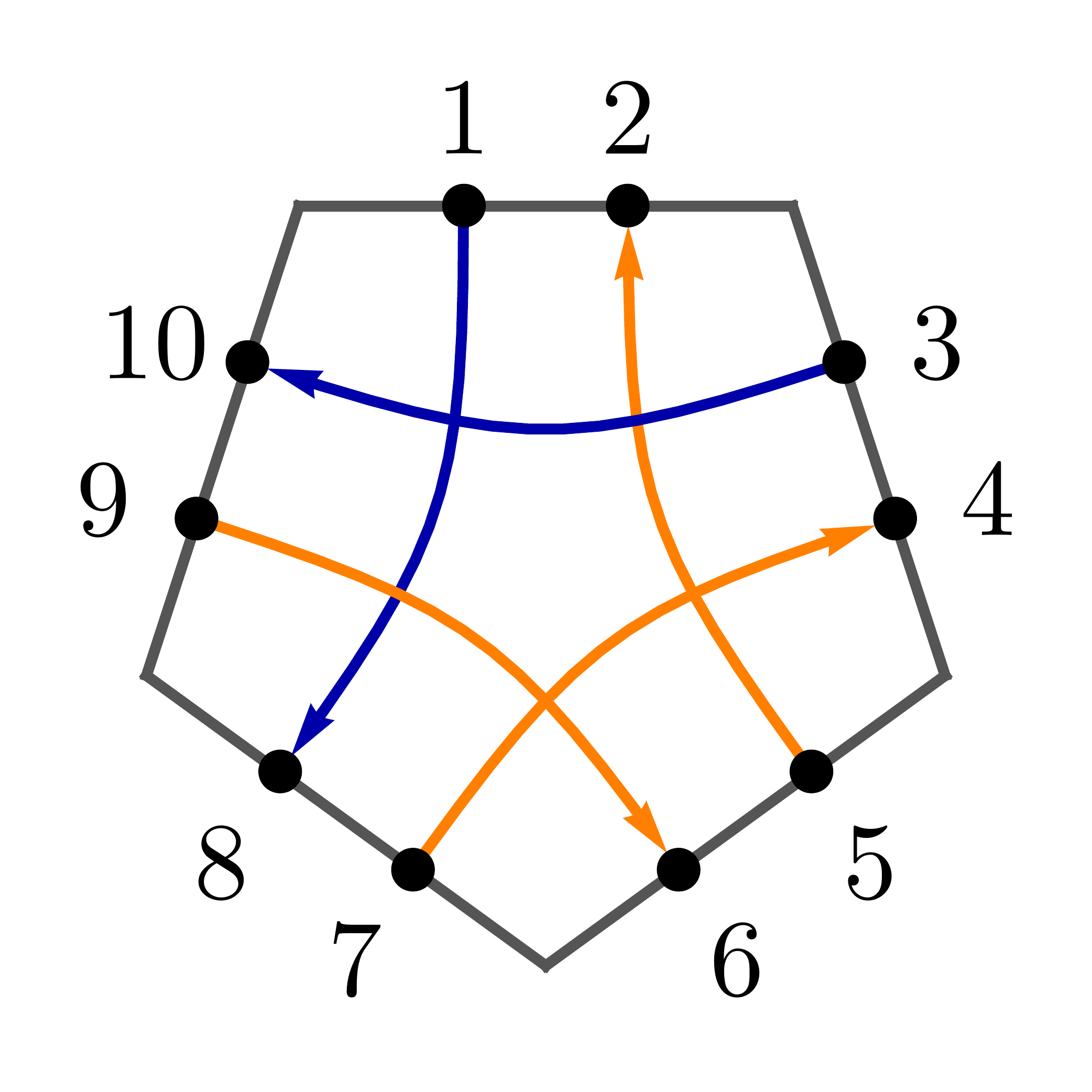}
\end{gathered} \ .
\end{align}
We will use this dual representation in all bulk plots, as it makes the dimer paths along the tiling more apparent.
The usefulness of the Majorana dimer picture comes from the simplicity of contracting tensors representing Majorana dimer states such as \eqref{EQ_HYPEC_STATES}: Contraction pairs up dimers along the contracted edges, with each new dimer's parity being the product of the old parities \cite{Jahn:2019nmz}. 
In addition, computing the entanglement entropy $S_A$ of a connected subsystem $A$ of a Majorana dimer state (or contraction thereof) reduces to simply counting the dimers between $A$ and its complement $A^\text{C}$, each contributing $\ln(2)/2$ to $S_A$.
While the HyPeC is generally composed of arbitrary logical bulk states, i.e., superpositions of $\bar{0}$ and $\bar{1}$, orthogonality conditions between the contracted states ensure that two-point correlation functions still exhibit a dimer structure, i.e., vanishing correlations between Majorana modes unconnected by dimers. Similarly, basis state superpositions affect the entanglement entropies of the HyPeC by corrections that depend on the logical states in \emph{residual bulk regions} only for certain boundary regions \cite{Pastawski2015}. 

Beyond the HyPeC, whose logical states on each tile are represented by \emph{perfect tensors} that maximally entangle each possible subsystem $A$ with the remaining sites, Majorana dimer states also represent \emph{block perfect tensors}, where this condition is relaxed to only hold for connected subsystems. A suitable pair of logical eigenstates $\bar{0}_n$ and $\bar{1}_n$ can be found for any $n=4m+1,m \in \mathbb{N}$. For instance, for $n=9$ the logical basis is given by 
\begin{align}
\label{EQ_HYNEC_STATES}
\ket{\bar{0}}_9\; &= \;
\begin{gathered}
\includegraphics[height=0.1\textheight]{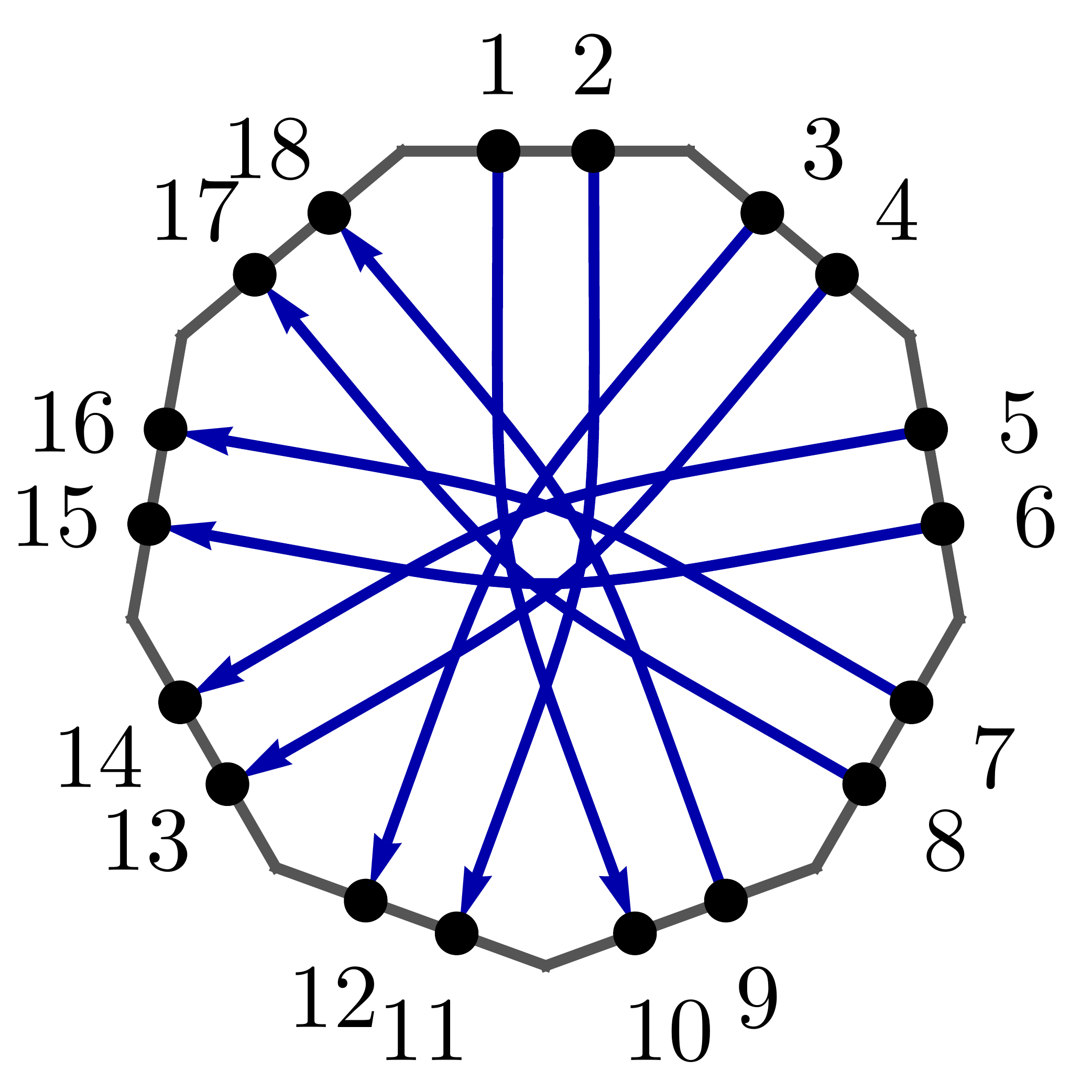}
\end{gathered} \ , &
\ket{\bar{1}}_9\; &= \;
\begin{gathered}
\includegraphics[height=0.1\textheight]{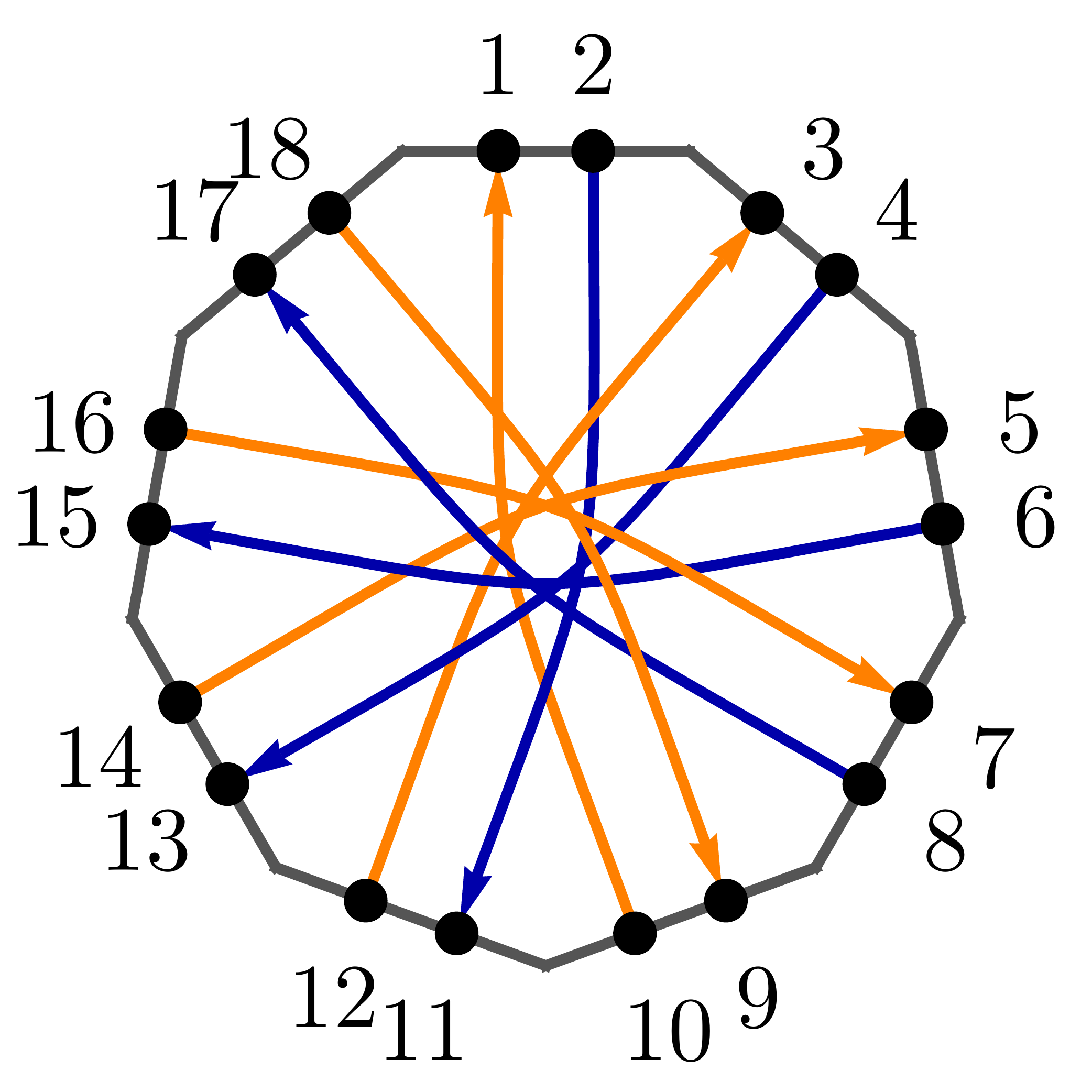}
\end{gathered} \ ,
\end{align}
or equivalently,
\begin{align}
\label{EQ_HYNEC_STATES2}
\ket{\bar{0}^\prime}_9\; &= \;
\begin{gathered}
\includegraphics[height=0.1\textheight]{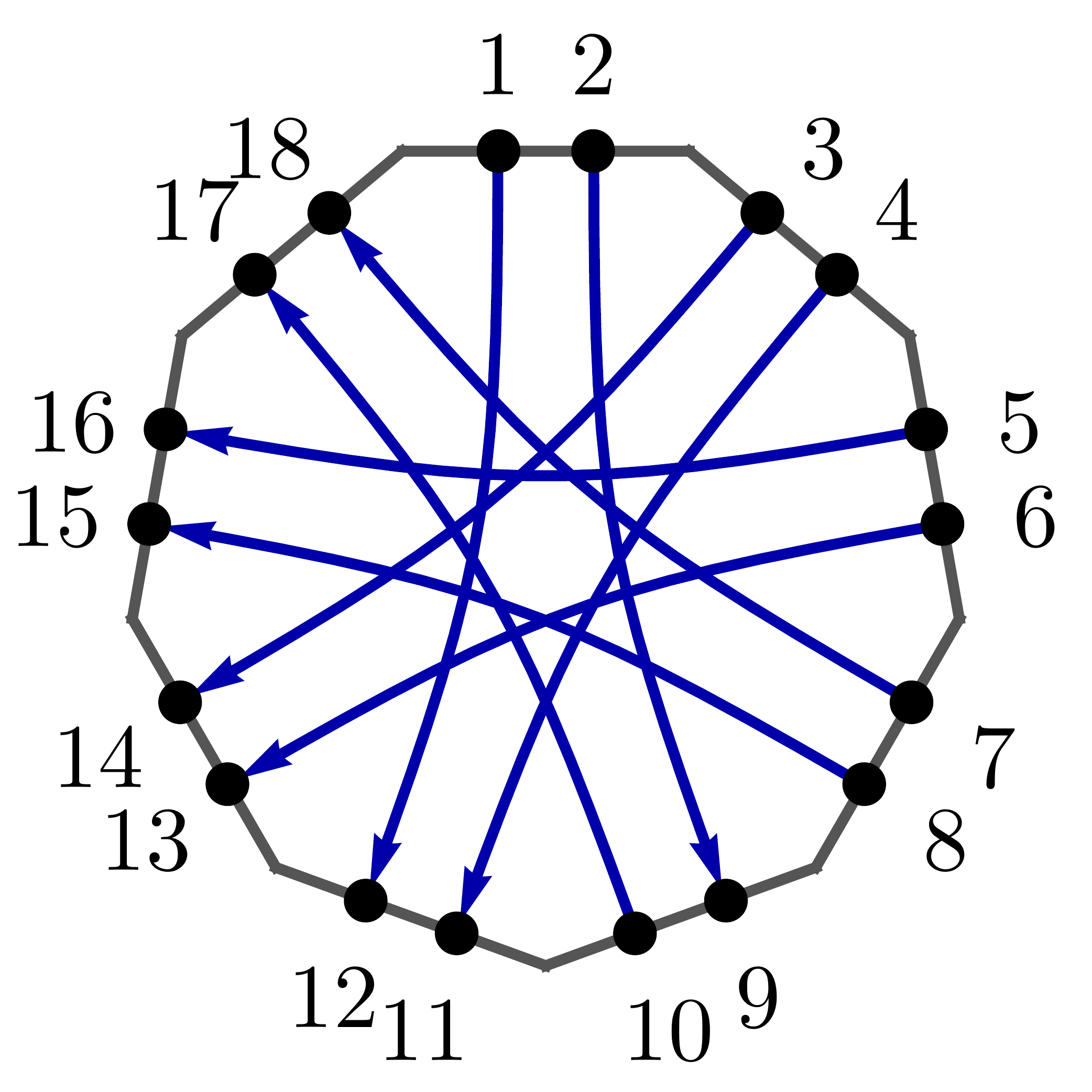}
\end{gathered} \ , &
\ket{\bar{1}^\prime}_9\; &= \;
\begin{gathered}
\includegraphics[height=0.1\textheight]{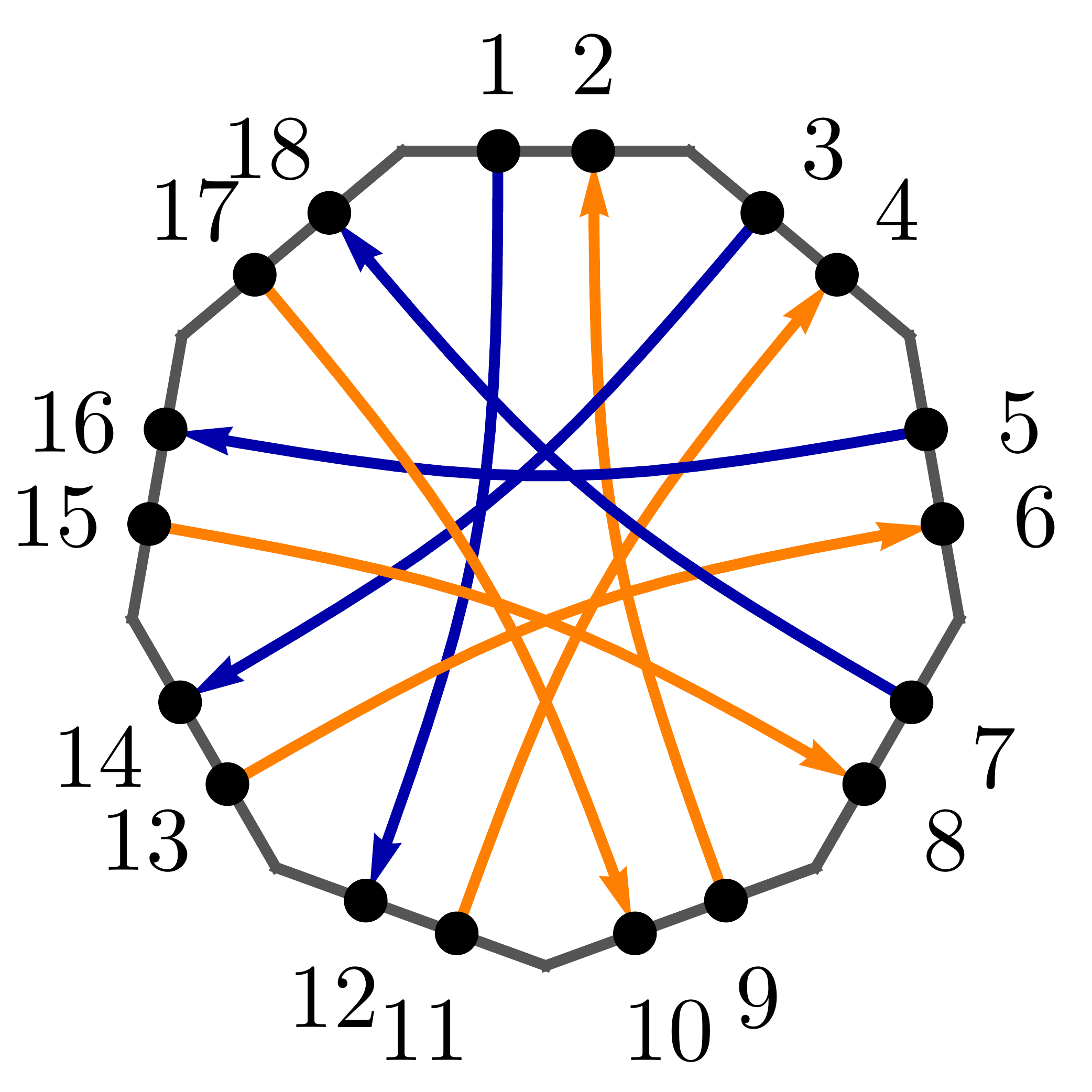}
\end{gathered} \ ,
\end{align}

\begin{figure}[tb]
\includegraphics[width=0.485\textwidth]{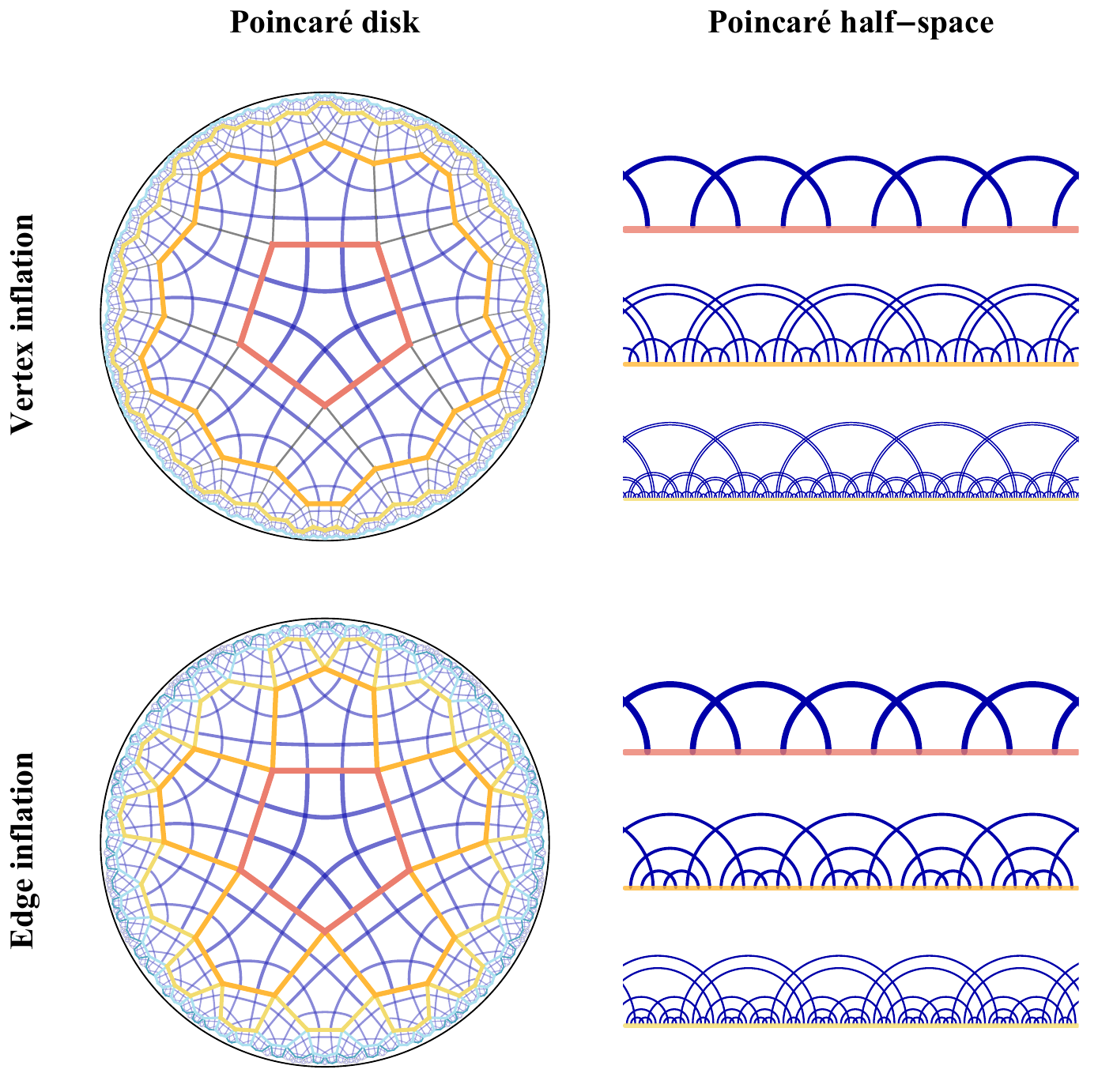}
\caption{Edge- and vertex-based inflation of the $\{5,4\}$ HyPeC in the form \eqref{EQ_HYPEC_STATES2}, with inflation layers colour-coded. The full tiling in the original Poincar\'e disk projection is shown on the left, while the dimers at the first three inflation layers are shown on the right, unfolded onto a line (Poincar\'e half-space projection).}
\label{FIG_DIMER_54_FULL}
\end{figure}

With the tools developed in the previous sections, the average entanglement entropy \cite{AreaReview}, 
and by extension the central charge, can be computed for any regular tiling based on Majorana dimer states.
We begin with the $n=5$ case of the HyPeC. For simplicity, we consider \emph{edge inflation} rather than vertex inflation in the following calculation: At each step, the tiles on all open edges are added to the contraction. 
The more complicated case of vertex inflation, which is more generalizable to arbitrary $\{n,k\}$ tilings, will be treated in Appendix \ref{APP_DIMERS}. 
The edge inflation rules for the $\{5,4\}$ tiling are
\begin{align}
a &\mapsto c a a b \ , &
b &\mapsto c a b \ , &
c &\mapsto \emptyset \ .
\end{align}
where the letters label boundary edges.
On the level of the $\{5,4\}$ tiling, these inflation rules can be visualized as follows:
\begin{align}
\begin{gathered}
\includegraphics[height=0.1\textwidth]{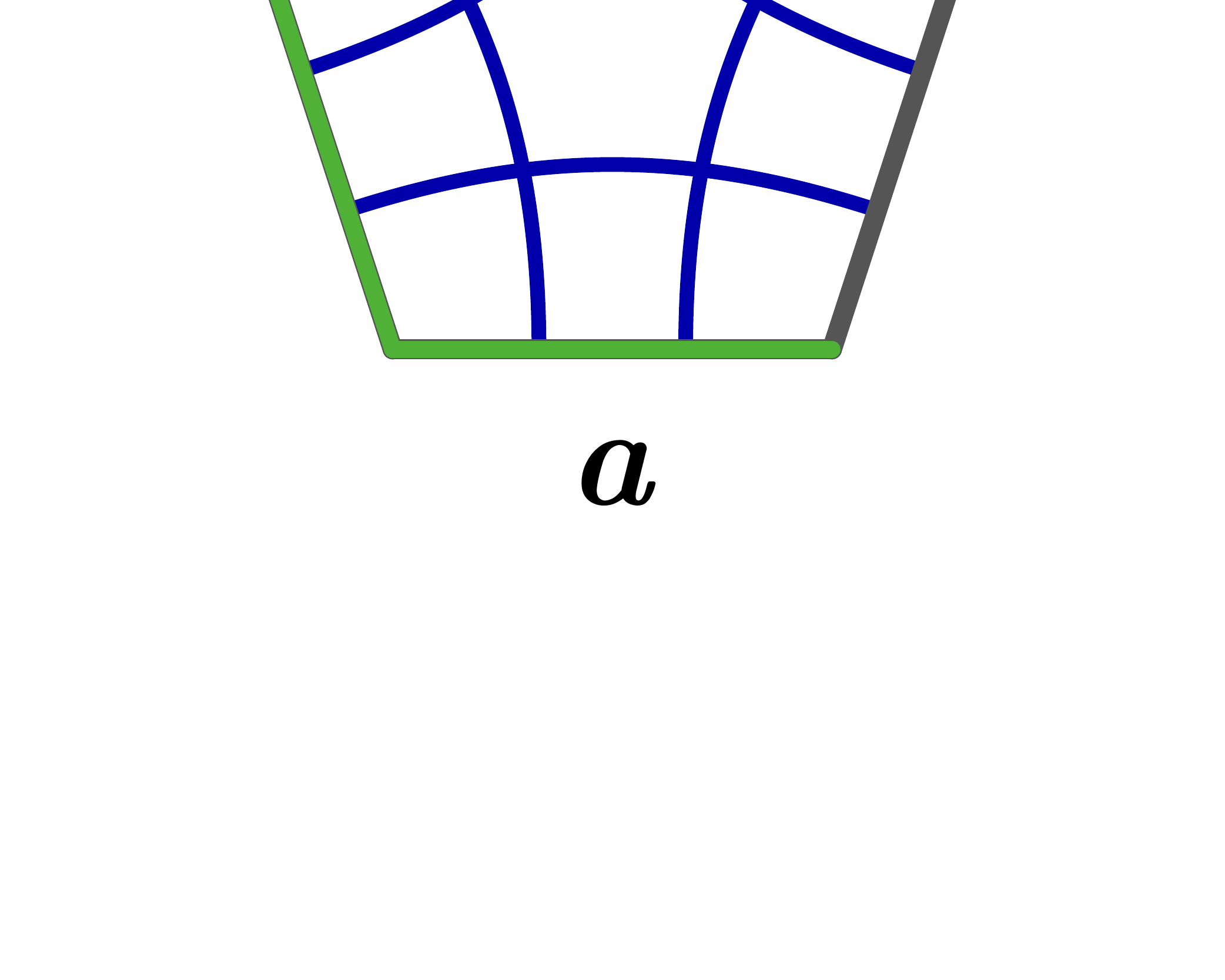}
\end{gathered}
&
\begin{gathered}
\vspace{20pt}
\scalebox{1.25}{$\quad\mapsto\quad\;$}
\end{gathered}
\begin{gathered}
\includegraphics[height=0.1\textwidth]{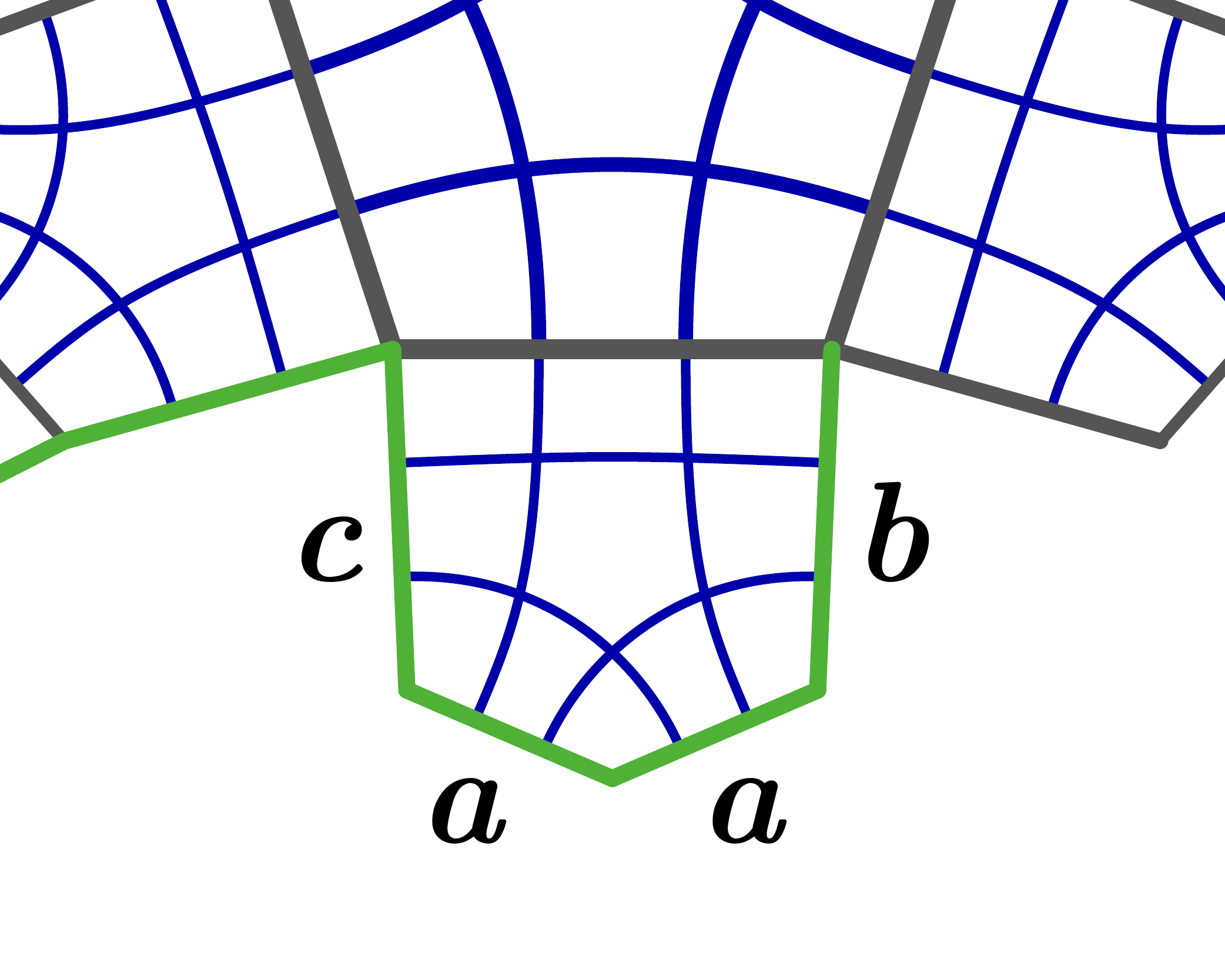}
\end{gathered} \; ,
\\
\begin{gathered}
\includegraphics[height=0.1\textwidth]{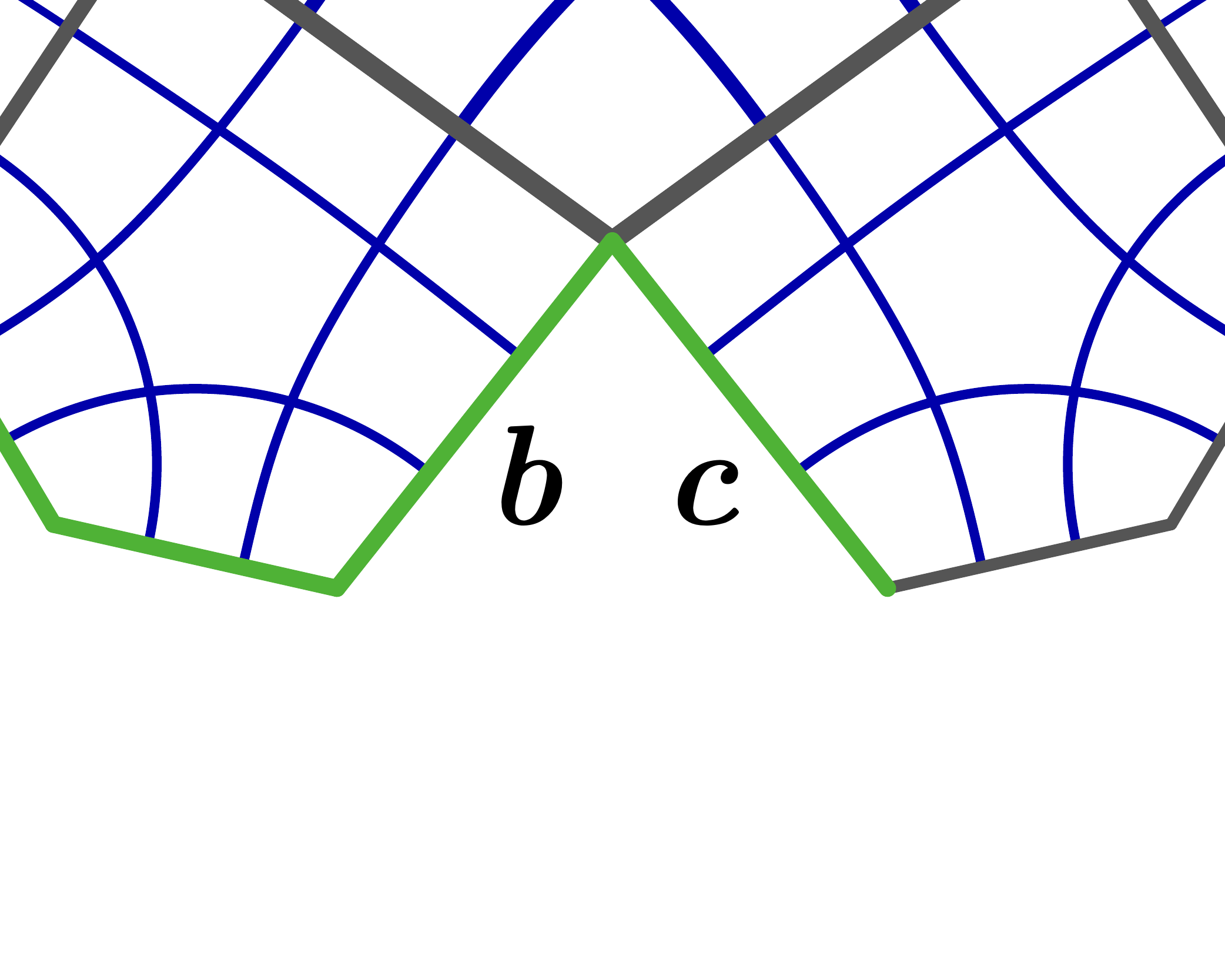}
\end{gathered}
&
\begin{gathered}
\vspace{20pt}
\scalebox{1.25}{$\quad\mapsto\quad\;$}
\end{gathered}
\begin{gathered}
\includegraphics[height=0.1\textwidth]{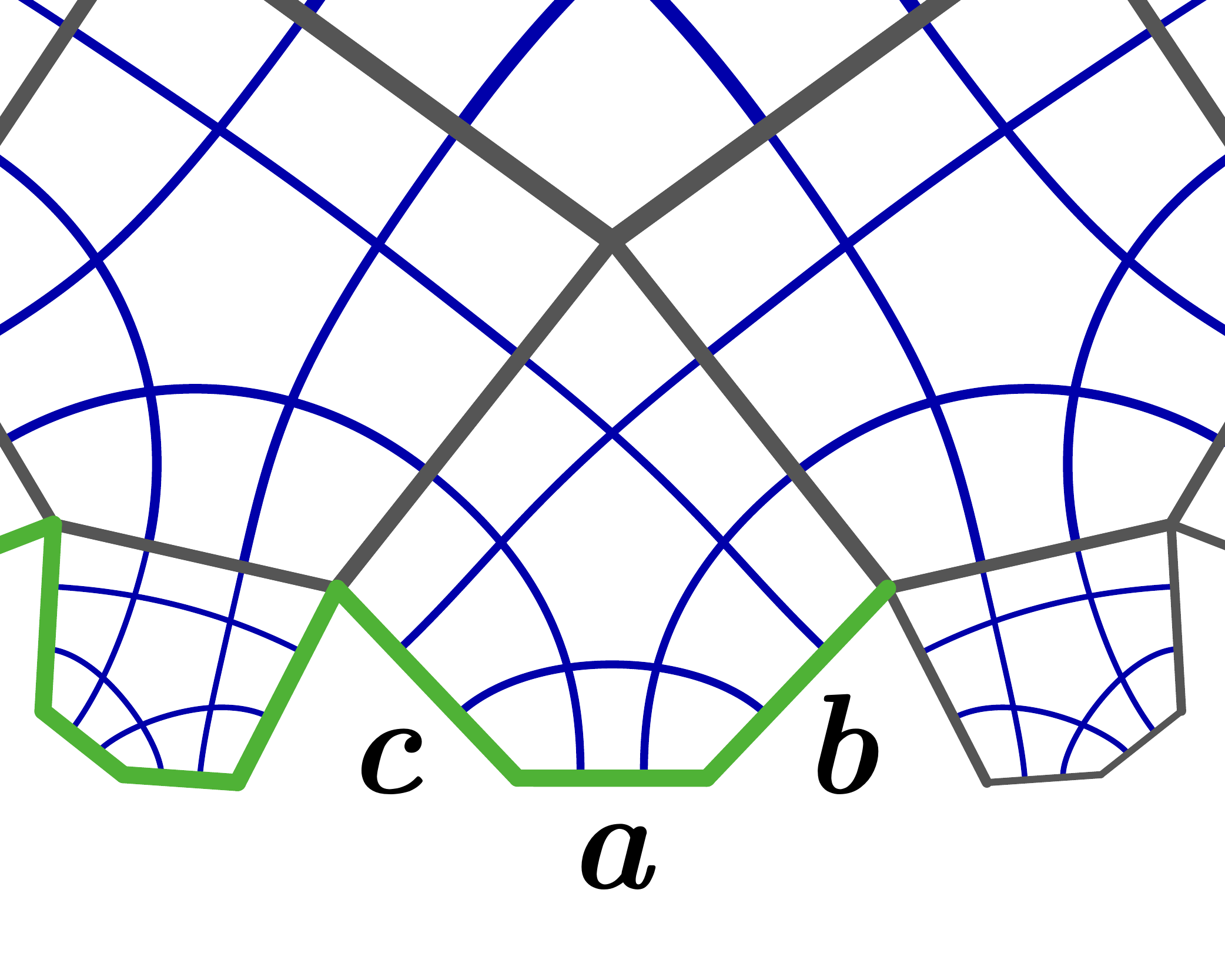}
\end{gathered} \; .
\end{align}
We have combined the rules for $b$ and $c$ as they always appear in the combination $b c$. A boundary region ending at the marked letter, as well as its inflated version, is highlighted in green. Fig.\ \ref{FIG_DIMER_54_FULL} (bottom) shows how these inflation rules act on the whole tiling, starting with a central pentagon (the sequence $a^5$).
To see the change in dimer structure more clearly, one can project the boundary onto a line. The inflation rules are then given by
\begin{align}
\begin{gathered}
\includegraphics[width=0.1\textwidth]{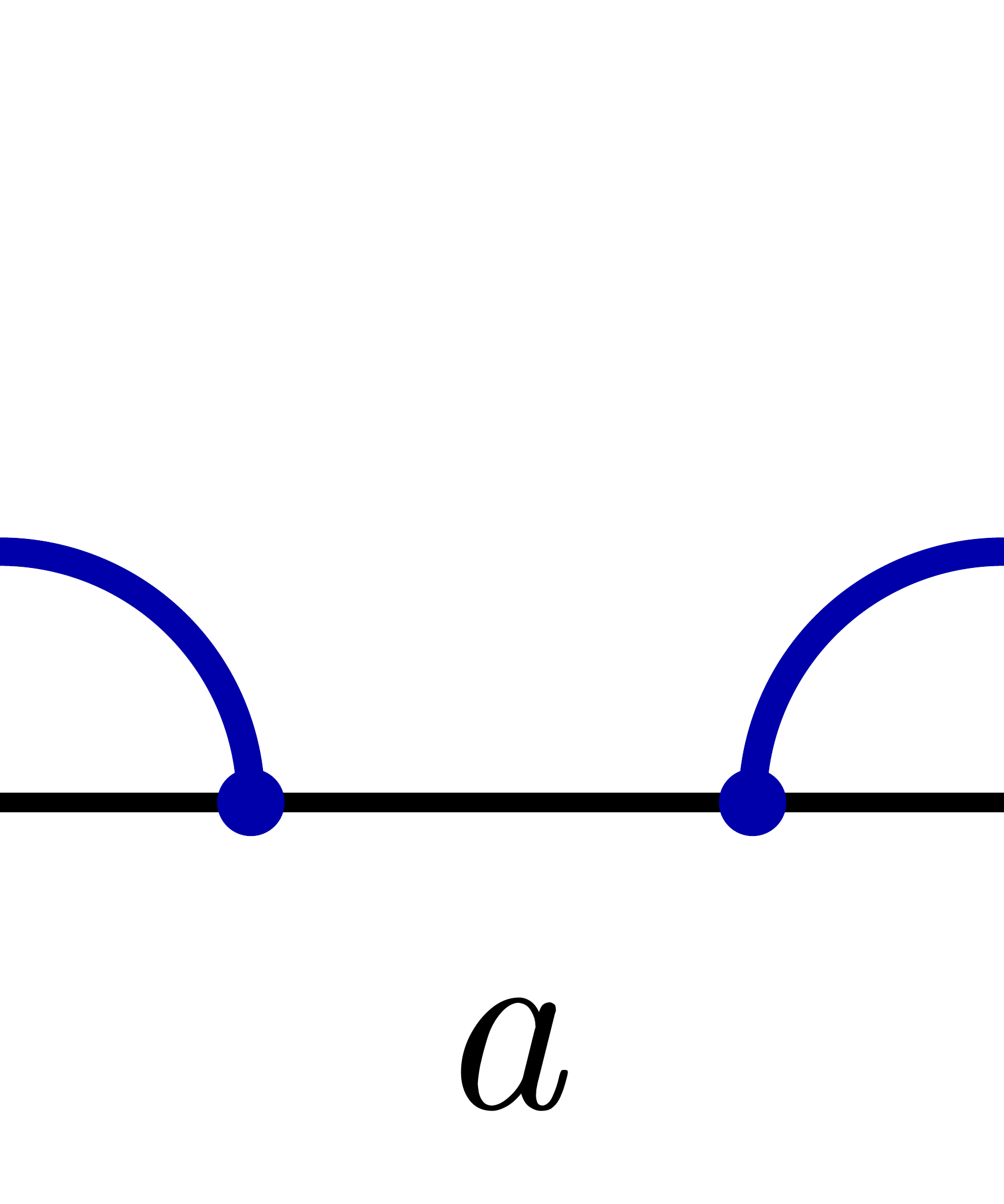}
\end{gathered}
&\scalebox{1.25}{$\;\quad\mapsto\quad\;$}
\begin{gathered}
\includegraphics[width=0.1\textwidth]{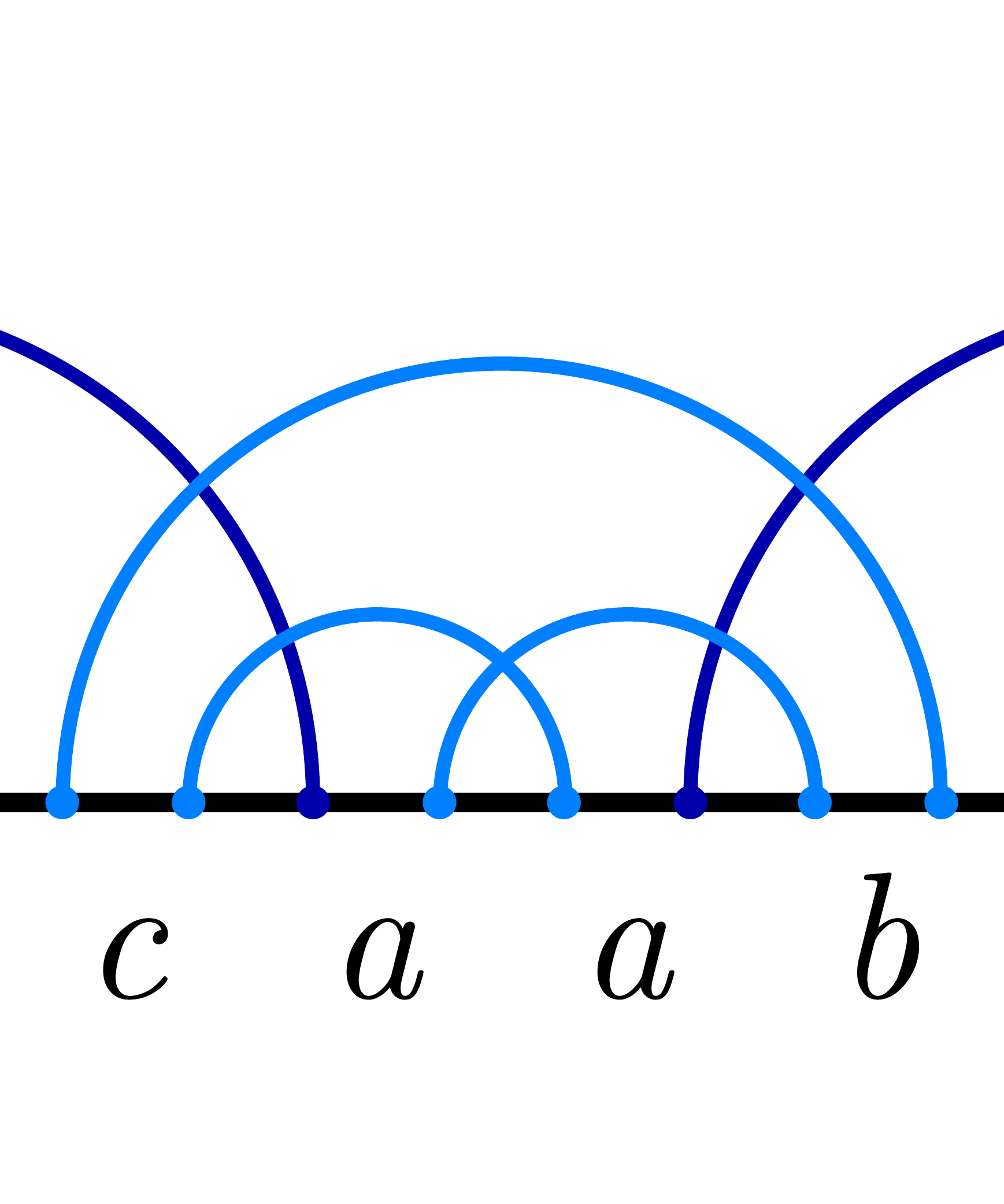}
\end{gathered} \ , \\
\begin{gathered}
\includegraphics[width=0.1\textwidth]{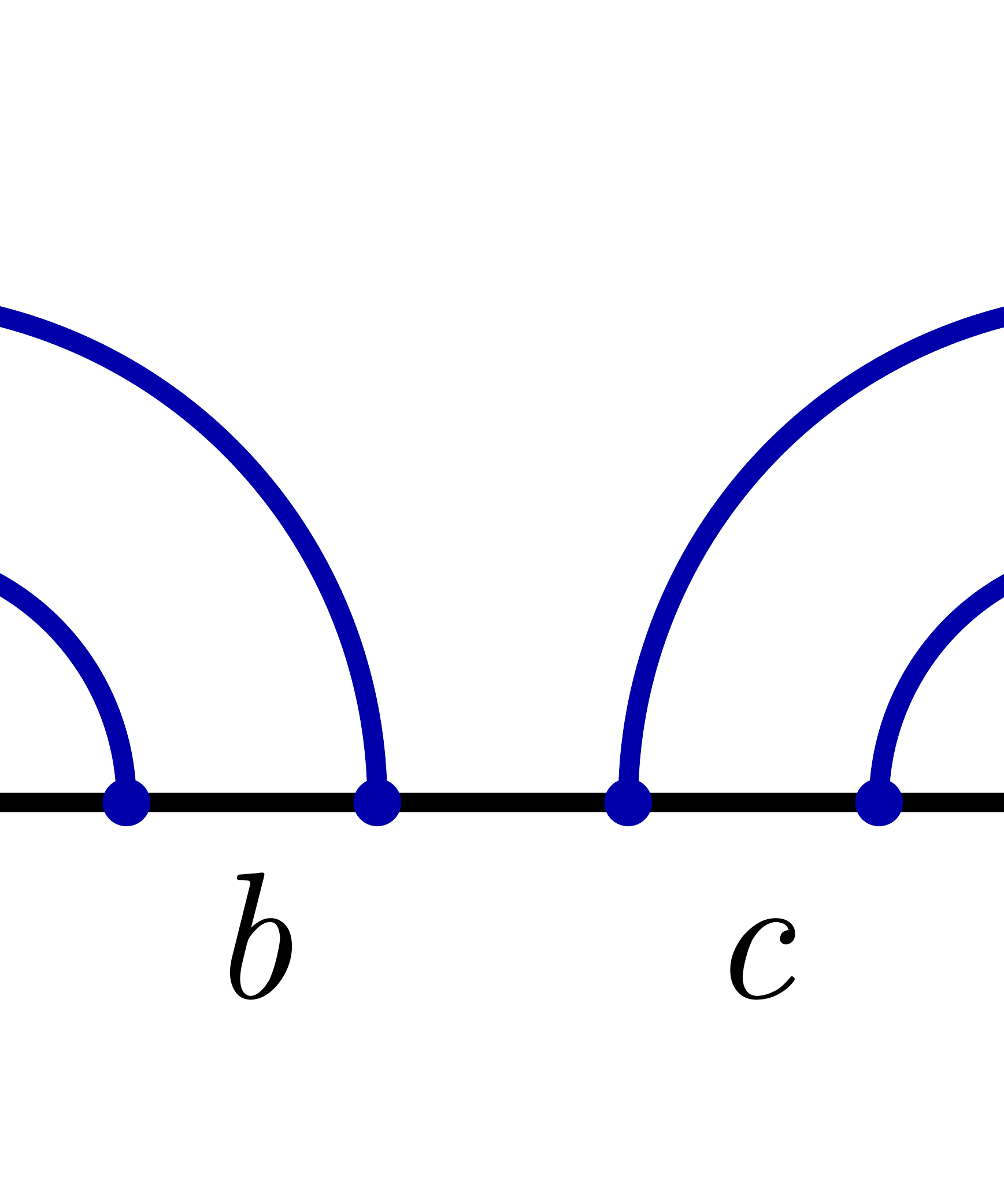}
\end{gathered}
&\scalebox{1.25}{$\;\quad\mapsto\quad\;$}
\begin{gathered}
\includegraphics[width=0.1\textwidth]{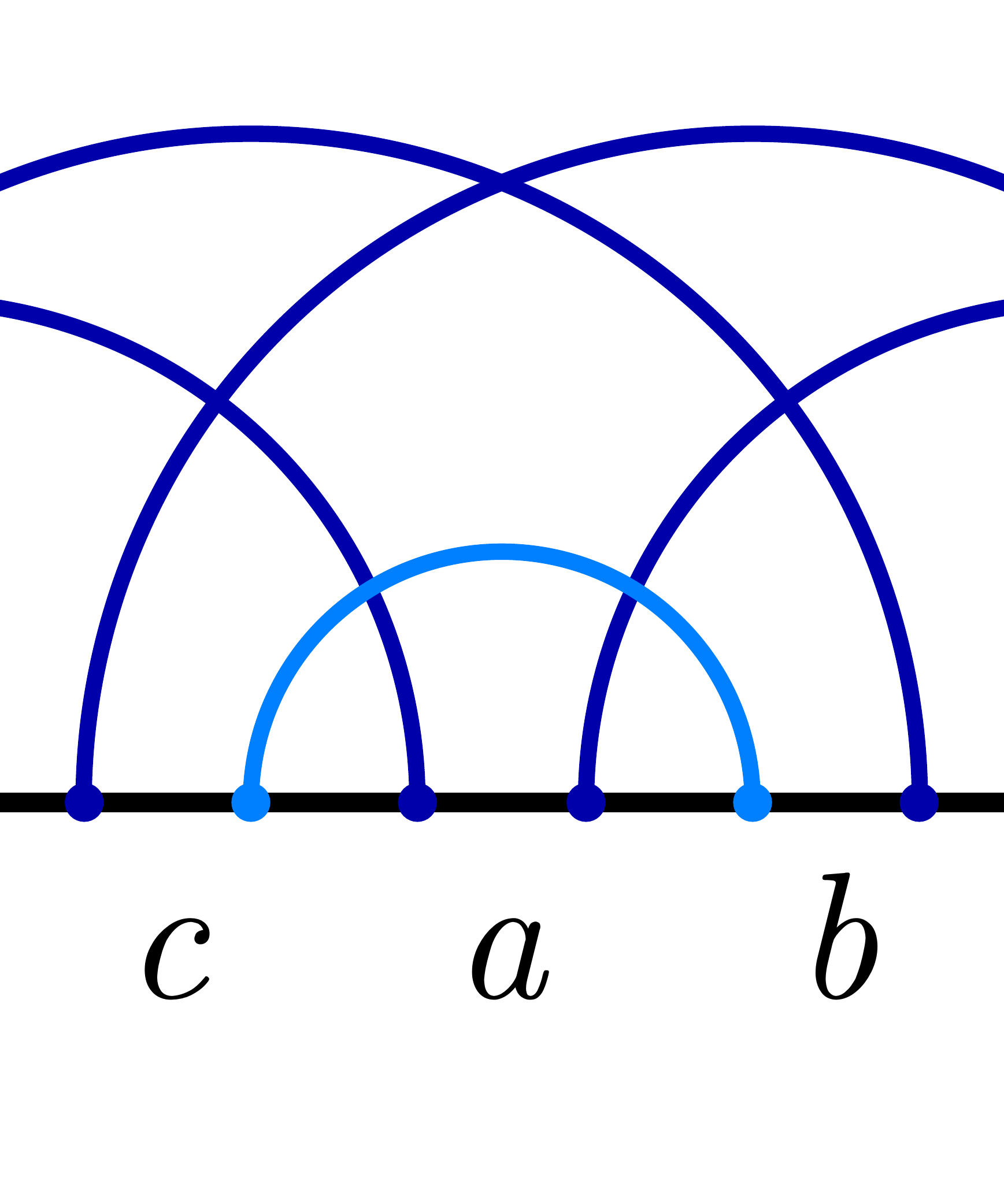}
\end{gathered} \ .
\end{align}
The new dimers added at each step are drawn in a lighter colour, while the ones that are extended from the previous layer are drawn darker. As we are interested in entanglement properties, the dimer parities (which we previously colour-coded) are irrelevant here.

\begin{figure}[t]
\includegraphics[width=0.47\textwidth]{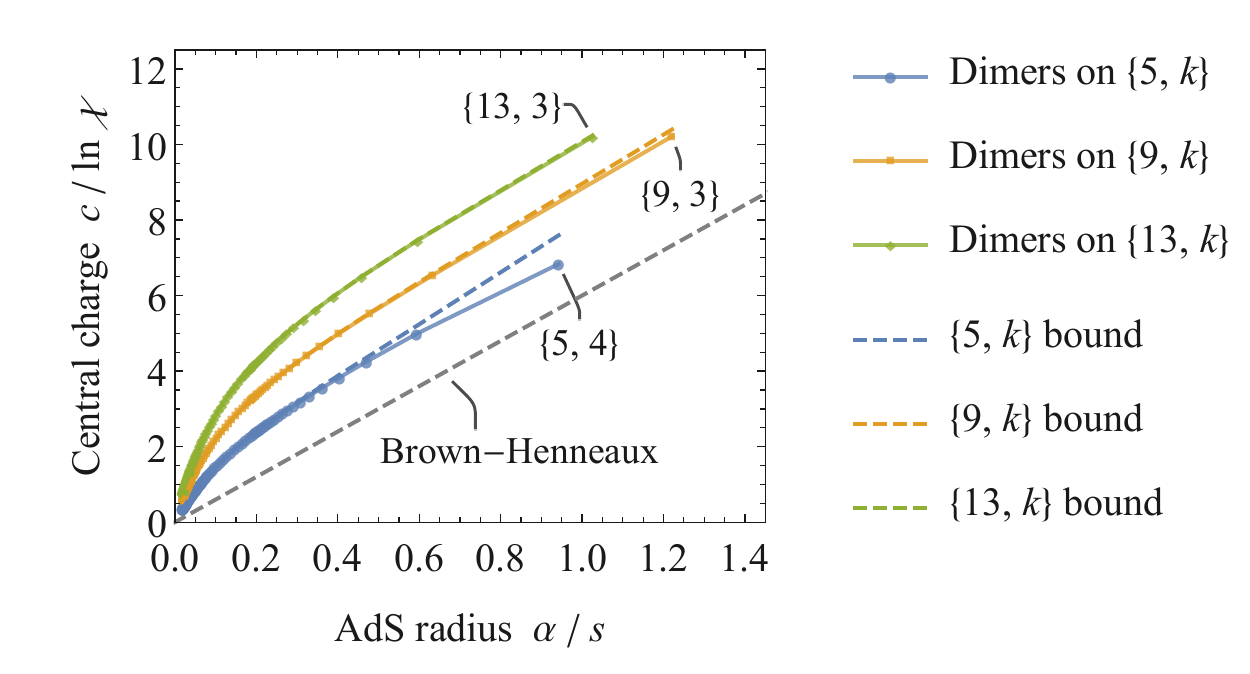}
\caption{Central charges for the $\{5,k\}$, $\{9,k\}$, and $\{13,k\}$ Majorana dimer models (solid curves, bottom to top) and corresponding geodesic bounds (dashed curves). The continuum Brown-Hennaux formula for $G=s/4 \ln\chi$ is shown as a dashed line.}
\label{FIG_CDIMERS}
\end{figure}

Having associated geometrical features of the 
inflated tiling with a specific dimer configuration, we 
can now exactly calculate the entanglement entropy that each inflation step produces.
As in the previous section, first consider a deflation or coarse-graining step that removes dimers and thus, entanglement entropy. Consider how a cut (green line) changes throughout a deflation step:
\begin{align}
\begin{gathered}
\includegraphics[width=0.1\textwidth]{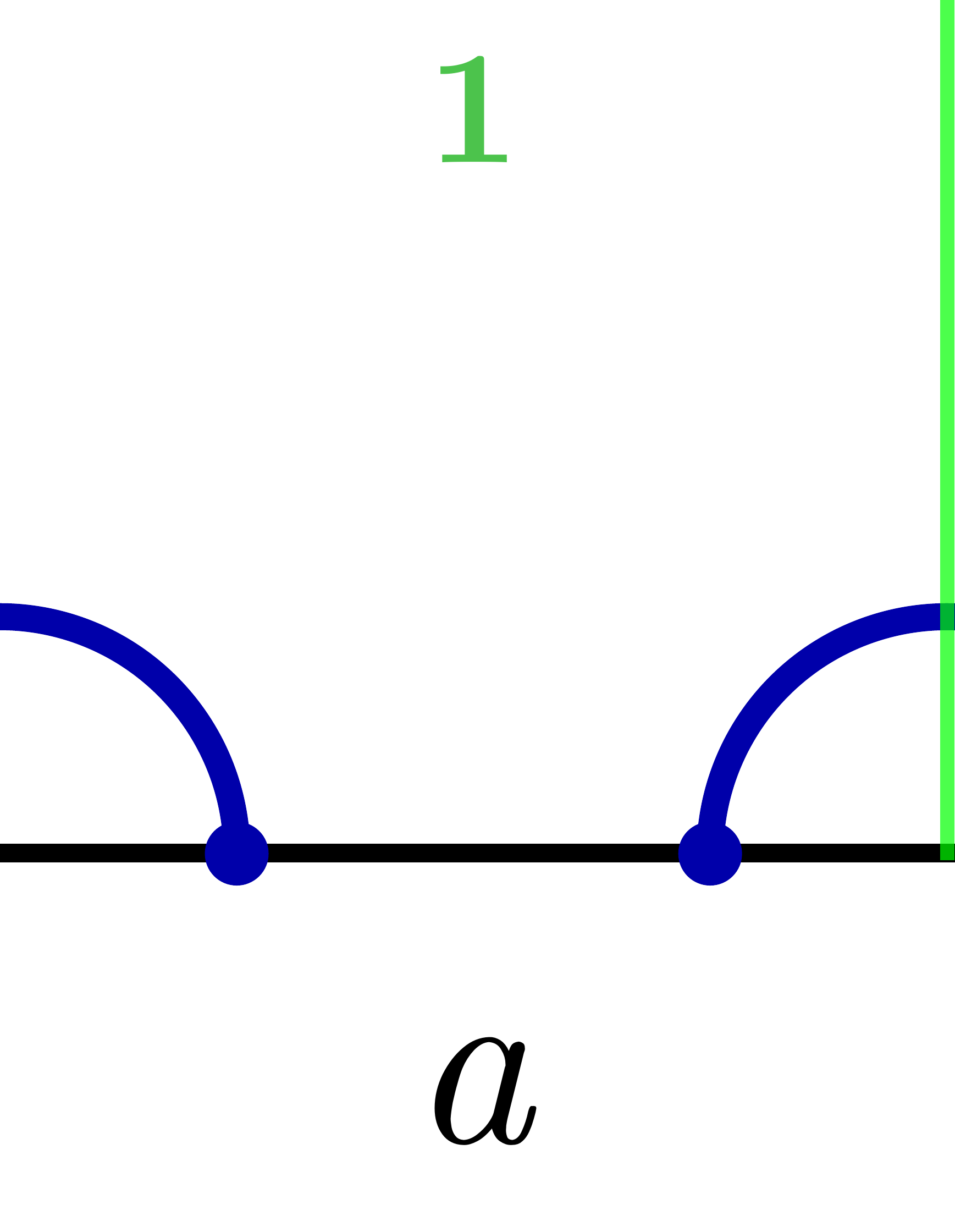}
\end{gathered}
&\scalebox{1.25}{$\;\quad\mapsfrom\quad\;$}
\begin{gathered}
\includegraphics[width=0.1\textwidth]{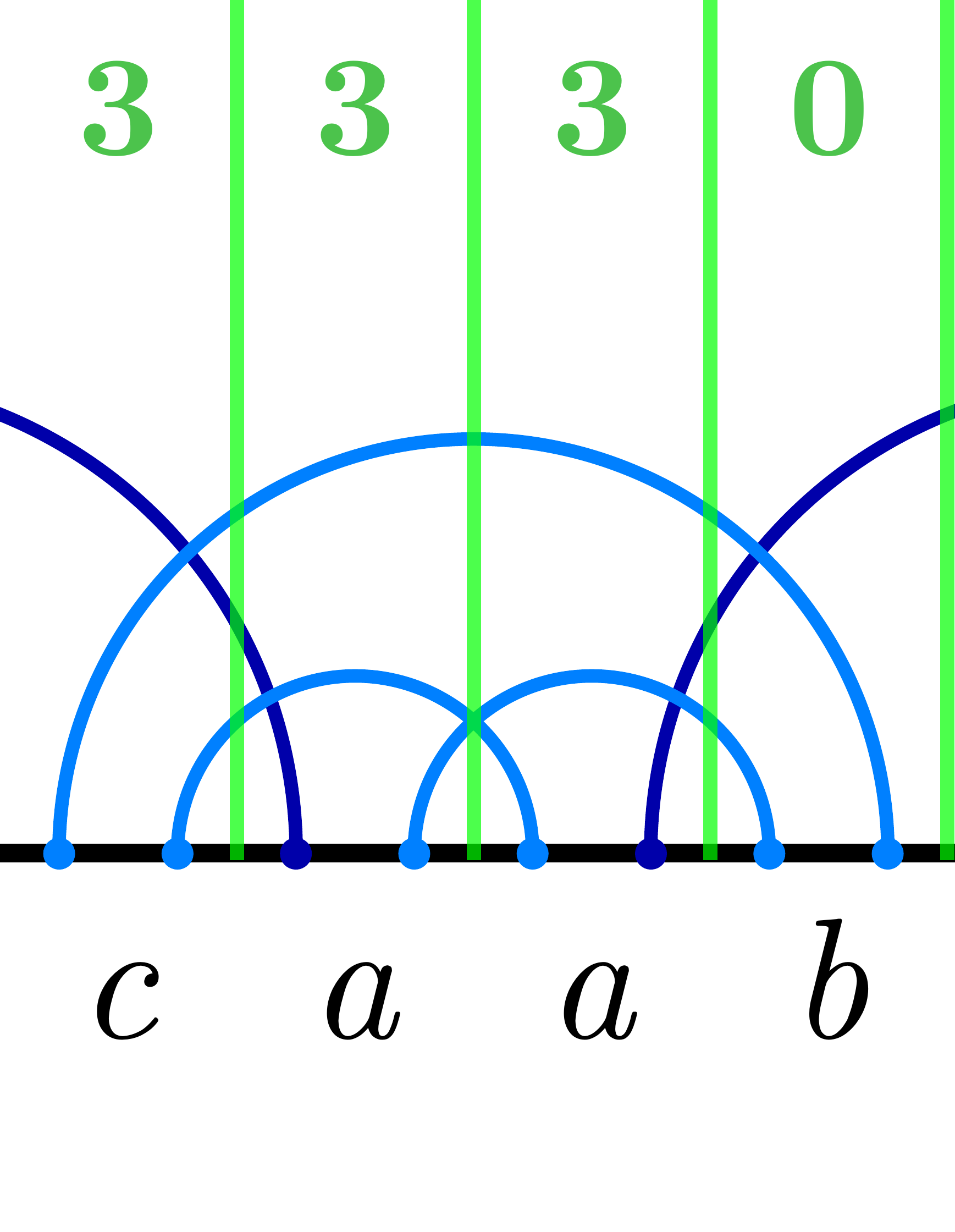}
\end{gathered} \ , \\
\begin{gathered}
\includegraphics[width=0.1\textwidth]{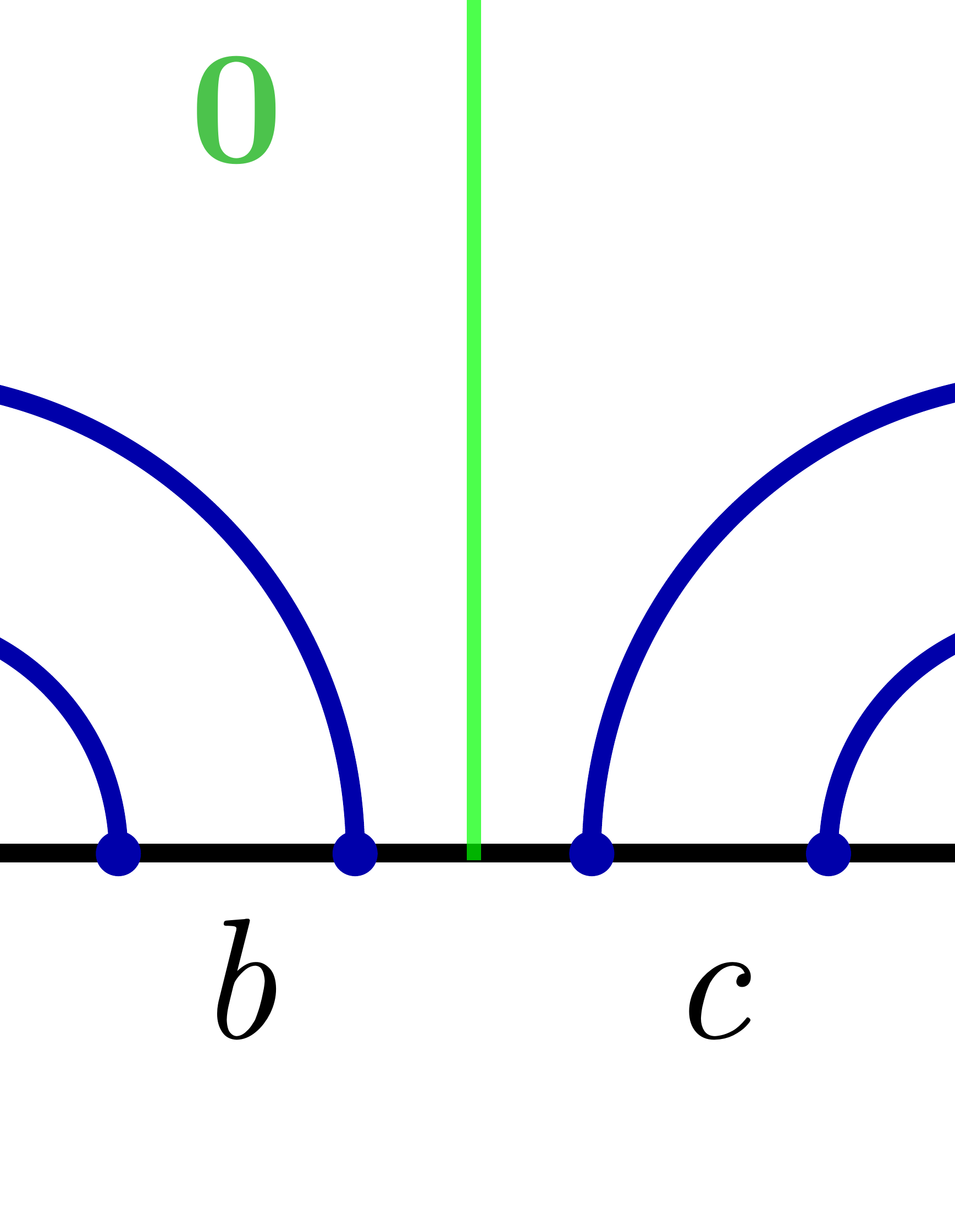}
\end{gathered}
&\scalebox{1.25}{$\;\quad\mapsfrom\quad\;$}
\begin{gathered}
\includegraphics[width=0.1\textwidth]{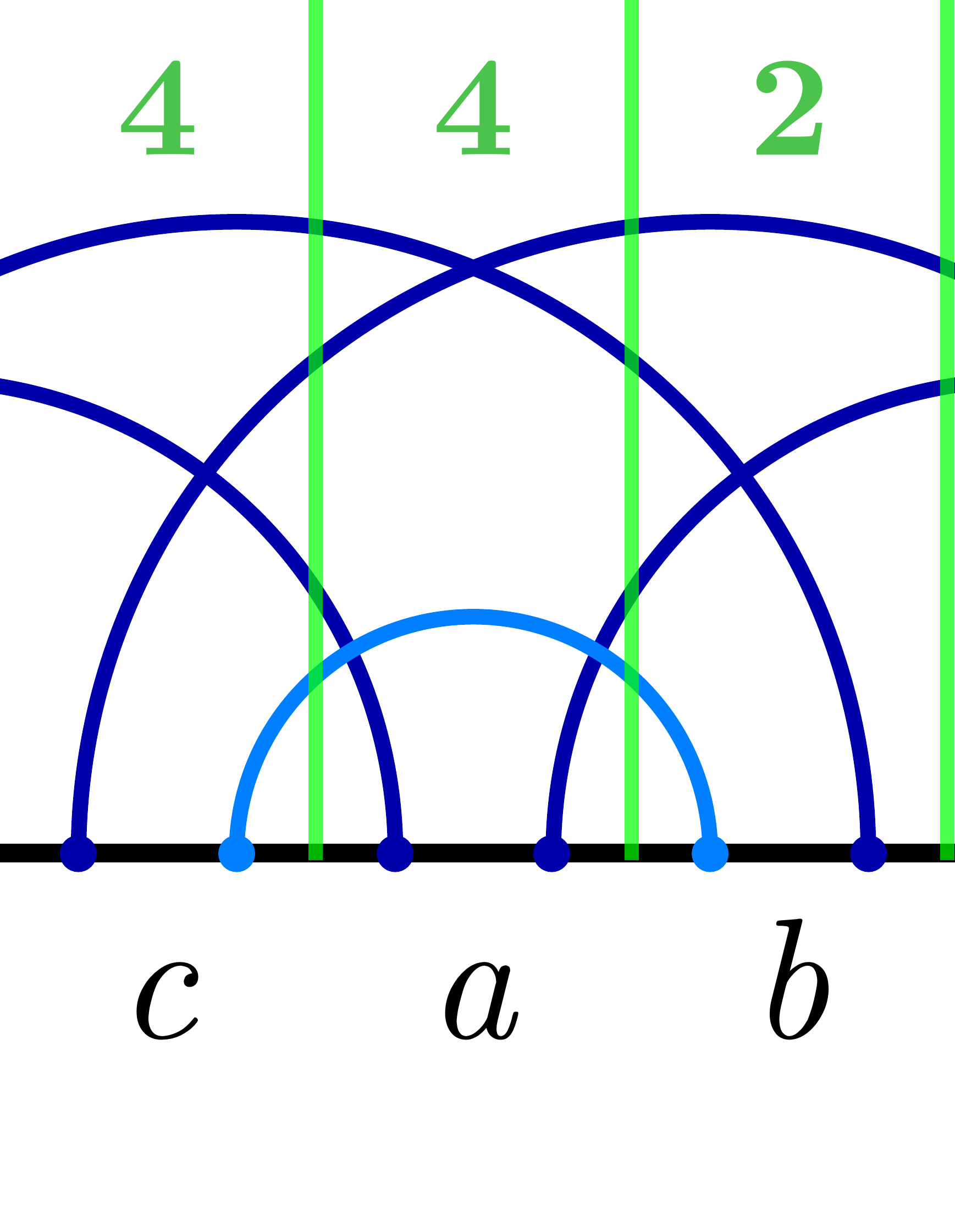}
\end{gathered} \ .
\end{align}
The green number counts the dimers that pass through the cut to the right to it, i.e., the local entanglement of a boundary region ending on a given edge.
From these diagrams, we now construct the substitution and entanglement matrices $M$ and $E$ that describe the Markov process underlying vertex inflation. While $M$ is constructed as before, the entries of $E$ are now composed of half the difference in dimer cuts between two inflation layers for a given substitution, as each dimer carries $\ln(2)/2$ entanglement. We thus find
\begin{align}
M &= 
\left(
\begin{array}{ccc}
 2 & 1 & 1 \\
 1 & 1 & 1 \\
 0 & 0 & 0
\end{array}
\right) \ , &
E &=
\left(
\begin{array}{ccc}
 1 & 0 & 1 \\
 2 & 1 & 2 \\
 0 & 0 & 0 
\end{array}
\right) \ .
\end{align}
Using \eqref{EQ_C_NK}, which now becomes an equality rather than an upper bound, this leads to an effective central charge
\begin{equation}
\label{EQ_C_54_EDGE}
c_{\{5,4\}_e}^\text{d} = \frac{6 \ln 2}{\ln\frac{3 + \sqrt{5}}{2}} \approx 4.32 \ ,
\end{equation}
a result in agreement with previous numerical studies \cite{Jahn:2019nmz}. Note that used the subscript of $\{5,4\}_e$ to denote edge inflation, while all values without such a subscript in this text refer to vertex inflation.
The computation of effective central charges follows analogously for vertex inflation, which we can generalize to arbitrary $\{5,k\}$ tiling with a finite number of letters. 
We can further extend this approach to the block perfect tensors associated with $\{4m+1,k\}$ tilings.
The central charges for these general cases are computed in Appendix \ref{APP_DIMERS}. 
The results for $m=1,2,3$, i.e., the hyperbolic pentagon code and the block-perfect \emph{nonagon} and \emph{tridecagon codes}, are shown in Fig.\ \ref{FIG_CDIMERS}. 
For all of these codes, the central charges saturate to their maximum allowed value at large $k$, with a slope at small $k$ similar to the Brown-Henneaux value.
Explicitly, at large $n$ and $k$ both the central charge bound and the exact Majorana dimer value scale as
\begin{equation}
c_{\{4m+1,k\}}^\text{d} = c_{\{4m+1,k\}}^\text{max} = \frac{(6 m + \frac{9}{2}) \ln\chi}{\ln\left( (4k - 8) m - k \right)} + O\left( m^{-1} \right) \ ,
\end{equation}
with a bond dimension $\chi=2$ for the dimer model. Thus, we conclude that for tilings with high curvature (large $n$ and $k$), our class of hyperbolic block perfect codes based on Majorana dimers produce maximal entanglement for any connected boundary region $A$. This is equivalent to a statement that \emph{residual bulk regions} become negligible in this limit, with a maximal flow of entanglement through the minimal cut $\gamma_A$.

\section{Discrete conformal transformations}

\begin{figure}[tb]
\includegraphics[width=0.47\textwidth]{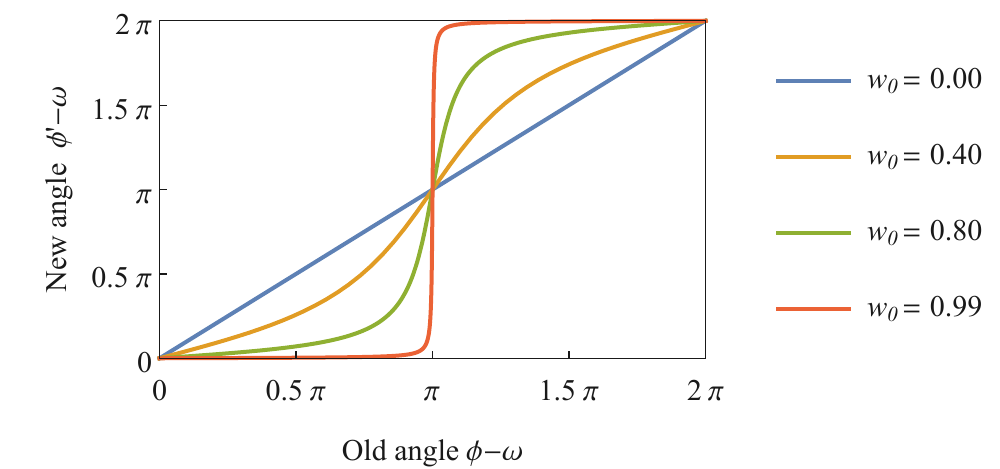}
\caption{Local scale transformation of Poincar\'e disk angle $\phi$ on the boundary under a translation in the bulk.}
\label{FIG_LOCAL_SCALING}
\end{figure}

In our analysis of central charges of discrete critical systems, we only considered the behavior of boundary states under \emph{global} scaling transformations, corresponding to an application of inflation rules on all boundary sites at once. 
However, global scaling transformations only form a subset of the conformal algebra. To study the remaining symmetries, we can equivalently consider the bulk symmetries \cite{Brown:1986nw}; in our time-slice case, these are the symmetries of the Poincar\'e disk \eqref{EQ_ADS_PDISK}.
Whereas the original AdS spacetime \eqref{EQ_ADS_GLOBAL} enjoys an $SO(2,2)$ symmetry (most apparent in its embedding as a hyperboloid in $2{+}2$-dimensional flat spacetime), the Poincar\'e disk is only invariant under $PSL(2,\mathbb{R})$ transformations, a subset of the \emph{M\"obius transformations}. If we represent a point in the disk as a complex number $z=\rho\, e^{\i \phi}$, then these transformations $M_{\theta,v}$ are given by
\begin{equation}
z \mapsto z^\prime = M_{\theta,v} (z) = e^{\i \theta} \frac{w + z}{1 + w^\star z} \ ,
\end{equation}
where $0\leq\theta<2\pi$ and $w = w_0\, e^{\i \omega}$ defines a point in the Poincar\'e disk, i.e., with $|w|<1$.
We can see how these transformations act on the AdS boundary by taking the $\rho=\rho_0 \to 1$ limit. We find:
\begin{itemize}
    \item A {\bf global scale transformation} determined by a change in cutoff $\rho_0 \to \rho_0^\prime$: The total length $L$ of the flat AdS boundary changes as 
    \begin{equation}
    \label{EQ_SCALING_GLOBAL}
        L \mapsto L^\prime = \frac{1-\rho_0}{1-\rho_0^\prime}\, L \ ,
    \end{equation}
    assuming that $\rho_0$ and $\rho_0^\prime$ are close to one.
    \item A {\bf translation} $T_\theta=M_{\theta,0}$ produced by a rotation of the Poincar\'e disk. Introducing the boundary coordinate $x=\alpha \phi / (1-\rho_0)$, this corresponds to a transformation
    \begin{equation}
        x \mapsto x^\prime = x + \frac{\alpha\, \theta}{1 - \rho_0} \ .
    \end{equation}
    \item A {\bf local scale transformation} $D_w=M_{0,w}$ by shifting the center of the Poincar\'e disk towards a point $w=w_0 e^{\i \omega}\neq 0$. Without loss of generality, we now assume that $\omega = 0$. At $\rho \to 1 $, we then find a transformation of the Poincar\'e angle $\phi$ of the form
    \begin{equation}
    \label{EQ_LOCAL_SCALING}
        \phi \mapsto \phi^\prime = \arctan\frac{(1-w_0^2) \sin\phi}{(1+w_0^2) \cos\phi + 2w_0} \ .
    \end{equation}
    This transformation is shown in Fig.\ \ref{FIG_LOCAL_SCALING} for different values of $w_0$ and general $\omega$.
    Lengths around the boundary point $x=\alpha \omega/(1-\rho_0)$ are stretched to 
      \begin{equation}
    x \mapsto x^\prime = \frac{1+w_0}{1-w_0} x, 
       \end{equation}while those around $y=\alpha (\pi + \omega)/(1-\rho_0)$ are contracted to 
      \begin{equation}
    y \mapsto y^\prime = \frac{1-w_0}{1+w_0} y.
       \end{equation}
\end{itemize}
We can rewrite any combination of translations and local scale transformations as a single M\"obius transformation $M_{\theta,v} = T_\theta \circ D_v$ by using the identities
\begin{subequations}
\begin{align}
T_\theta \circ T_\phi &= T_{\theta+\phi} \ , \\
D_v \circ D_w &= T_{\arg\frac{1 + v w^\star}{1 + v^\star w}} \circ D_{\frac{v+w\,}{1+v w^\star}} \ , \\
D_v \circ T_\theta &= T_\theta \circ D_{e^{-\i \theta} v} \ .
\end{align}
\end{subequations}
All transformation considered so far preserve orientation. If we drop this constraint, we also find
the following:
\begin{itemize}
\item An \textbf{inversion} $I_{v,\theta}$ through a bulk reflection around a geodesic through a point $v$ at normal angle $\theta$, expressed with the complex conjugate $C(z) = z^\star$ as 
\begin{align}
z \mapsto I_{v,\theta} (z) &= D_v \circ T_\theta \circ C \circ T_{-\theta} \circ D_{-v} (z) \nonumber \\
 &= D_{-v} \circ T_{-2\theta} \circ D_{v^\star} \circ C (z) \ .
\end{align}
Note that $I_{v,\theta}^2 = \id$, and that any inversion is equivalent to complex conjugation and a M\"obius transformation.
For a boundary coordinate system centered around a boundary angle $\phi=0$ and ``infinity'' identified as $\phi=\pi$, we choose $v=x$ with $x\in\mathbb{R}$ and $\theta=-\pi/2$, leading to a \emph{canonical} inversion
\begin{align}
z \mapsto I_x (z) &= D_{\frac{2x}{1+x^2}} \circ T_\pi \circ C (z) \nonumber\\
 &=  C \circ T_\pi \circ D_{\frac{-2x}{1+x^2}} (z) \ .
\end{align}
\item By combining inversion and translation, we can also construct the \textbf{special conformal transformation}
\begin{align}
z \mapsto K_{\theta,x}(z) &= I_x \circ T_\theta \circ I_x (z) \nonumber\\
 &= D_{\frac{2x}{1+x^2}} T_{-\theta} D_{\frac{-2x}{1+x^2}} (z) \ ,
\end{align}
which is just an (orientation-preserving) M\"obius transformation.
\end{itemize}

\begin{figure}[tb]
\centering
\includegraphics[width=0.5\textwidth]{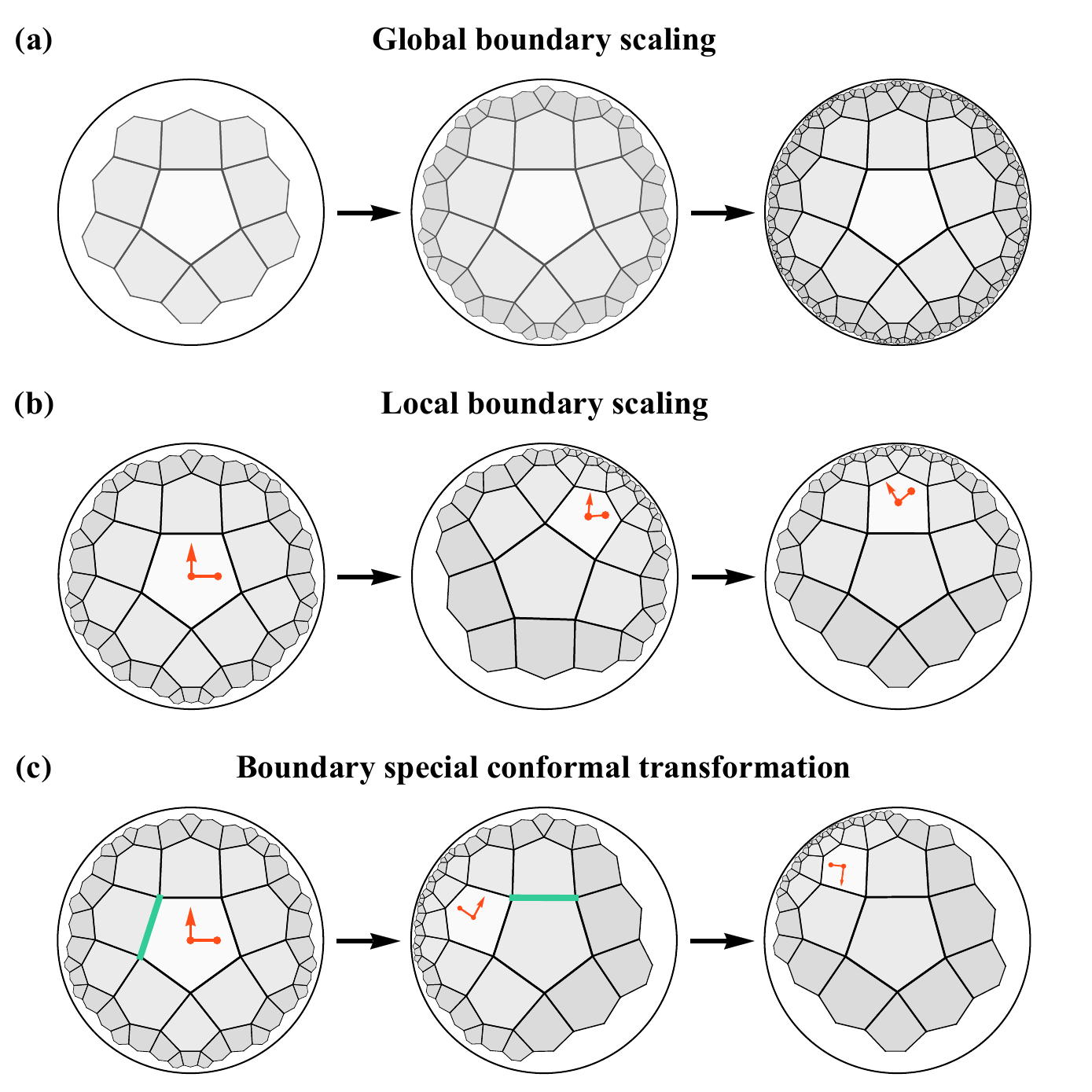}
\caption{(a) Global scale transformation by growing the hyperbolic bulk tiling through \emph{vertex inflation}.
(b) Local scale transformation by a M\"obius transformation composed of a bulk translation (first step) and a rotation (second step).
(c) Successive reflections around a bulk edge and its tiling-symmetric rotation (green lines), with the same effect as a M\"obius transformation.
}
\label{FIG_TILING_TRANSFORM}
\end{figure}

Discretizing the Poincar\'e disk with a (regular) tiling breaks these continuous symmetries. First consider global and local scaling transformations. 
Rather than a continuous transformation \eqref{EQ_SCALING_GLOBAL}, global inflation (Fig.\ \ref{FIG_TILING_TRANSFORM}, top) rescales the subsystem by an asymptotic constant $\lambda$, the eigenvalue of the substitution matrix for the given tiling. 
When a tensor network is embedded into a regular tiling, choosing identical tensors that are invariant under cyclic permutations of indices, thus preserving the tiling symmetries, leads to boundary states that can be fine- or coarse-grained by any power of $\lambda$ under inflation or deflation. 

Next, consider the local scale transformations: A regular tiling is invariant only under those bulk M\"obius transformation that map tiles onto tiles. As shown in Fig.\ \ref{FIG_TILING_TRANSFORM} (center), this requires a combination of bulk translation and rotation. At finite cutoff, i.e., finite number of tiles, this reduces the density of boundary edges in one region of the boundary while increasing it in the opposing region, just as for the continuous case shown in Fig.\ \ref{FIG_LOCAL_SCALING}.

Finally, special conformal transformations are broken down in two parts, as they can be composed of inversions and translations: We previously constructed inversions through a bulk reflection around a geodesic, while in a $\{n,k\}$ tiling such transformations only leave the lattice invariant if we reflect around its (geodesic) \emph{edges}. Furthermore, translations are broken down to $\mathbb{Z}_n$ and $\mathbb{Z}_k$ rotations when centering the lattice around tiles and vertices, respectively.
The resulting transformation, shown in Fig.\ \ref{FIG_TILING_TRANSFORM} (bottom), is again equivalent to a bulk translation and rotation, yielding no new symmetries.
Note that while exact translation invariance is broken, the quasiregular structure of the boundary still exhibits \emph{self-similarity} between any sufficiently large subsystems \cite{Boyle:2018uiv}.

\begin{figure}[tb]
\includegraphics[width=0.5\textwidth]{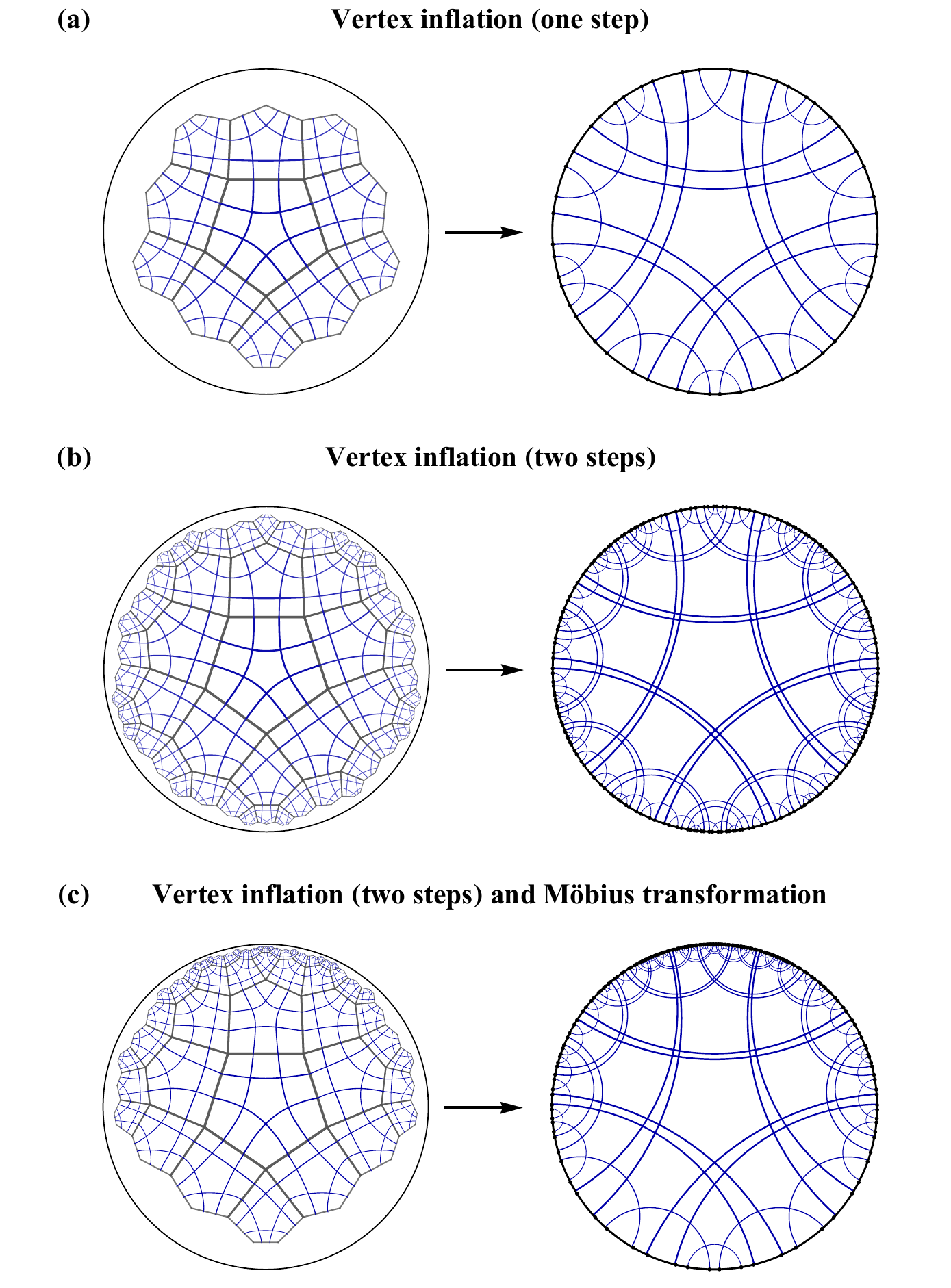}
\caption{(a) A contraction of a hyperbolic tensor network built from Majorana dimers (left) leads to a boundary Majorana dimer state (right).
(b) An inflation step on the tiling (right) leads to a global scale transformation on the boundary state (right).
(c) Certain combinations of Poincar\'e disk translations and rotations in the bulk (left) produce a local scaling transformation on the boundary state (right).
}
\label{FIG_DIMER_CONTR}
\end{figure}

To exemplify these invariance properties with actual states, consider the $\{5,4\}$ HyPeC in Majorana dimers. By projecting the dimer endpoints on the Poincar\'e disk boundary, we can directly compare the states at different cutoffs, as shown in Fig.\ \ref{FIG_DIMER_CONTR}. A global scale transformation increases the resolution of the dimer pattern evenly on the boundary, while a local scale transformation changes it unevenly (Fig.\ \ref{FIG_DIMER_CONTR}, bottom). At the same time, the correlation structure of the boundary states is preserved.
The local scale transformation produced by a bulk M\"obius transformation can be seen as a special case of a local application of inflation and deflation rules. Suitable combinations of such local and global scalings are equivalent to an inflation rule applied only to a subsystem of the boundary, leaving the remainder unchanged.
Note that these transformations are independent from the the actual choice of inflation rule, which fixes the boundary central charge. 

\section{Discussion}

In this work we have studied the entanglement entropy scaling of boundary states of generic hyperbolic tensor networks based on regular tilings. This has 
allowed us to derive a maximal central charge $c^\text{max}$ that such boundary states can possess, with a saturation of this bound corresponding to maximal entanglement through the bulk for any connected boundary region.
We have then related $c^\text{max}$ to the radius of curvature $\alpha$ of the metric into which the tiling is embedded, leading to a discrete analogue of the continuum Brown-Henneaux (BH) formula, where we have identified the gravitational constant $G$ via the Ryu-Takayanagi (RT) prescription.
We find that these bounds are always \emph{above} the continuum value, i.e., that bulk entanglement through a regular hyperbolic tensor network can be as large as through a continuum AdS time-slice. We have further identified two 
distinctly different regimes:
At large AdS radius $\alpha$ and central charge $c^\text{max}$, where the RT identification of $G$ is expected to hold,  we find an approximate relationship
\begin{equation}
c^\text{max} \approx c_0 + 6\, \frac{\alpha \ln\chi}{s} \ ,
\end{equation}
where $s$ is the geodesic length of each edge in the tiling and $\chi$ the bond dimension of the tensor network embedded into the tiling. The constant $c_0$, which produces an offset compared to the BH formula, depends on the $n$-gon tiling and increases with $n$. As the hyperbolic area of a single $n$-gon increases with $n$ as well but remains finite at large $k$, we may interpret $c_0$ as counting the additional degrees in each $n$-gon in the coarse-grained lattice compared to the continuous Poincar\'e disk.
In the opposite limit at small $\alpha$ and $c$, however, we identify a linear relationship without an offset,
\begin{equation}
c^\text{max} \approx f_n\, \frac{\alpha \ln\chi}{s} \ ,
\end{equation}
where the tiling-dependent constant $f_n$ increases with $n$, taking its lowest value $f_3=12$ for triangular tilings.

Furthermore, we find a specific holographic tensor network model that saturates these bounds: The \emph{hyperbolic pentagon code} (HyPeC), a toy model for quantum error correction in AdS/CFT. This model as well as its generalizations can be expressed in the fermionic language of Majorana dimers, which allows for an exact treatment of its entanglement structure in terms of paired Majorana modes.
Using this picture, we showed how successively larger contractions of the tensor network produce a strong disorder renormalization group flow. 
This allowed us to endow a class of models of holographic quantum error correction with the notion of a discretized conformal field theory with aperiodic structure.
The exact central charges resulting from this physical CFT interpretation were derived and shown to saturate to $c^\text{max}$ at large curvature.

Our approach advances the understanding of boundary states of holographic tensor network models, with bounds on central charges for any model based on a regular bulk geometry, which includes the HaPPY holographic codes \cite{Pastawski2015}, block perfect CSS codes \cite{PhysRevA.98.052301}, holographic codes on ideal regular tilings \cite{Osborne:2017woa}, hyper-invariant tensor networks \cite{PhysRevLett.119.141602}, random tensor networks on fixed backgrounds \cite{Hayden2016}, and $p$-adic AdS/CFT models \cite{Gubser:2016guj}, whose Bruhat-Tits tree is identified with a regular tiling \cite{Heydeman:2016ldy}.
Note that our procedure reverses the approach of dynamically reconstructing a discrete bulk geometry from the entanglement structure of a given boundary state, such as considered in Refs.\ \cite{Bao:2018pvs,Bao:2019fpq}; instead, we started with a bulk geometry and derived constraints on the boundary entanglement.
We have also shown that the formulation of quantum error correcting codes in terms of Majorana dimers is essential for understanding their boundary states and RG flow.
The boundary states of these dimer models extend the properties of widely studied aperiodic singlet models to fractional fermionic modes with similar entanglement and RG properties, suggesting that both approaches are representations of a more general class of aperiodic critical theories.
While the Majorana dimer states are \emph{non-interacting}, their use as a code basis in a quantum error correction code such as the HyPeC generally leads to \emph{interacting boundary states} whose correlations and entanglement structure follow those of the non-interacting basis states.
Our results thus show that entanglement renormalization of CFTs can be performed with tensor network approaches other than the MERA, realizing geometries that can be more naturally embedded into an AdS time-slice. 
Understanding their discrete symmetries will be crucial for the development of more powerful tensor network models of AdS/CFT.

\emph{Acknowledgements.} We thank Marek Gluza, Xiaoliang Qi, Sukhbinder Singh, 
Tadashi Takayanagi, and Charlotte Verhoeven for helpful comments and discussions. This work has been
supported by the  Studienstiftung des Deutschen Volkes, the Templeton Foundation, the J\'anos Bolyai Scholarship, the DFG (CRC 183, EI 519/15-1), the NKFIH (K124351, K124152, K124176), the Hungarian Quantum Technology National Excellence Program (Proj. No. 2017-1.2.1-NKP-2017-00001), and the FQXi.

\appendix

\begin{widetext}
\newpage
\section{Geodesic inflation}
\label{APP_GEODESICS}

\begin{table}[b]
\renewcommand{\arraystretch}{1.5}

\begin{tabular}{@{} c @{\hspace{0.8cm}} c @{$\quad$}  c @{$\quad$} c @{$\quad$} c @{$\quad$} c @{$\quad$} c @{$\quad$} c @{$\quad$} c @{}}
\widerule
Type  & $a_1$ & $a_2$ & $a_3$ & $a_4$ & $a_5$ & $b=b_1$ & $b_2$ & $b_3$  \\
\colrule
$d_L$ & $d-1$ & $d-1$ & $d+1$ & $d-1$ & $d$   & $d+1$   & $d+1$ & $d$   \\
$d_R$ & $d-1$ & $d+1$ & $d-1$ & $d$   & $d-1$ & $d+1$   & $d$   & $d+1$ \\
\widerule
\end{tabular}
\caption{Relative depth of vertex neighbours to the left and right of a given vertex with depth $d$.}
\label{TAB_VERTEXTYPES}
\end{table}

In order to build inflation rules for regular tilings that inflate vertices on radial geodesics, we need to label vertices by their graph distance $d$ to the center of the tiling, or \emph{depth}. For an $\{n,k\}$ tiling with $n>5$ and $k>3$, we first distinguish between $a$ vertices, which have two neighbouring vertices (up to the given inflation layer), and $b$ vertices, which have three. Within the sequence of vertices at a given layer, we consider the depths $d_L$ and $d_R$ of the neighbours to the left and right with respect to the depth $d$ of a given vertex. For $b$ vertices, $(d_L,d_R)=(d+1,d+1)$. However, we need to distinguish five types of $a$ vertices, listed in Tab.\ \ref{TAB_VERTEXTYPES}. For even $n=2m$, only $a_1$ to $a_3$ appear, leading to an inflation rule
\begin{align}
a_1 &\mapsto a_3^{m-3} b \left( a_2^{m-2} a_1 a_3^{m-2} b \right)^{k-3} a_2^{m-3} a_1 \ ,\\
a_2 &\mapsto a_3^{m-3} b \left( a_2^{m-2} a_1 a_3^{m-2} b \right)^{k-3} a_2^{m-2} a_1 \ ,\\
a_3 &\mapsto a_3^{m-2} b \left( a_2^{m-2} a_1 a_3^{m-2} b \right)^{k-3} a_2^{m-3} a_1 \ ,\\
b &\mapsto a_3^{m-2} b \left( a_2^{m-2} a_1 a_3^{m-2} b \right)^{k-4} a_2^{m-2} a_1 \ ,
\end{align}
and a corresponding substitution matrix
\begin{equation}
M_{\{ 2m,k \}} = 
\left(
\begin{array}{cccc}
 k-2 & k (m-2)-2 m+3 & k (m-2)-2 m+3 & k-2 \\
 k-2 & (k-2) (m-2) & k (m-2)-2 m+3 & k-2 \\
 k-2 & k (m-2)-2 m+3 & (k-2) (m-2) & k-2 \\
 k-3 & (k-3) (m-2) & (k-3) (m-2) & k-3 \\
\end{array}
\right) \ .
\end{equation}
The edge increase from inflation onto a $b$ vertex is always $1$, and increases with distance from the nearest $b$ vertex. This is summarized in the entanglement matrix
\begin{equation}
E_{\{ 2m,k \}} = 
\left(
\begin{array}{cccc}
 m+\frac{1}{2-k} & \frac{-2 m^2+k (m-2) (m+1)+6}{2 k (m-2)-4 m+6} & \frac{-2 m^2+k (m-2) (m+1)+6}{2 k (m-2)-4 m+6} & 1 \\
 m & \frac{m+1}{2} & \frac{-2 m^2+k (m-2) (m+1)+6}{2 k (m-2)-4 m+6} & 1 \\
 m+\frac{1}{2-k} & \frac{-2 m^2+k (m-2) (m+1)+6}{2 k (m-2)-4 m+6} & \frac{m+1}{2} & 1 \\
 m & \frac{m+1}{2} & \frac{m+1}{2} & 1 \\
\end{array}
\right)\ .
\end{equation}
Applying \eqref{EQ_C_NK} leads to the central charge bound
\begin{equation}
\label{EQ_C_EVEN_N}
c_{\{2m,k\}} \leq c_{\{2m,k\}}^\text{max} = \frac{3 (m+1) \ln\chi}{\ln \left(k (m-1)+\sqrt{(k-2) (m-1) ((k-2) m-k)}-2 m+1\right)} \ ,
\end{equation}
where $\chi$ is the bond dimension of the underlying tensor network embedded into the $\{2m,k\}$ tiling.
For odd $n=2m{+}1$, the inflation rule is more complicated and includes all five types of $a$ vertices,
\begin{align}
a_1 &\mapsto a_5 a_3^{m-3} b \left( a_2^{m-1} a_4 a_5 a_3^{m-1} b \right)^{k-3} a_2^{m-3} a_4 \ ,\\
a_2 &\mapsto a_5 a_3^{m-3} b \left( a_2^{m-1} a_4 a_5 a_3^{m-1} b \right)^{k-3} a_2^{m-1} a_4 \ ,\\
a_3 &\mapsto a_5 a_3^{m-1} b \left( a_2^{m-1} a_4 a_5 a_3^{m-1} b \right)^{k-3} a_2^{m-3} a_4 \ ,\\
a_4 &\mapsto a_3^{m-3} b \left( a_2^{m-1} a_4 a_5 a_3^{m-1} b \right)^{k-3} a_2^{m-1} a_1 \ ,\\
a_5 &\mapsto a_5 a_3^{m-1} b \left( a_2^{m-1} a_4 a_5 a_3^{m-1} b \right)^{k-3} a_2^{m-3} a_4 \ ,\\
b   &\mapsto a_3^{m-1} b \left( a_2^{m-1} a_4 a_5 a_3^{m-1} b \right)^{k-4} a_2^{m-1} a_4 \ .
\end{align}
This leads to a substitution matrix
\begin{equation}
M_{\{ 2m+1,k \}} = 
\left(
\begin{array}{cccccc}
 0 & k (m-2)-2 m+3 & k (m-2)-2 m+3 & k-2 & k-2 & k-2 \\
 0 & (k-2) (m-2) & k (m-2)-2 m+3 & k-2 & k-2 & k-2 \\
 0 & k (m-2)-2 m+3 & (k-2) (m-2) & k-2 & k-2 & k-2 \\
 1 & (k-2) (m-2) & k (m-2)-2 m+3 & k-3 & k-2 & k-2 \\
 0 & k (m-2)-2 m+3 & (k-2) (m-2) & k-2 & k-3 & k-2 \\
 0 & (k-3) (m-2) & (k-3) (m-2) & k-3 & k-3 & k-3 \\
\end{array}
\right) \ .
\end{equation}
The entanglement matrix is given by
\begin{equation}
E_{\{ 2m+1,k \}} = 
\left(
\begin{array}{cccccc}
 0 & \frac{-2 m^2+k (m-2) (m+1)+6}{2 k (m-2)-4 m+6} & \frac{-2 m^2+k (m-2) (m+1)+6}{2 k (m-2)-4 m+6} & m+\frac{1}{2-k} & m+\frac{1}{2-k} & 1 \\
 0 & \frac{m+1}{2} & \frac{-2 m^2+k (m-2) (m+1)+6}{2 k (m-2)-4 m+6} & m & m+\frac{1}{2-k} & 1 \\
 0 & \frac{-2 m^2+k (m-2) (m+1)+6}{2 k (m-2)-4 m+6} & \frac{m+1}{2} & m+\frac{1}{2-k} & m & 1 \\
 m & \frac{m+1}{2} & \frac{-2 m^2+k (m-2) (m+1)+6}{2 k (m-2)-4 m+6} & m & m+\frac{1}{2-k} & 1 \\
 0 & \frac{-2 m^2+k (m-2) (m+1)+6}{2 k (m-2)-4 m+6} & \frac{m+1}{2} & m+\frac{1}{2-k} & m & 1 \\
 0 & \frac{m+1}{2} & \frac{m+1}{2} & m & m & 1 \\
\end{array}
\right) \ .
\end{equation}
The resulting central charge bound is
\begin{equation}
\label{EQ_C_ODD_N}
c_{\{2m+1,k\}}^\text{max} = \frac{3 \left(m-\frac{1}{4 m-2}+\frac{3}{2}\right) \ln\chi}{\ln \frac{2 k m+\sqrt{(-2 k m+k+4 m)^2-4}-k-4 m}{2}} \ .
\end{equation}
Note that for large $n$, \eqref{EQ_C_EVEN_N} and \eqref{EQ_C_ODD_N} lead to the same asymptotic behavior,
\begin{equation}
c_{\{n,k\}}^\text{max} = \frac{(6+3n) \ln\chi}{2\ln\left( 2-2 k + (k-2) n \right)} + O\left( n^{-1} \right) \ .
\end{equation}
For $\{ n,3 \}$ tilings (hyperbolic for $n>6$), we also need to distinguish between even and odd $n$. In the case $n=2m$, we find the inflation rule
\begin{align}
a_1 &\mapsto a_3^{m-3} b a_2^{m-3} a_1 \ ,\\
a_2 &\mapsto a_3^{m-3} b a_2^{m-2} a_1 \ ,\\
a_3 &\mapsto a_3^{m-2} b a_2^{m-3} a_1 \ ,\\
b   &\mapsto \emptyset \ .
\end{align}
and the substitution and entanglement matrices 
\begin{align}
M_{\{ 2m,3 \}} &= 
\left(
\begin{array}{cccc}
 1 & m-3 & m-3 & 1 \\
 1 & m-2 & m-3 & 1 \\
 1 & m-3 & m-2 & 1 \\
 0 & 0 & 0 & 0 \\
\end{array}
\right) \ , &
E_{\{ 2m,3 \}} &=
\left(
\begin{array}{cccc}
 m-1 & \frac{m}{2} & \frac{m}{2} & 1 \\
 m & \frac{m+1}{2} & \frac{m}{2} & 1 \\
 m-1 & \frac{m}{2} & \frac{m+1}{2} & 1 \\
 0 & 0 & 0 & 0 \\
\end{array}
\right) \ . 
\end{align}
This yields a maximum central charge
\begin{equation}
\label{EQ_C_EVEN_N_K3}
c_{\{2m,3\}}^\text{max} = \frac{3 (m+1) \ln\chi}{\ln \left(\sqrt{m^2-4 m+3}+m-2\right)} \ .
\end{equation}
For odd $n=2m+1$, inflation again involves $a_1$ to $a_5$:
\begin{align}
a_1 &\mapsto a_5 a_3^{m-3} b a_2^{m-3} a_4 \ ,\\
a_2 &\mapsto a_5 a_3^{m-3} b a_2^{m-2} a_4 \ ,\\
a_3 &\mapsto a_5 a_3^{m-2} b a_2^{m-3} a_4 \ ,\\
a_4 &\mapsto a_5 a_3^{m-3} b a_2^{m-2} a_1 \ ,\\
a_5 &\mapsto a_3^{m-2} b a_2^{m-3} a_4 \ ,\\
b   &\mapsto \emptyset \ .
\end{align}
This corresponds to 
\begin{align}
M_{\{ 2m+1,3 \}} &= 
\left(
\begin{array}{cccccc}
 0 & m-3 & m-3 & 1 & 1 & 1 \\
 0 & m-2 & m-3 & 1 & 1 & 1 \\
 0 & m-3 & m-2 & 1 & 1 & 1 \\
 1 & m-2 & m-3 & 0 & 1 & 1 \\
 0 & m-3 & m-2 & 1 & 0 & 1 \\
 0 & 0 & 0 & 0 & 0 & 0 \\
\end{array}
\right) \ , \\
E_{\{ 2m+1,3 \}} &= 
\left(
\begin{array}{cccccc}
 0 & \frac{m}{2} & \frac{m}{2} & m-1 & m-1 & 1 \\
 0 & \frac{m+1}{2} & \frac{m}{2} & m & m-1 & 1 \\
 0 & \frac{m}{2} & \frac{m+1}{2} & m-1 & m & 1 \\
 m & \frac{m+1}{2} & \frac{m}{2} & 0 & m-1 & 1 \\
 0 & \frac{m}{2} & \frac{m+1}{2} & m-1 & 0 & 1 \\
 0 & 0 & 0 & 0 & 0 & 0 \\
\end{array}
\right) \ ,
\end{align}
and gives a central charge bound of
\begin{equation}
\label{EQ_C_ODD_N_K3}
c_{\{2m+1,3\}}^\text{max} = \frac{3 \left(m-\frac{1}{4 m-2}+\frac{3}{2}\right) \ln\chi}{\ln \frac{\sqrt{4 m^2-12 m+5}+2 m-3}{2}} \ .
\end{equation}
Note that even though the inflation rules are different, the bounds \eqref{EQ_C_EVEN_N_K3} and \eqref{EQ_C_ODD_N_K3} agree with the generic $\{n,k\}$ bounds \eqref{EQ_C_EVEN_N} and \eqref{EQ_C_ODD_N} derived earlier.
Similarly, the $\{n,k\}$ inflation rules for $n=4$ and $n=5$ are special, as well, but lead to the same bounds. The $n=4$ case was already covered in the main text. For $n=5$, we need to split up $b$ vertices into three categories $b_1$, $b_2$, and $b_3$. 
For $n=5$, we find the inflation rules
\begin{align}
a_1 &\mapsto b_3 (a_2 a_3 b_1)^{k-4} a_2 a_3 b_2 \ ,\\
a_2 &\mapsto b_3 (a_2 a_3 b_1)^{k-4} a_2 a_3 b_1 a_1\ ,\\
a_3 &\mapsto b_1 (a_2 a_3 b_1)^{k-4} a_2 a_3 b_2 \ ,\\
b_1 &\mapsto a_3 b_1 (a_2 a_3 b_1)^{k-4} a_2 \ ,\\
b_2 &\mapsto a_3 b_1 (a_2 a_3 b_1)^{k-4} a_1 \ ,\\
b_3 &\mapsto b_1 (a_2 a_3 b_1)^{k-4} a_2 \ ,
\end{align}
leading to substitution and entanglement matrices
\begin{align}
M_{\{ 5,k \}} &= 
\left(
\begin{array}{cccccc}
 0 & k-3 & k-3 & k-4 & 1 & 1 \\
 1 & k-3 & k-3 & k-3 & 0 & 1 \\
 0 & k-3 & k-3 & k-3 & 1 & 0 \\
 0 & k-3 & k-3 & k-3 & 0 & 0 \\
 1 & k-4 & k-3 & k-3 & 0 & 0 \\
 0 & k-3 & k-4 & k-3 & 0 & 0 \\
\end{array}
\right) \ , &
E_{\{ 5,k \}} &= 
\left(
\begin{array}{cccccc}
 0 & 2 & 2 & 1 & 1 & 1 \\
 2 & 2 & 2 & 1 & 0 & 1 \\
 0 & 2 & 2 & 1 & 1 & 0 \\
 0 & 2 & 2 & 1 & 0 & 0 \\
 2 & 2 & 2 & 1 & 0 & 0 \\
 0 & 2 & 2 & 1 & 0 & 0 \\
\end{array}
\right) \ .
\end{align}
This yields the expected maximum central charge
\begin{equation}
c_{\{5,k\}}^\text{max} = \frac{10 \ln\chi}{\ln \frac{ \sqrt{9 k^2-48 k+60}+3 k-8}{2}} \ .
\end{equation}

\newpage
\section{Majorana dimer polygon models}
\label{APP_DIMERS}

We can construct block perfect Majorana dimer models for an $\{n,k\}$ tiling for $n=4m{+}1,m \in \mathbb{N}$. The $n=5$ case is simply the HyPeC model.
In the main text, we already computed its central charge under edge inflation, which we now generalize to vertex inflation.
From \eqref{EQ_REGINF_NK}, we find the inflation rule
\begin{align}
a &\mapsto abaab \ , &
b &\mapsto ab \ .
\end{align}
Without loss of generality, we identify each letter with the edge on the left/clockwise to the vertex it stands for. To distinguish dimer content, we need to designate four sub-letters $a_1,a_2,b_1,b_2$. In terms of dimer diagrams, using the convention of \eqref{EQ_HYPEC_STATES}, the inflation rule is given by
\begin{align}
\begin{gathered}
\includegraphics[height=0.1\textwidth]{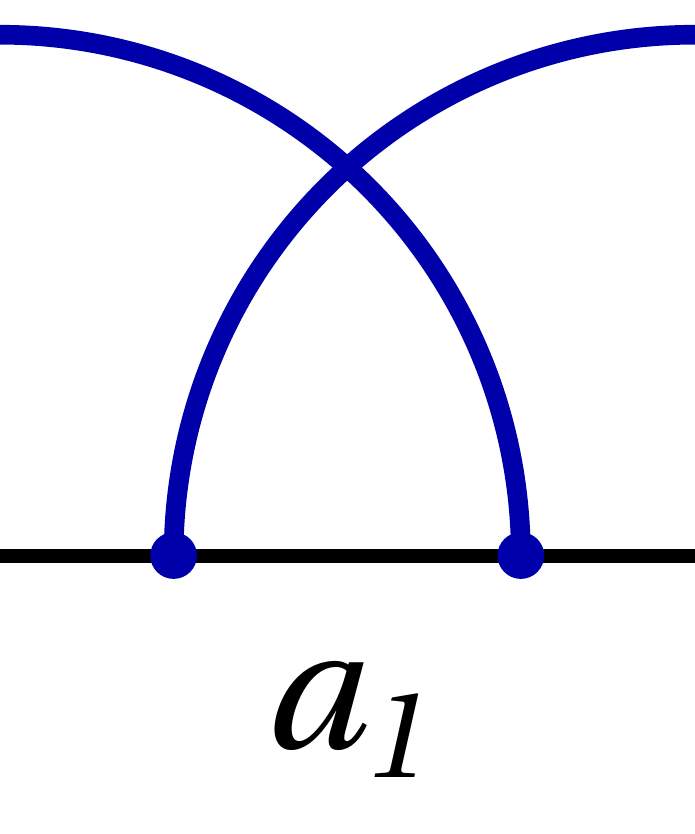}
\end{gathered}
&\scalebox{1.25}{$\quad\mapsto\quad\;$}
\begin{gathered}
\includegraphics[height=0.1\textwidth]{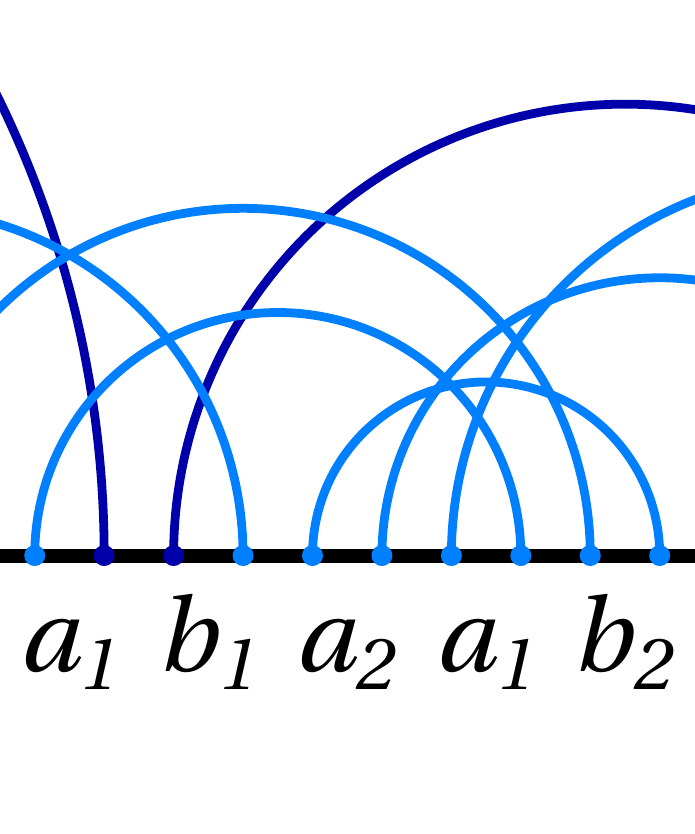}
\end{gathered} \ , &
\begin{gathered}
\includegraphics[height=0.1\textwidth]{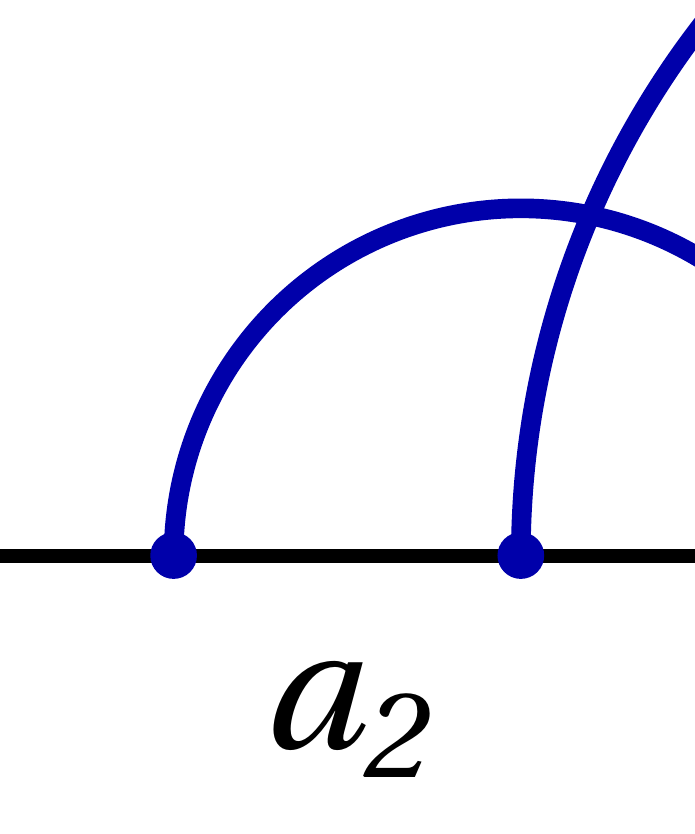}
\end{gathered}
&\scalebox{1.25}{$\quad\mapsto\quad\;$}
\begin{gathered}
\includegraphics[height=0.1\textwidth]{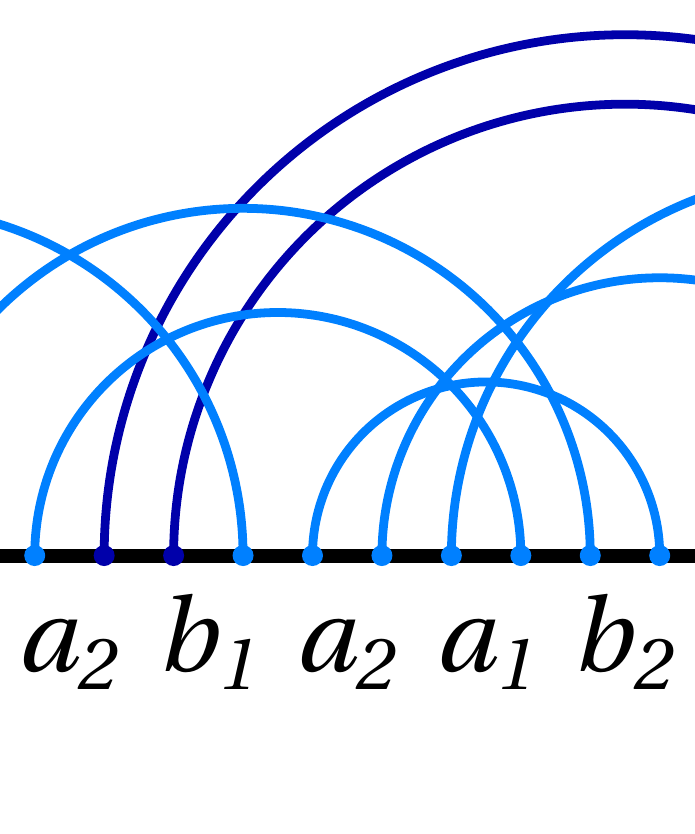}
\end{gathered} \ , \\
\begin{gathered}
\includegraphics[height=0.1\textwidth]{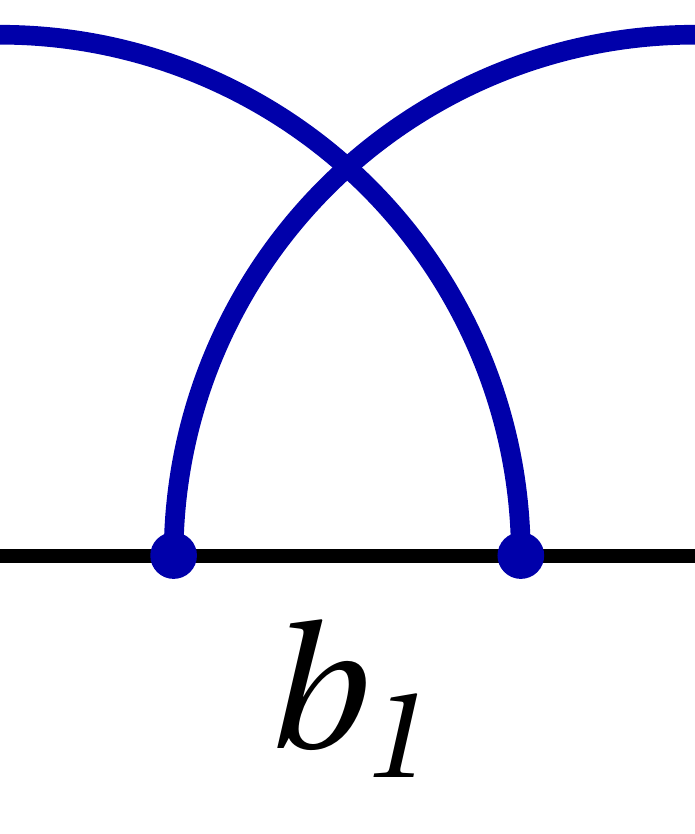}
\end{gathered}
&\scalebox{1.25}{$\quad\mapsto\quad\;$}
\begin{gathered}
\includegraphics[height=0.1\textwidth]{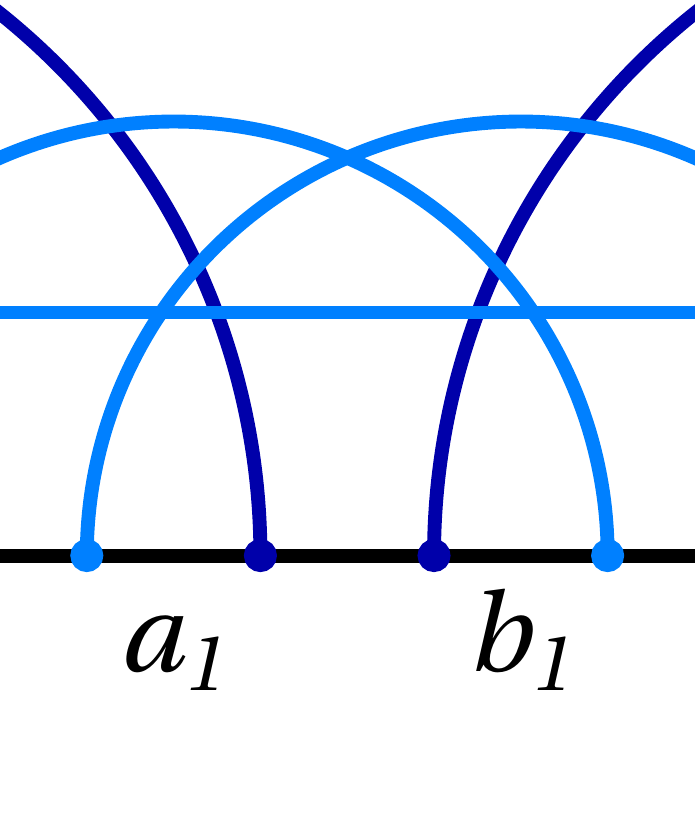}
\end{gathered} \ , &
\begin{gathered}
\includegraphics[height=0.1\textwidth]{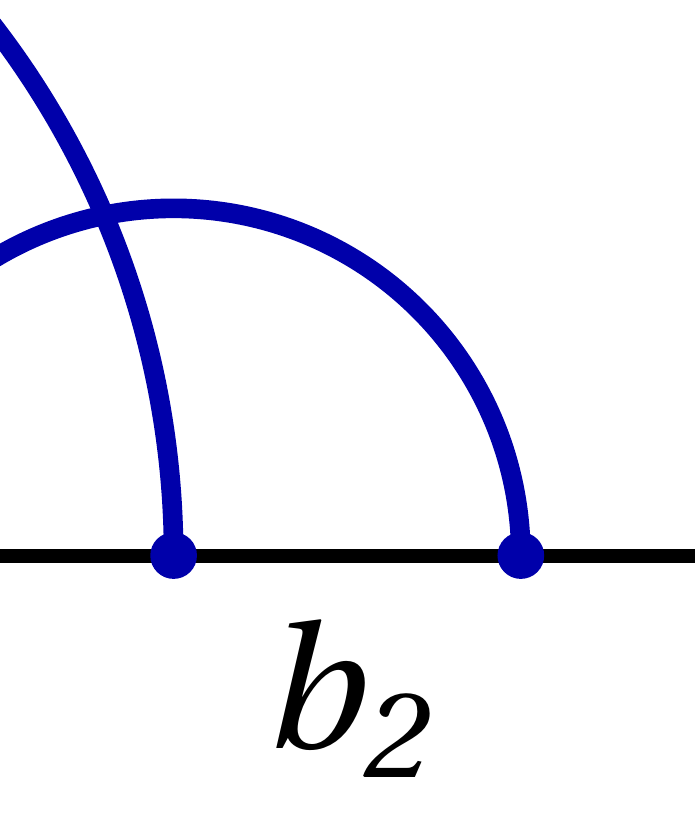}
\end{gathered}
&\scalebox{1.25}{$\quad\mapsto\quad\;$}
\begin{gathered}
\includegraphics[height=0.1\textwidth]{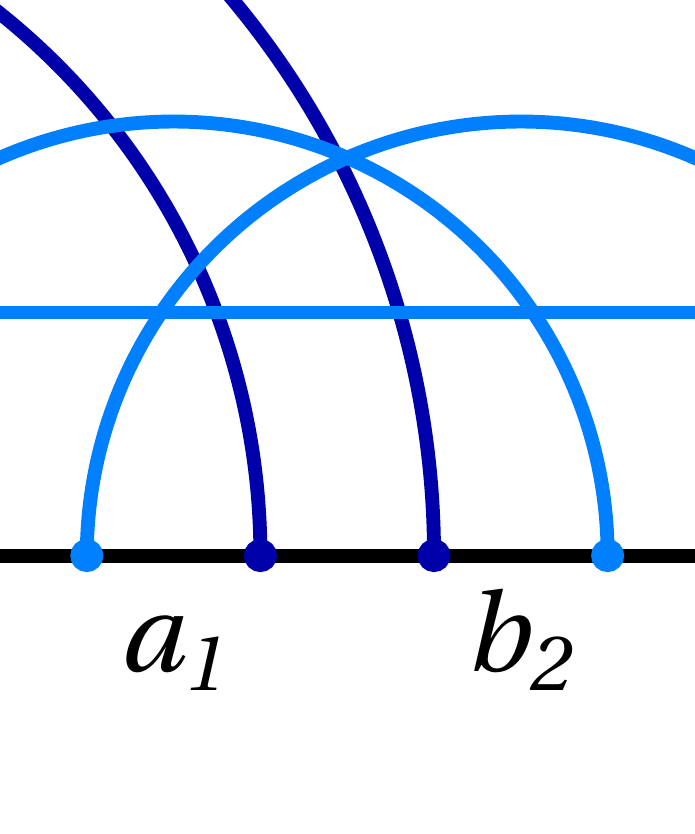}
\end{gathered} \ .
\end{align}
As in the main text, new dimers added at each step are drawn in a lighter colour than those that are extended from the previous layer. 
Note that each inflated dimer configuration contains two open dimers on either end of the sequence that connect to the previous and following sequence within the layer added in a vertex inflation step.
The full dimer configuration in the Poincar\'e disk is shown in Fig.\ \ref{FIG_DIMER_54_FULL} (top) along with the dimers at the first three inflation layers. When starting from the central pentagon, the initial sequence is given by $(a_1)^5$.

Again, we calculate the central charge by considering the loss of local entanglement through deflation.
The corresponding cuts (green lines) and the number of dimers passing through it (green number) are given by
\begin{align}
\begin{gathered}
\includegraphics[height=0.1\textwidth]{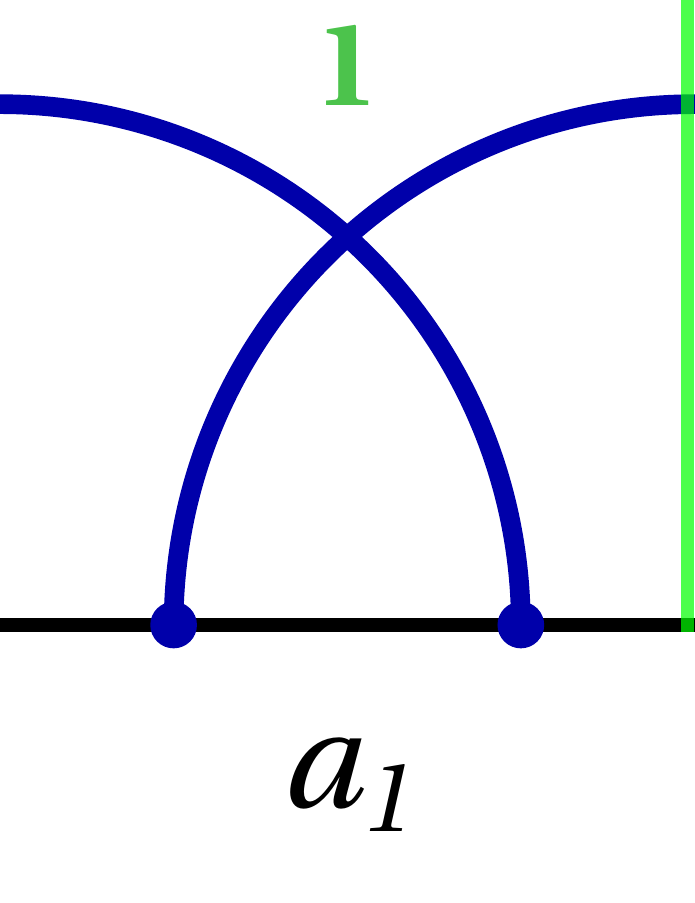}
\end{gathered}
&\scalebox{1.25}{$\quad\mapsfrom\quad\;$}
\begin{gathered}
\includegraphics[height=0.1\textwidth]{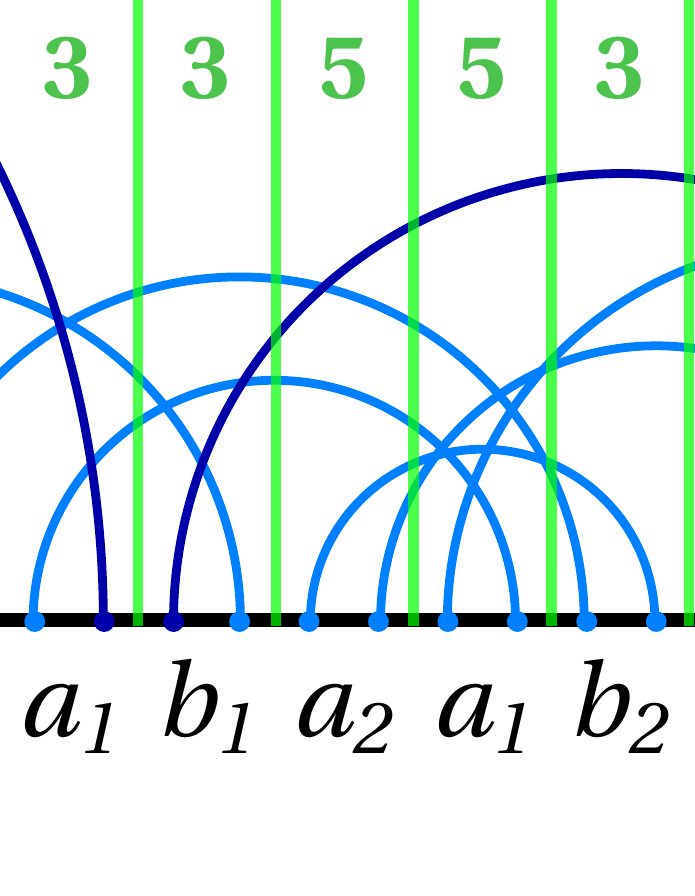}
\end{gathered} \ , &
\begin{gathered}
\includegraphics[height=0.1\textwidth]{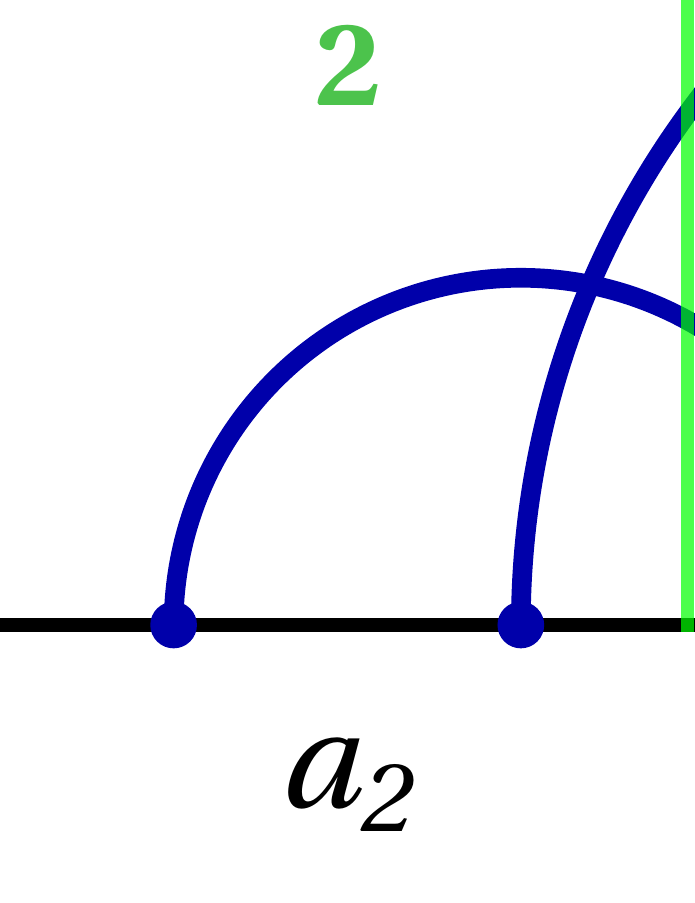}
\end{gathered}
&\scalebox{1.25}{$\quad\mapsfrom\quad\;$}
\begin{gathered}
\includegraphics[height=0.1\textwidth]{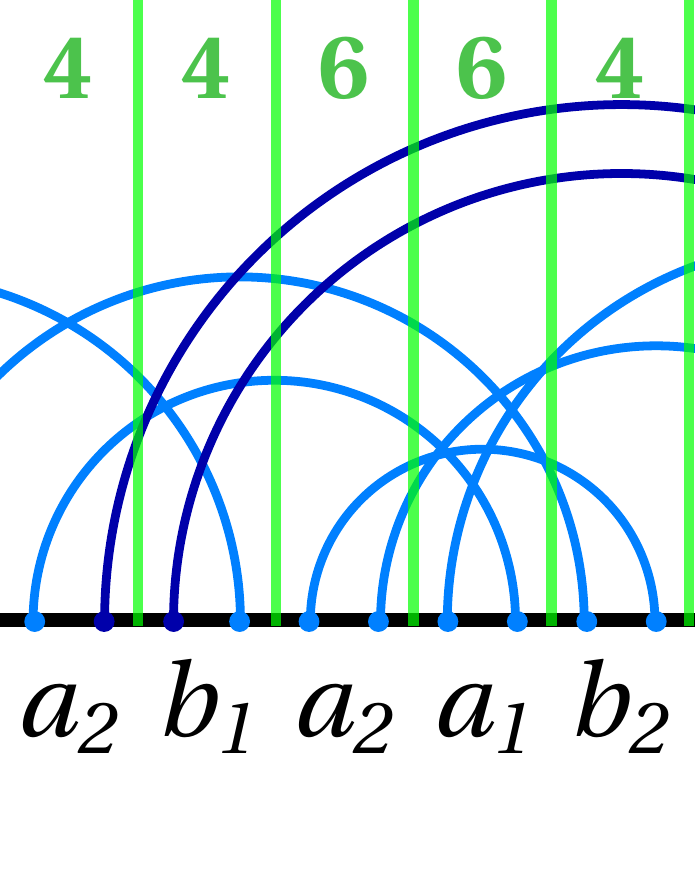}
\end{gathered} \ ,\\
\begin{gathered}
\includegraphics[height=0.1\textwidth]{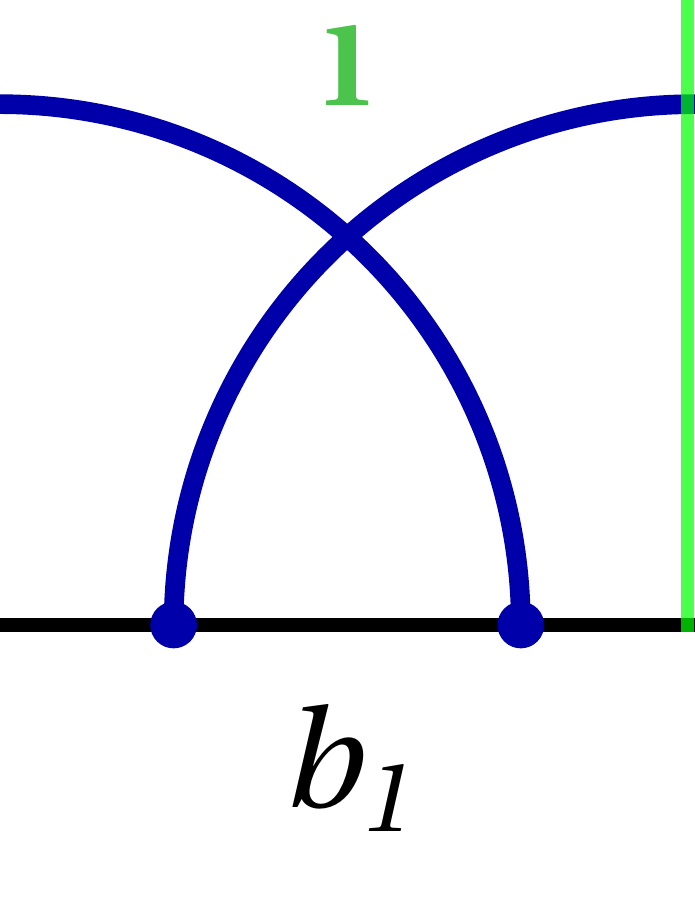}
\end{gathered}
&\scalebox{1.25}{$\quad\mapsfrom\quad\;$}
\begin{gathered}
\includegraphics[height=0.1\textwidth]{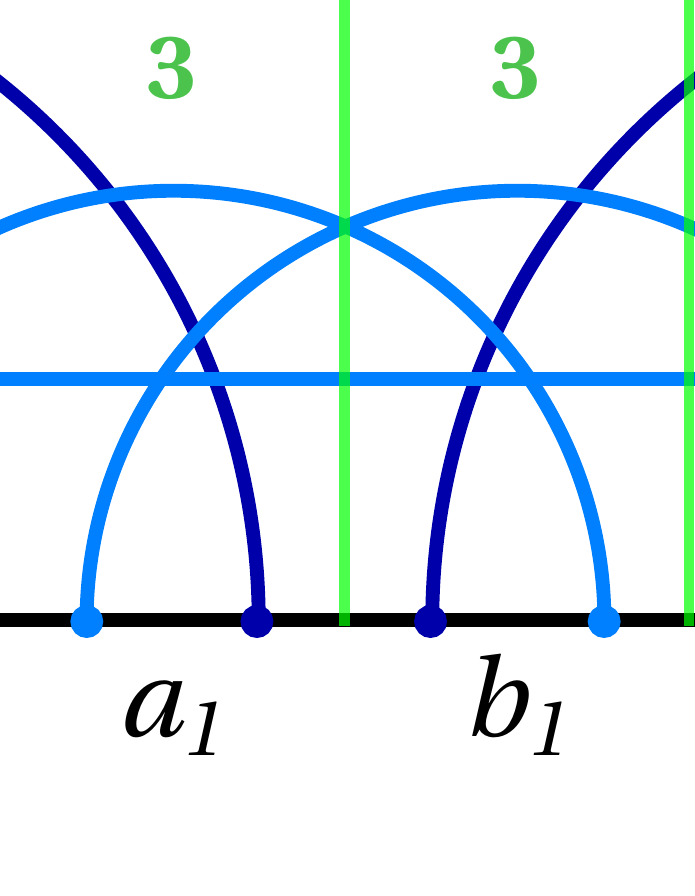}
\end{gathered} \ , &
\begin{gathered}
\includegraphics[height=0.1\textwidth]{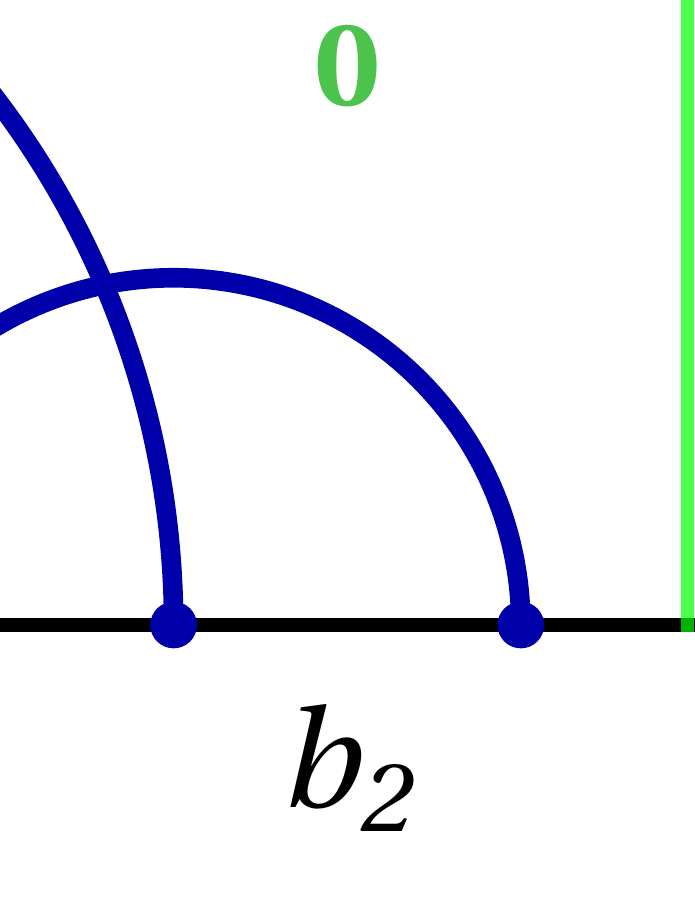}
\end{gathered}
&\scalebox{1.25}{$\quad\mapsfrom\quad\;$}
\begin{gathered}
\includegraphics[height=0.1\textwidth]{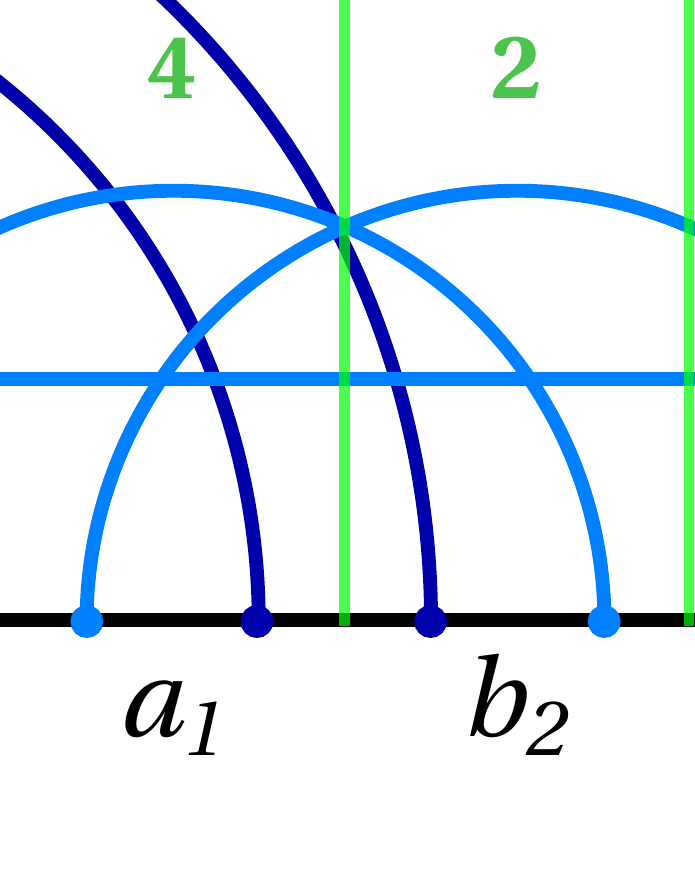}
\end{gathered} \ .
\end{align}
These diagrams lead us to the substitution and entanglement matrices $M$ and $E$ of the Markov process of the form
\begin{align}
M &= 
\left(
\begin{array}{cccc}
 2 & 1 & 1 & 1 \\
 1 & 2 & 1 & 1 \\
 1 & 0 & 1 & 0 \\
 1 & 0 & 0 & 1 \\
\end{array}
\right) \ , &
E &=
\left(
\begin{array}{cccc}
 \frac{3}{2} & 2 & 1 & 1 \\
 2 & \frac{3}{2} & 1 & 1 \\
 1 & 0 & 1 & 0 \\
 2 & 0 & 0 & 1 \\
\end{array}
\right) \ .
\end{align}
Turning \eqref{EQ_C_NK} into an equality, we find the effective central charge
\begin{equation}
\label{EQ_C_54_VERTEX}
c_{\{5,4\}}^\text{d} = \frac{9 \ln 2}{\ln\left( \sqrt{3}+2 \right)} \approx 4.74 \ .
\end{equation}
Note that this result is \emph{larger} than $c_{\{5,4\}_e}^\text{d}$ from \eqref{EQ_C_54_EDGE}, our result for edge inflation. 
Instead of a $\{5,4\}$ tiling, we can also consider a general $\{5,k\}$ tiling with $k>3$, using the same perfect tensors on each tile. This corresponds to a vertex inflation rule
\begin{align}
a_1 &\mapsto a_1 b_1 \left( a_2 a_1 b_2 \right)^{k-3} \ , &
b_1 &\mapsto a_1 b_1 \left( a_2 a_1 b_2 \right)^{k-4} \ , \\
a_2 &\mapsto a_2 b_1 \left( a_2 a_1 b_2 \right)^{k-3} \ , &
b_2 &\mapsto a_1 b_2 \left( a_2 a_1 b_2 \right)^{k-4} \ .
\end{align}
The substitution and entanglement matrices then take the more general form
\begin{align}
M &= 
\left(
\begin{array}{cccc}
 k-2 & k-3 & 1 & k-3 \\
 k-3 & k-2 & 1 & k-3 \\
 k-3 & k-4 & 1 & k-4 \\
 k-3 & k-4 & 0 & k-3 \\
\end{array}
\right) \ ,  
\end{align}
and
\begin{align}
E &=
\left(
\begin{array}{cccc}
 \frac{1 + 2 (k-3)}{1+k-3} & 2 & 1 & 1 \\
 2 & \frac{1 + 2 (k-3)}{1+k-3} & 1 & 1 \\
 \frac{1+ 2 (k-4)}{1+k-4} & 2 & 1 & 1 \\
 2 & 2 & 0 & 1 \\
\end{array}
\right) \ .
\end{align}
This leads us to the central charge
\begin{equation}
c_{\{5,k\}}^\text{d} = \frac{ \left(\frac{2}{10-3 k}+10\right) \ln 2}{\ln \left(\frac{1}{2} \left(\sqrt{9 k^2-48 k+60}+3 k-8\right)\right)} \ .
\end{equation}
Note that this model corresponds to a bond dimension $\chi=2$, hence the $\ln 2$ term in the numerator. Considering the large $k$ limit, we find 
\begin{equation}
c_{\{5,k\}}^\text{d}  = \frac{10 \ln 2}{\ln\left( 3k - 8 \right)}  + O\left( k^{-1} \right) \ ,
\end{equation}
which is exactly the same limit as the geodesic bound on central charges (Tab.\ \ref{TAB_C}). As shown in Fig.\ \ref{FIG_CDIMERS}, this saturation occurs quickly as $k$ is increased.

For $n=9$ and more complex polygons, we have to distinguish two cases: If $k=3$, the inflation rule requires five different types of letters, while only four are needed in the $k>3$ case. 
The inflation rule for the $\{9,3\}$ tiling with dimer states \eqref{EQ_HYNEC_STATES} follows from \eqref{EQ_REGINF_K3} and is given by the following dimer substitutions:
\begin{align}
\begin{gathered}
\includegraphics[height=0.1\textwidth]{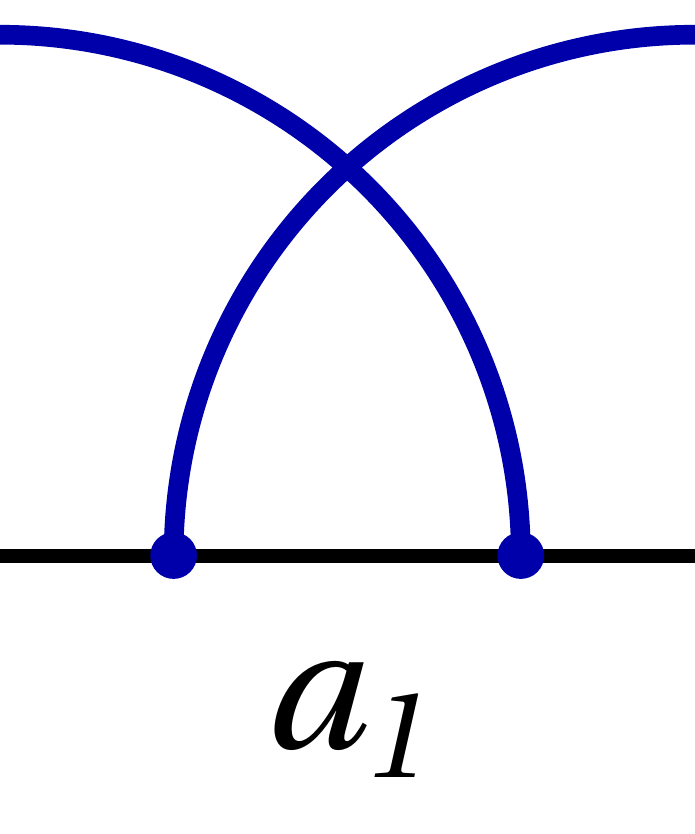}
\end{gathered}
&\scalebox{1.25}{$\quad\mapsto\quad\;$}
\begin{gathered}
\includegraphics[height=0.1\textwidth]{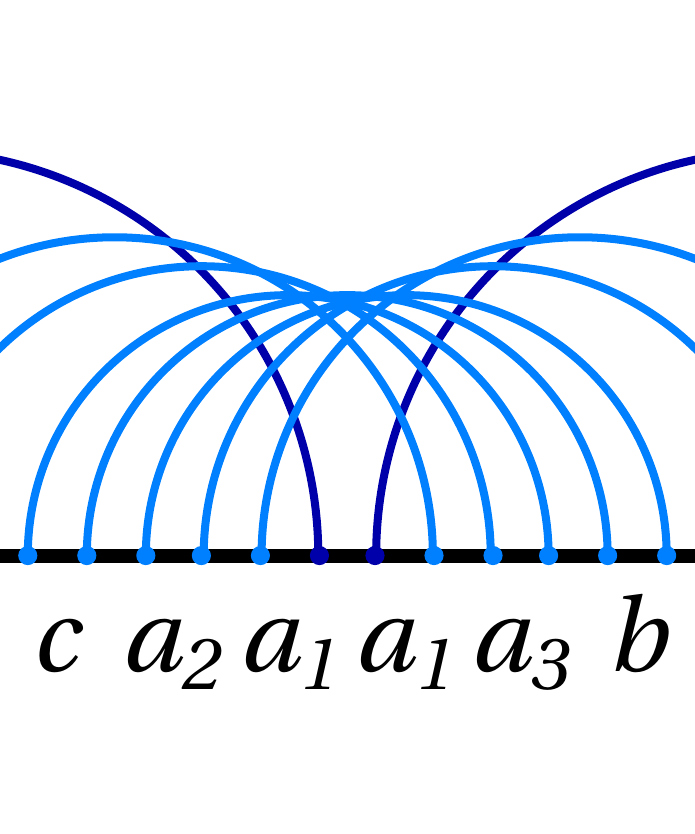}
\end{gathered} \ , &
\begin{gathered}
\includegraphics[height=0.1\textwidth]{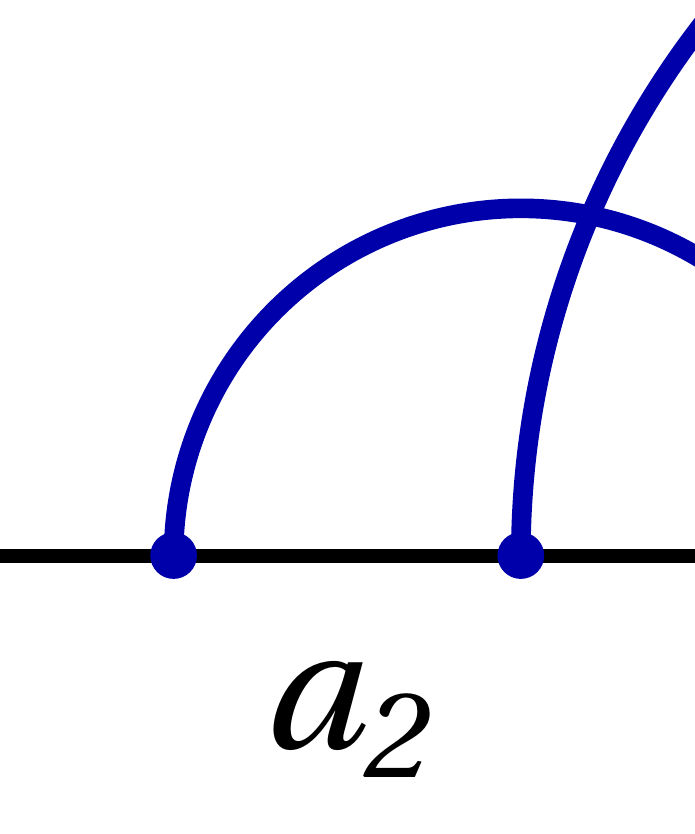}
\end{gathered}
&\scalebox{1.25}{$\quad\mapsto\quad\;$}
\begin{gathered}
\includegraphics[height=0.1\textwidth]{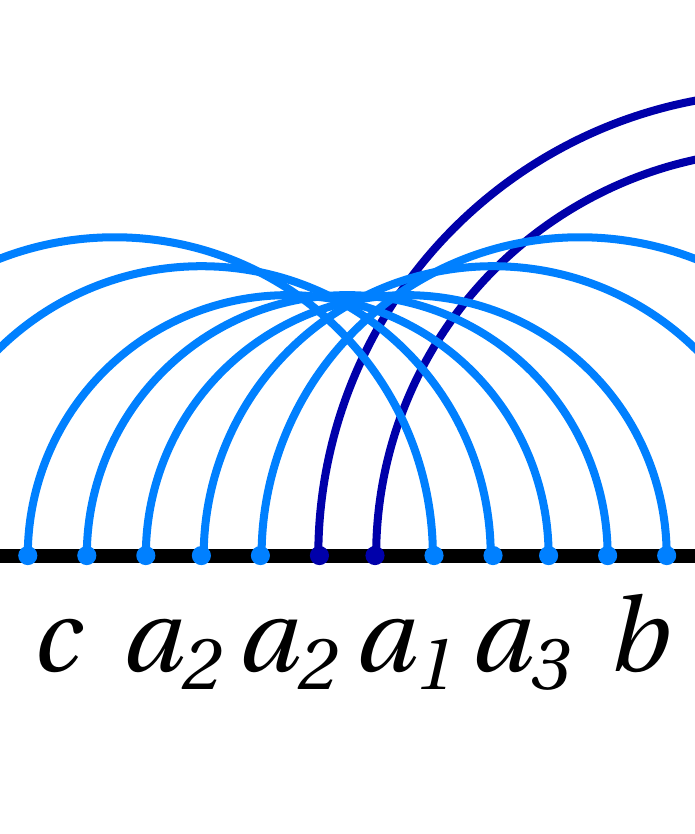}
\end{gathered} \ , \\
\begin{gathered}
\includegraphics[height=0.1\textwidth]{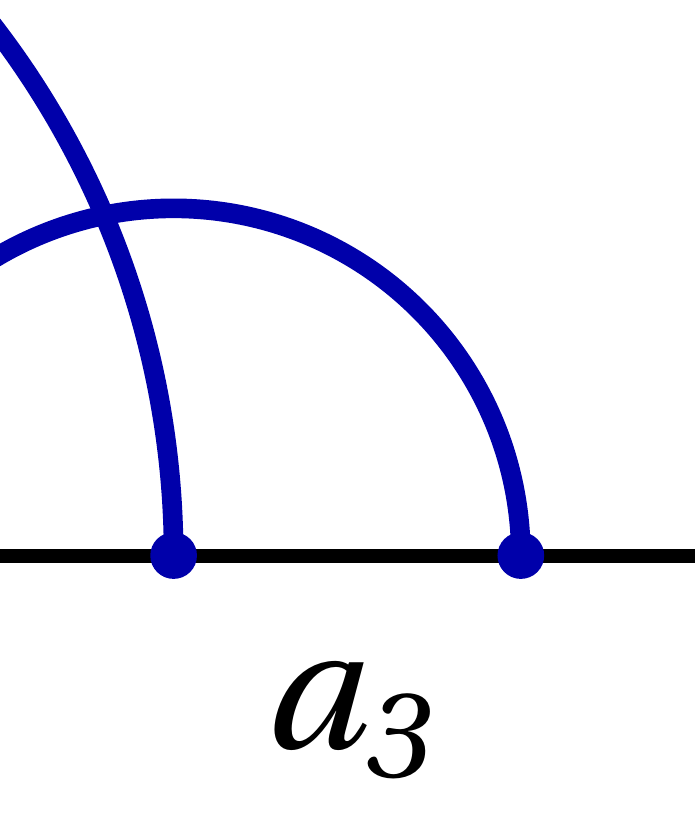}
\end{gathered}
&\scalebox{1.25}{$\quad\mapsto\quad\;$}
\begin{gathered}
\includegraphics[height=0.1\textwidth]{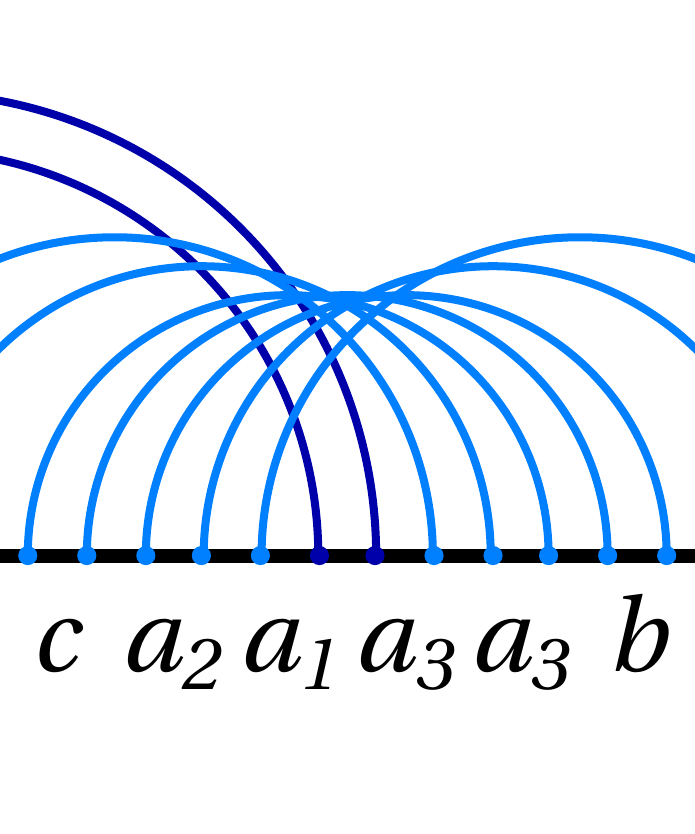}
\end{gathered} \ , &
\begin{gathered}
\includegraphics[height=0.1\textwidth]{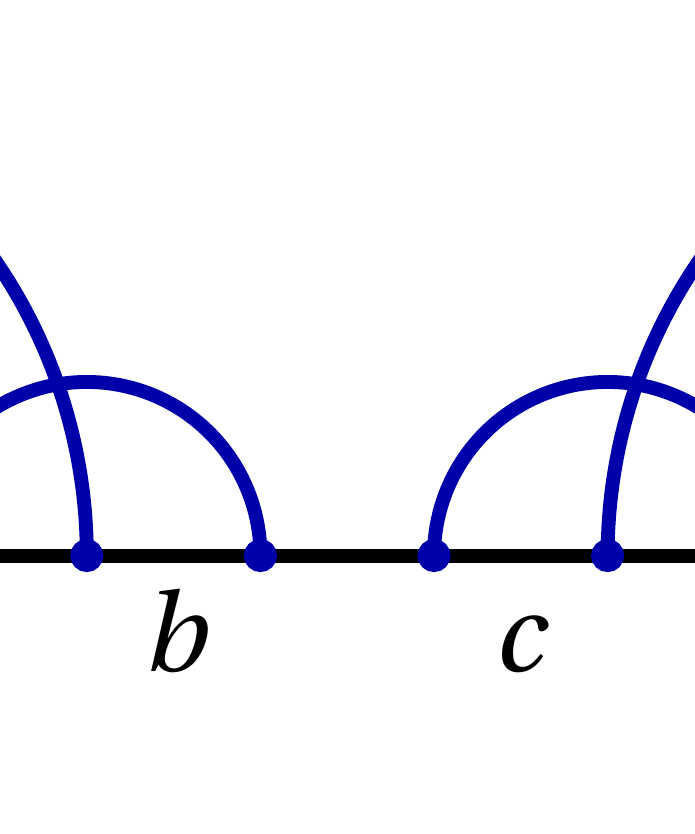}
\end{gathered}
&\scalebox{1.25}{$\quad\mapsto\quad\;$}
\begin{gathered}
\includegraphics[height=0.1\textwidth]{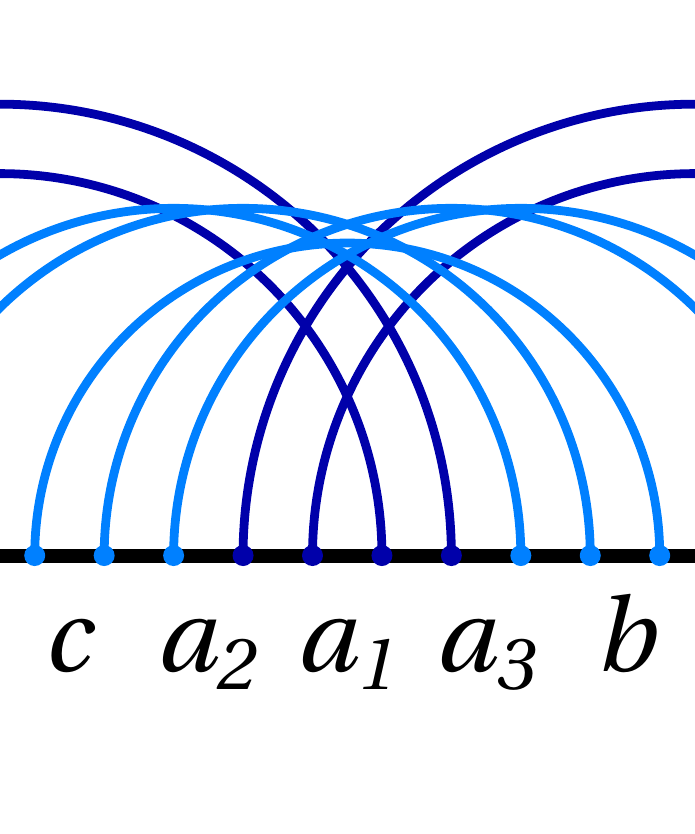}
\end{gathered} \ .
\end{align}
The inflation rule for the letters $b$ and $c$ has been combined for the sake of simplicity. The entanglement change under deflation depends on the cut and is given by
\begin{align}
\begin{gathered}
\includegraphics[height=0.1\textwidth]{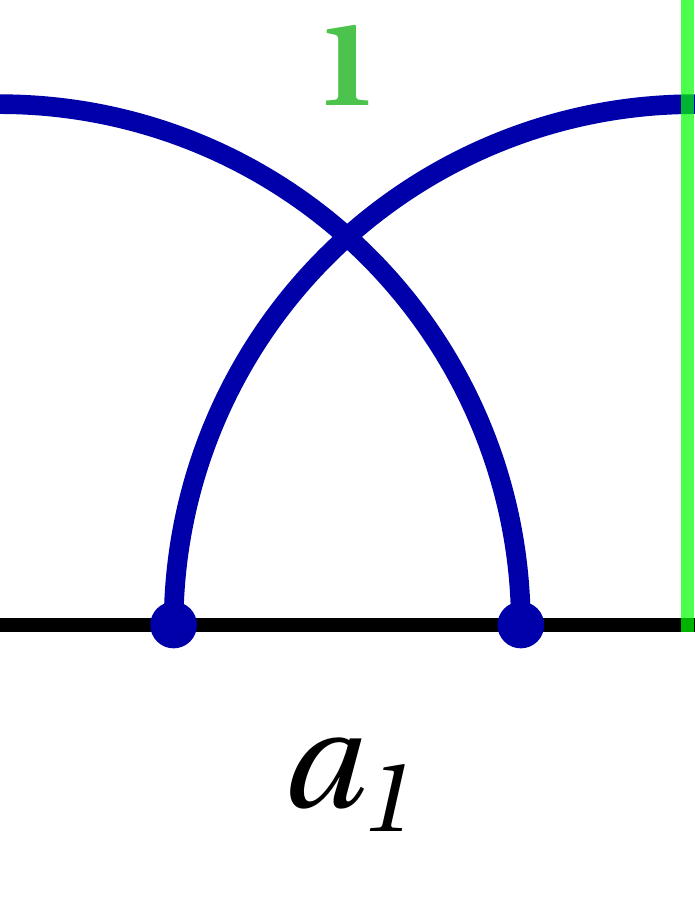}
\end{gathered}
&\scalebox{1.25}{$\quad\mapsfrom\quad\;$}
\begin{gathered}
\includegraphics[height=0.1\textwidth]{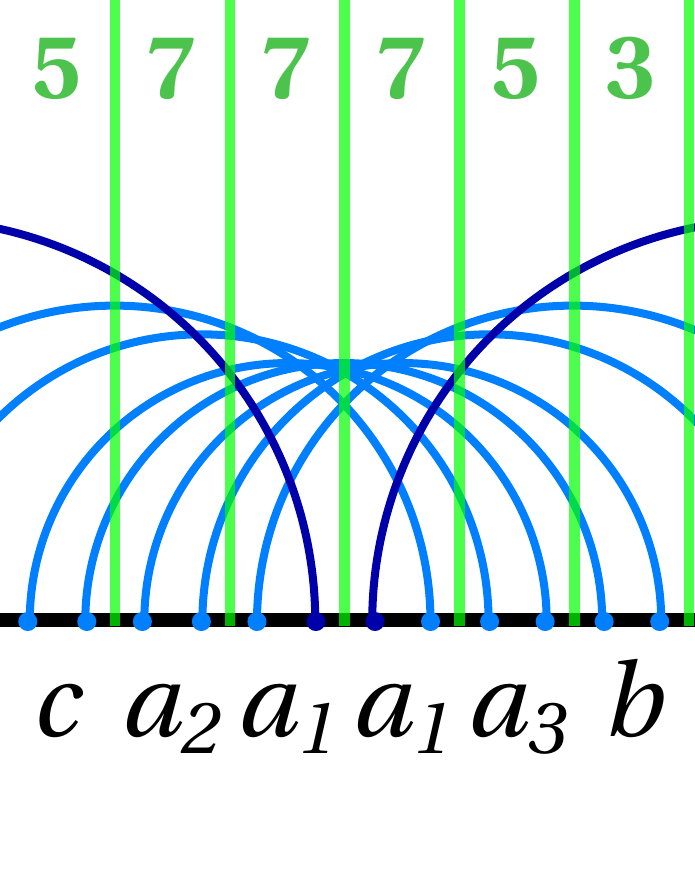}
\end{gathered} \ , &
\begin{gathered}
\includegraphics[height=0.1\textwidth]{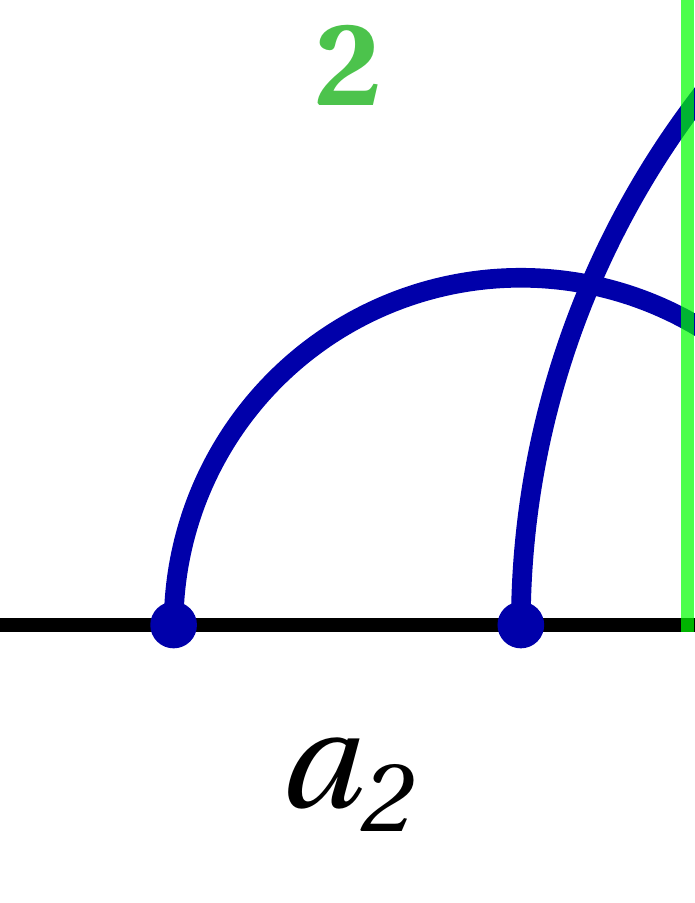}
\end{gathered}
&\scalebox{1.25}{$\quad\mapsfrom\quad\;$}
\begin{gathered}
\includegraphics[height=0.1\textwidth]{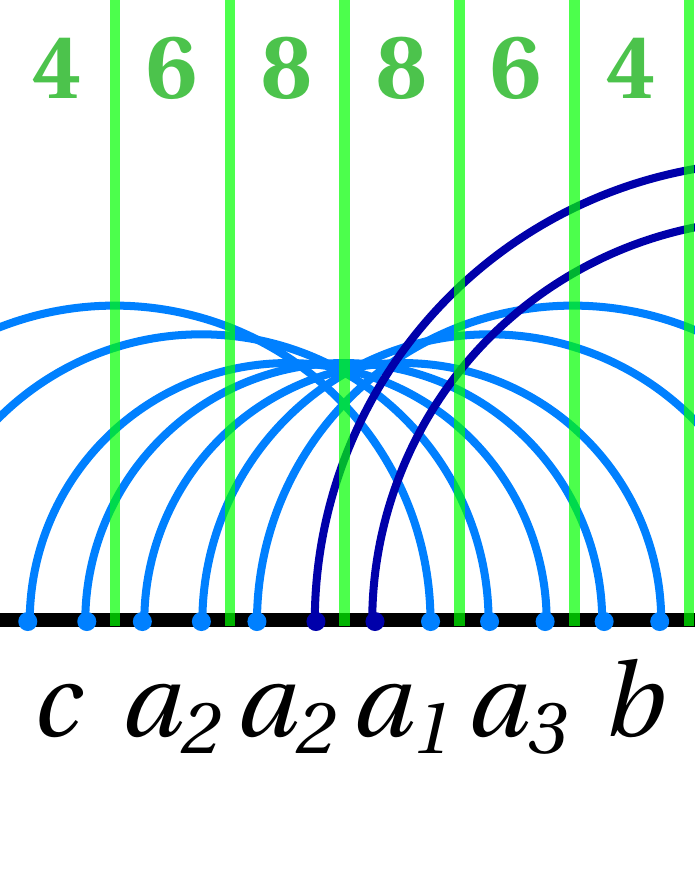}
\end{gathered} \ , \\
\begin{gathered}
\includegraphics[height=0.1\textwidth]{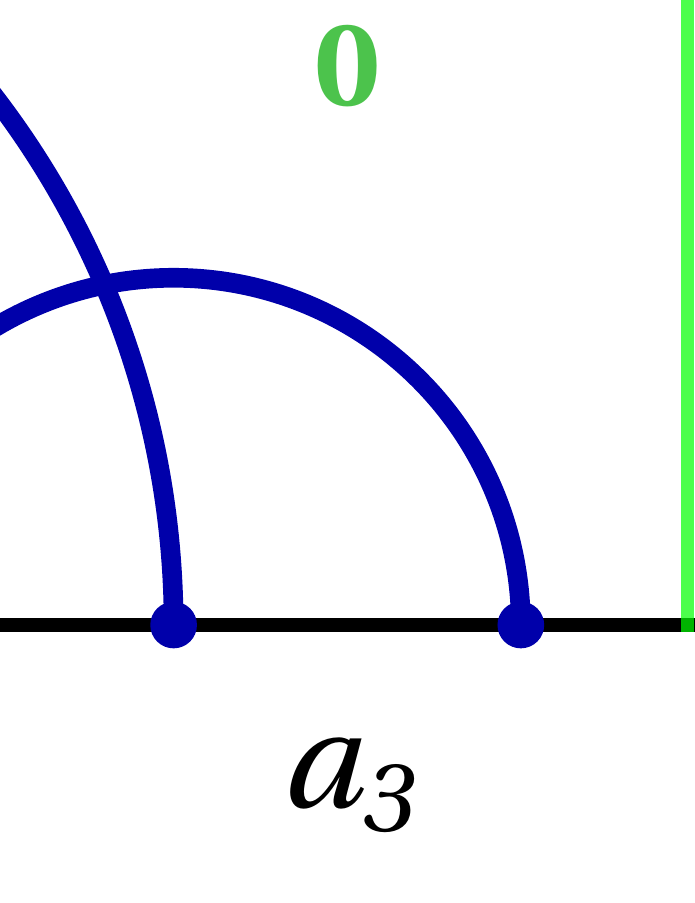}
\end{gathered}
&\scalebox{1.25}{$\quad\mapsfrom\quad\;$}
\begin{gathered}
\includegraphics[height=0.1\textwidth]{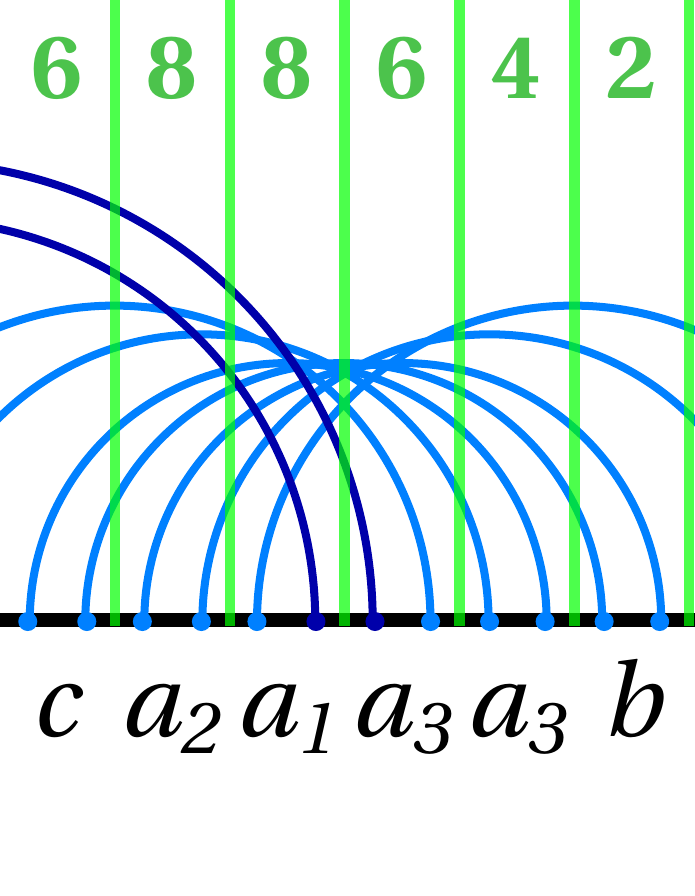}
\end{gathered} \ , &
\begin{gathered}
\includegraphics[height=0.1\textwidth]{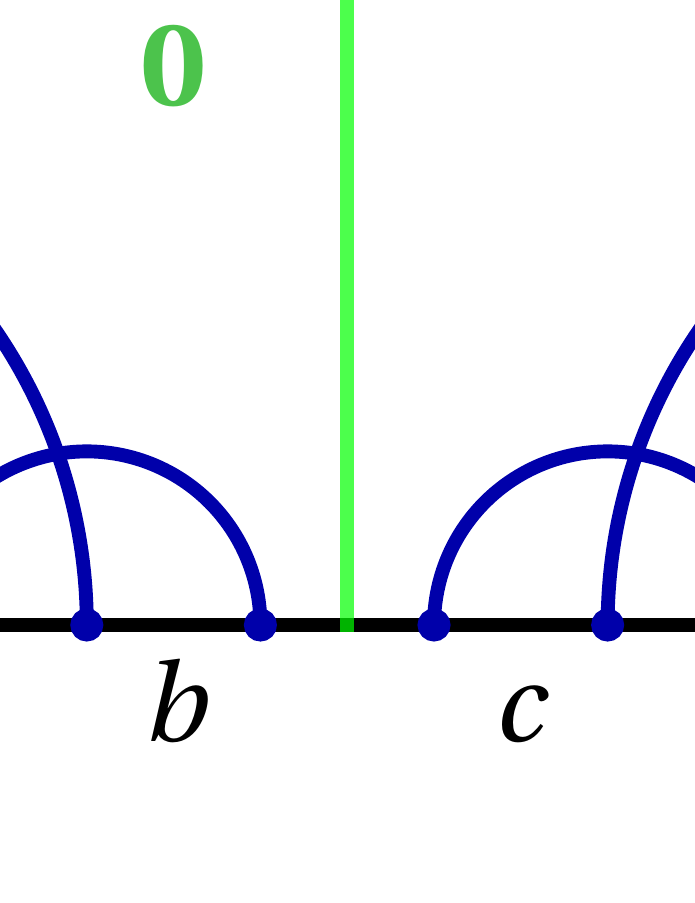}
\end{gathered}
&\scalebox{1.25}{$\quad\mapsfrom\quad\;$}
\begin{gathered}
\includegraphics[height=0.1\textwidth]{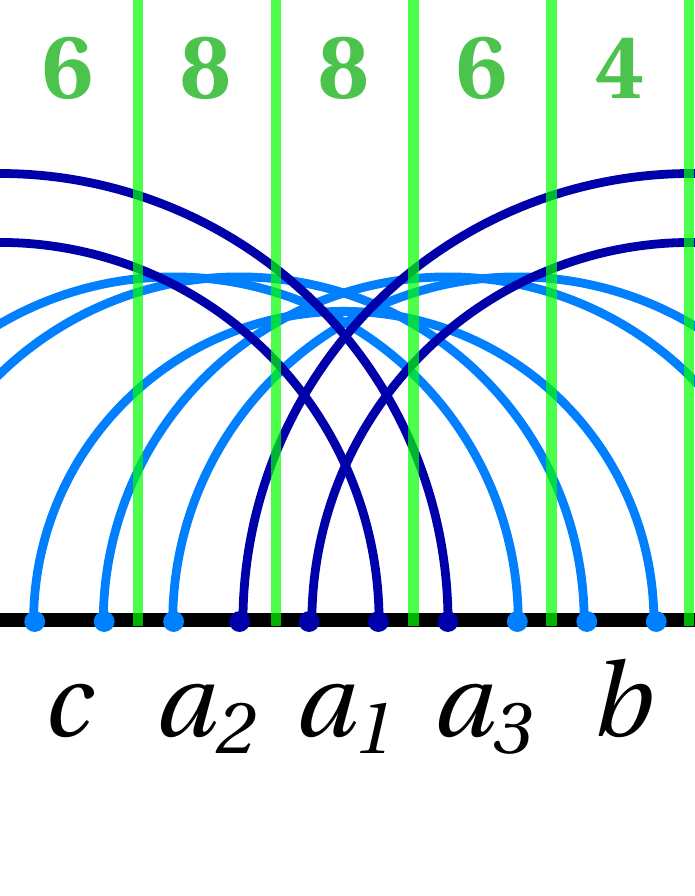}
\end{gathered} \ .
\end{align}
The substitution and entanglement matrices follow accordingly,
\begin{align}
M &=
\left(
\begin{array}{ccccc}
 2 & 1 & 1 & 1 & 1 \\
 1 & 2 & 1 & 1 & 1 \\
 1 & 1 & 2 & 1 & 1 \\
 1 & 1 & 1 & 1 & 1 \\
 0 & 0 & 0 & 0 & 0 \\
\end{array}
\right) \ , &
E &= \left(
\begin{array}{ccccc}
 3 & 3 & 2 & 1 & 2 \\
 3 & \frac{5}{2} & 2 & 1 & 1 \\
 4 & 4 & \frac{5}{2} & 1 & 3 \\
 4 & 4 & 3 & 2 & 3 \\
 0 & 0 & 0 & 0 & 0 \\
\end{array}
\right) \ ,
\end{align}
which leads to a central charge
\begin{equation}
c_{\{9,3\}}^\text{d} = \frac{16 \ln 2}{\ln \frac{\sqrt{21}+5}{2}} \approx 7.08 \ .
\end{equation}
We can generalize this result to tilings at higher $n=4m{+}1$, which correspond to an inflation rule
\begin{align}
a_1 &\mapsto c a_2^{2m-3} a_1 a_1 a_3^{2m-3} b \ , &
b &\mapsto c a_2^{2m-3} a_1 a_3^{2m-3} b \ , \\
a_2 &\mapsto c a_2^{2m-2} a_1 a_3^{2m-3} b \ , &
c &\mapsto \emptyset \ , \\
a_3 &\mapsto c a_2^{2m-3} a_1 a_3^{2m-2} b \ .
\end{align}
The matrices $M$ and $E$ then take the form
\begin{align}
M &=
\left(
\begin{array}{ccccc}
 2 & 2 m-3 & 2 m-3 & 1 & 1 \\
 1 & 2 m-2 & 2 m-3 & 1 & 1 \\
 1 & 2 m-3 & 2 m-2 & 1 & 1 \\
 1 & 2 m-3 & 2 m-3 & 1 & 1 \\
 0 & 0 & 0 & 0 & 0 \\
\end{array}
\right) \ , & 
E &=
\left(
\begin{array}{ccccc}
 2 m-1 & m+1 & m & 1 & 2 \\
 2 m-1 & m+\frac{1}{2} & m & 1 & 1 \\
 2 m & m+2 & m+\frac{1}{2} & 1 & 3 \\
 2 m & m+2 & m+1 & 2 & 3 \\
 0 & 0 & 0 & 0 & 0 \\
\end{array}
\right) \ .
\end{align}
From this we find the central charge
\begin{equation}
c_{\{4m+1,3\}}^\text{d} = \frac{\left(6 m+\frac{3}{10-8 m}+\frac{9}{2}\right) \ln 2}{\ln\frac{\sqrt{16 m^2-24 m+5}+4 m-3}{2}} \ .
\end{equation}
Now consider the cases $n=4m{+}1,k>3$, which correspond to the inflation rules
\begin{align}
a_1 &\mapsto a_2^{2m-2} a_1 a_1 a_3^{2m-3} b \left( a_2^{2m-1} a_1 a_3^{2m-2} b \right)^{k-3} \ , \\
a_2 &\mapsto a_2^{2m-1} a_1 a_3^{2m-3} b \left( a_2^{2m-1} a_1 a_3^{2m-2} b \right)^{k-3} \ , \\
a_3 &\mapsto a_2^{2m-2} a_1 a_3^{2m-2} b \left( a_2^{2m-1} a_1 a_3^{2m-2} b \right)^{k-3} \ , \\
b &\mapsto a_2^{2m-2} a_1 a_3^{2m-2} b \left( a_2^{2m-1} a_1 a_3^{2m-2} b \right)^{k-4} \ .
\end{align}
We explicitly compute the $\{9,4\}$ tiling, which can be expressed by the dimer inflation rules
\begin{align}
\begin{gathered}
\includegraphics[height=0.1\textwidth]{dimers_93_a1.pdf}
\end{gathered}
&\scalebox{1.25}{$\quad\mapsto\quad\;$}
\begin{gathered}
\includegraphics[height=0.2\textwidth]{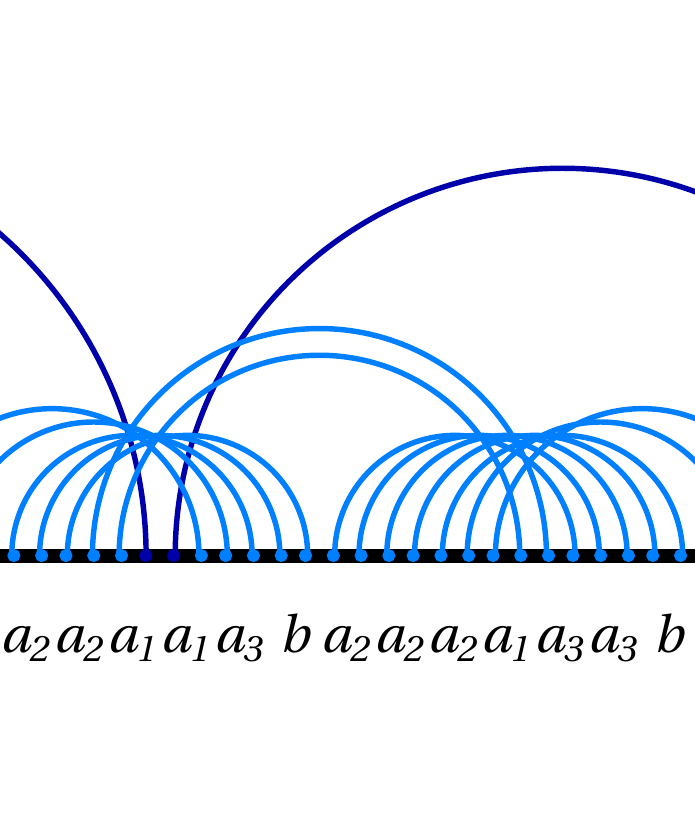}
\end{gathered} \ , &
\begin{gathered}
\includegraphics[height=0.1\textwidth]{dimers_93_a2.pdf}
\end{gathered}
&\scalebox{1.25}{$\quad\mapsto\quad\;$}
\begin{gathered}
\includegraphics[height=0.2\textwidth]{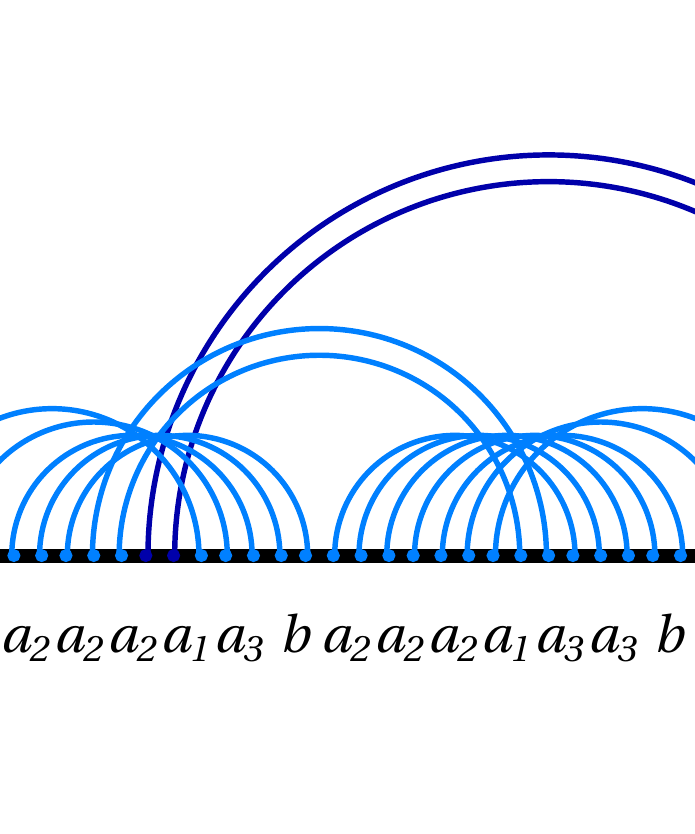}
\end{gathered} \ , \\
\begin{gathered}
\includegraphics[height=0.1\textwidth]{dimers_93_a3.pdf}
\end{gathered}
&\scalebox{1.25}{$\quad\mapsto\quad\;$}
\begin{gathered}
\includegraphics[height=0.2\textwidth]{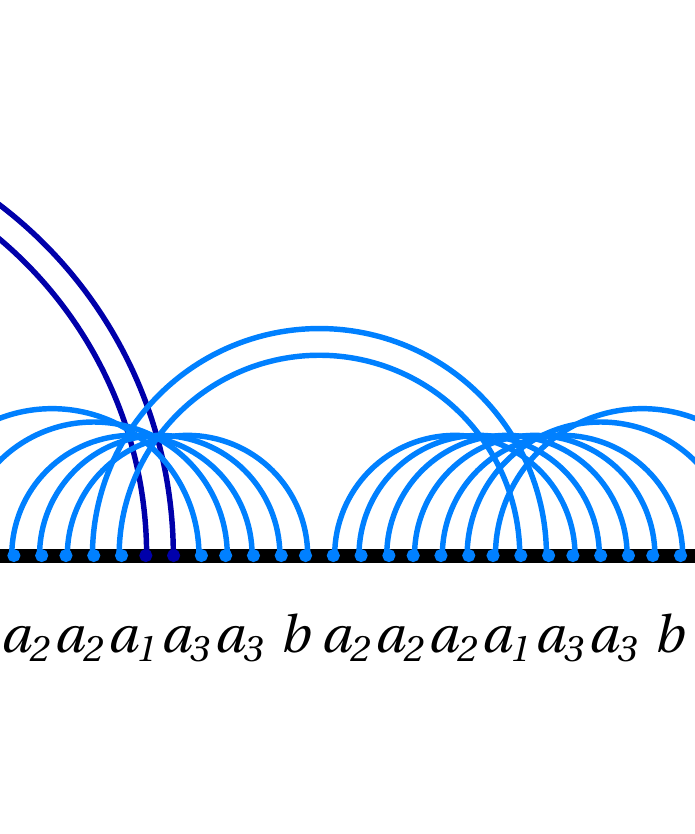}
\end{gathered} \ , &
\begin{gathered}
\includegraphics[height=0.1\textwidth]{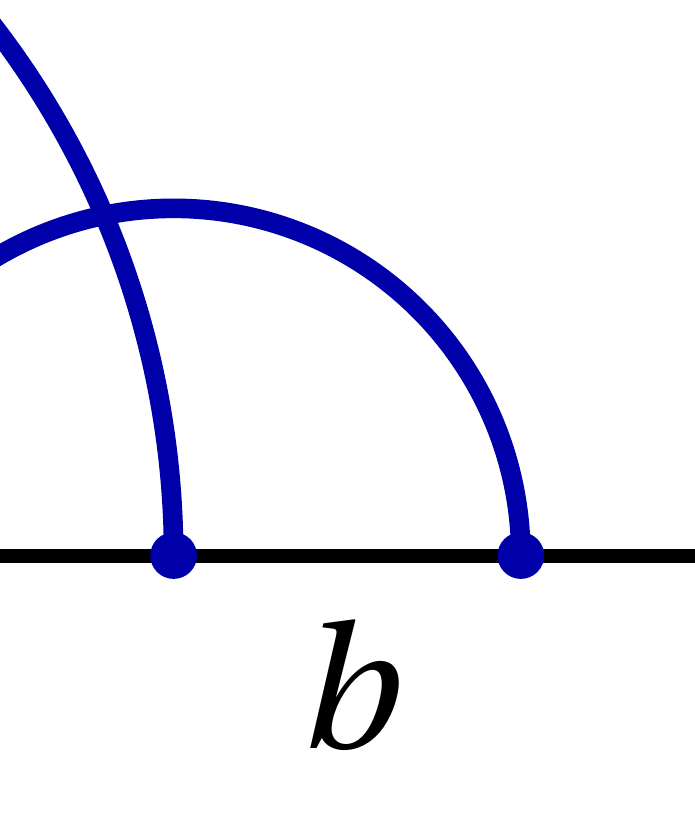}
\end{gathered}
&\scalebox{1.25}{$\quad\mapsto\quad\;$}
\begin{gathered}
\includegraphics[height=0.2\textwidth]{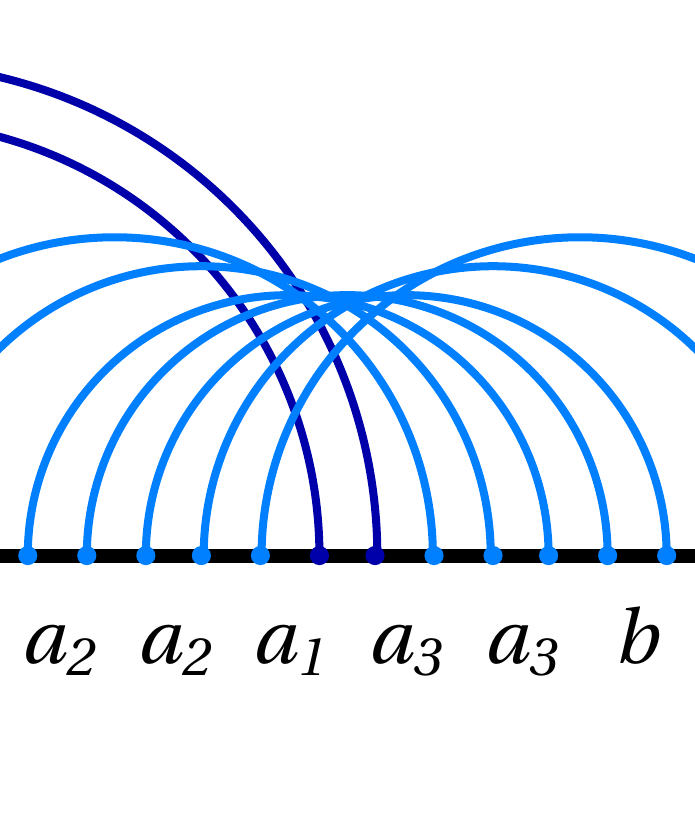}
\end{gathered} \ .
\end{align}
Under deflation, the letters correspond to the following cuts:

\begin{align}
\begin{gathered}
\includegraphics[height=0.1\textwidth]{dimers_93_a1_cut.pdf}
\end{gathered}
&\scalebox{1.25}{$\quad\mapsfrom\quad\;$}
\begin{gathered}
\includegraphics[height=0.2\textwidth]{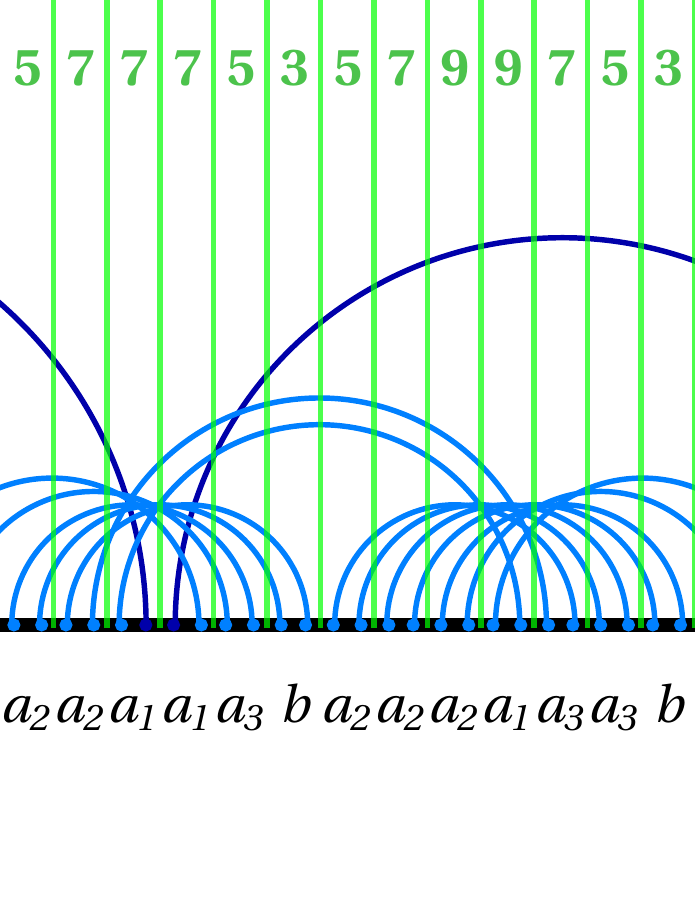}
\end{gathered} \ , &
\begin{gathered}
\includegraphics[height=0.1\textwidth]{dimers_93_a2_cut.pdf}
\end{gathered}
&\scalebox{1.25}{$\quad\mapsfrom\quad\;$}
\begin{gathered}
\includegraphics[height=0.2\textwidth]{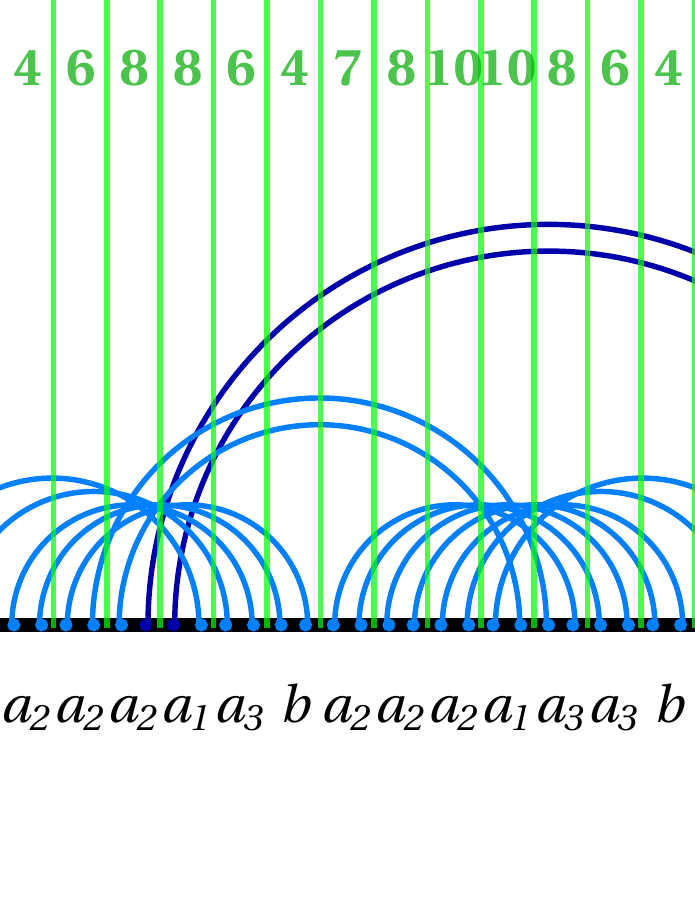}
\end{gathered} \ , \\
\begin{gathered}
\includegraphics[height=0.1\textwidth]{dimers_93_a3_cut.pdf}
\end{gathered}
&\scalebox{1.25}{$\quad\mapsfrom\quad\;$}
\begin{gathered}
\includegraphics[height=0.2\textwidth]{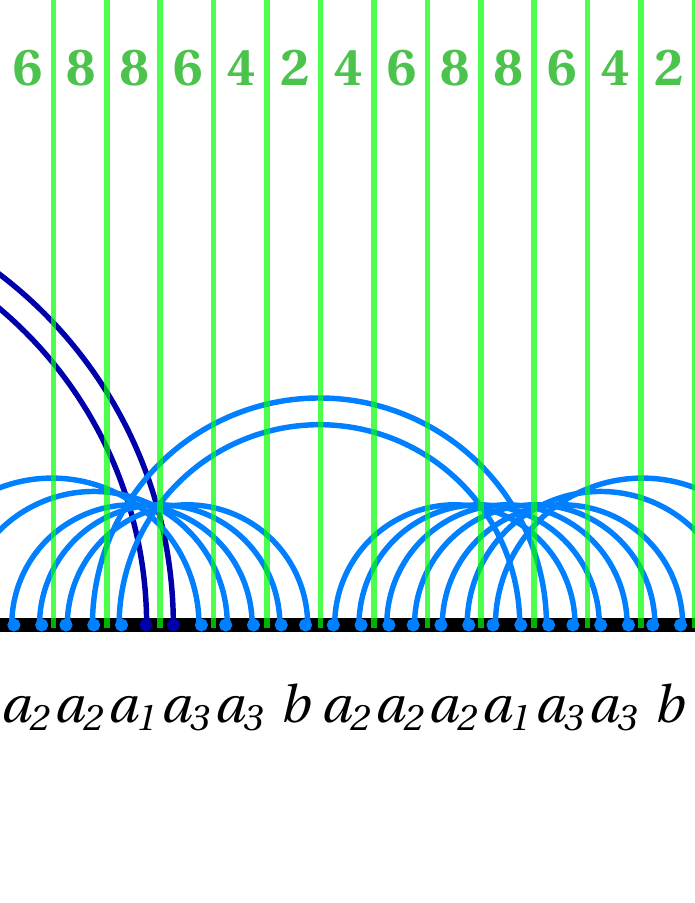}
\end{gathered} \ , &
\begin{gathered}
\includegraphics[height=0.1\textwidth]{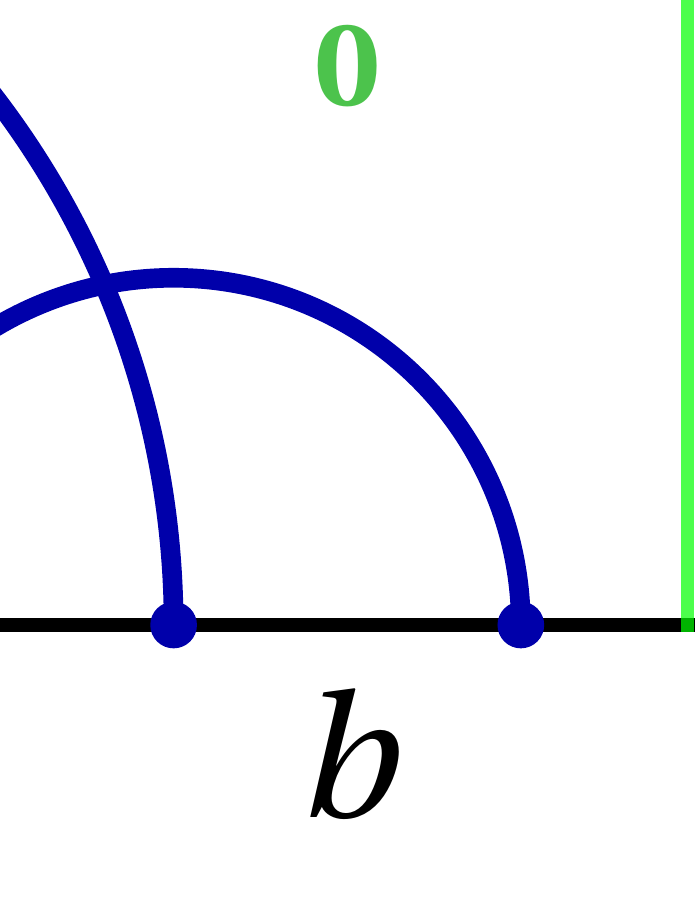}
\end{gathered}
&\scalebox{1.25}{$\quad\mapsfrom\quad\;$}
\begin{gathered}
\includegraphics[height=0.2\textwidth]{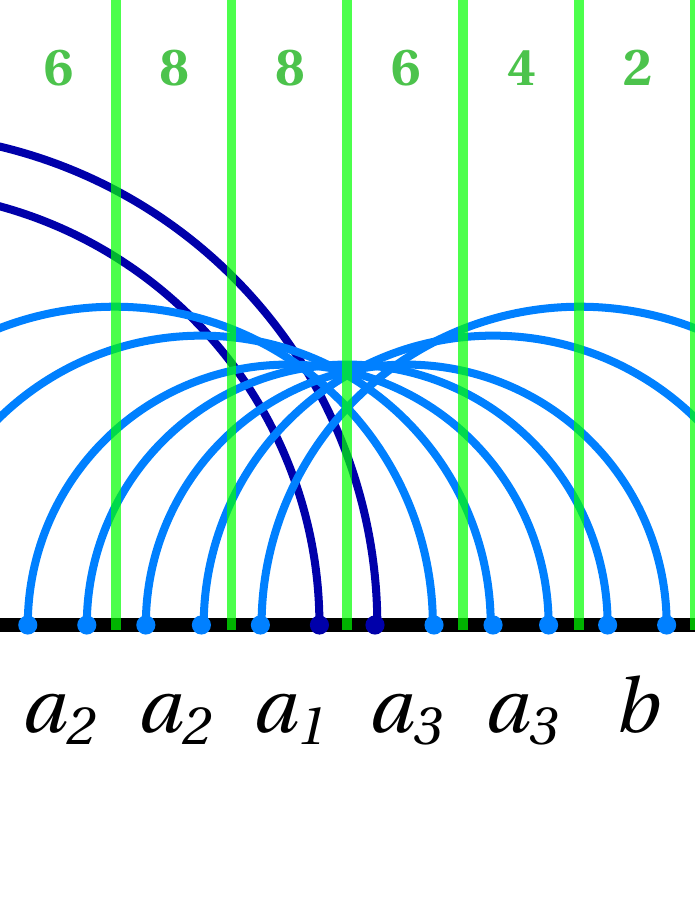}
\end{gathered} \ .
\end{align}
From this we construct the entanglement and substitution matrices
\begin{align}
M &=
\left(
\begin{array}{cccc}
 3 & 5 & 3 & 2 \\
 2 & 6 & 3 & 2 \\
 2 & 5 & 4 & 2 \\
 1 & 2 & 2 & 1 \\
\end{array}
\right) \ , &
E &=
\left(
\begin{array}{cccc}
 \frac{10}{3} & \frac{14}{5} & \frac{7}{3} & 1 \\
 \frac{7}{2} & \frac{5}{2} & \frac{7}{3} & 1 \\
 4 & \frac{16}{5} & \frac{5}{2} & 1 \\
 4 & \frac{7}{2} & \frac{5}{2} & 1 \\
\end{array}
\right) \ .
\end{align}
We then find the central charge
\begin{equation}
c_{\{9,4\}}^\text{d} = \frac{81 \ln 2}{5 \ln \left(\sqrt{35}+6\right)} \approx 4.53 \ .
\end{equation}
For arbitrary $k$, we find
\begin{align}
M &=
\left(
\begin{array}{cccc}
 k-1 & 3 k-7 & 2 k-5 & k-2 \\
 k-2 & 3 (k-2) & 2 k-5 & k-2 \\
 k-2 & 3 k-7 & 2 (k-2) & k-2 \\
 k-3 & 3 k-10 & 2 (k-3) & k-3 \\
\end{array}
\right) \ , &
E &=
\left(
\begin{array}{cccc}
 \frac{4 k-6}{k-1} & \frac{22-9 k}{7-3 k} & \frac{13-5 k}{5-2 k} & 1 \\
 \frac{4 k-9}{k-2} & \frac{3 k-7}{k-2} & \frac{13-5 k}{5-2 k} & 1 \\
 4 & \frac{20-9 k}{7-3 k} & \frac{5}{2} & 1 \\
 4 & \frac{29-9 k}{10-3 k} & \frac{5}{2} & 1 \\
\end{array}
\right) \ ,
\end{align}
leading to 
\begin{equation}
c_{\{9,k\}}^\text{d} = \frac{6 \frac{19 k-49}{7 k-18} \ln 2}{\ln\frac{\sqrt{49 k^2-224 k+252}+7 k-16}{2}} \ .
\end{equation}
Generalizing even further to arbitary $n=4m{+}1$ yields the matrices
\begin{align}
M &=
\left(
\begin{array}{cccc}
 k-1 & -4 m+k (2 m-1)+1 & 2 k (m-1)-4 m+3 & k-2 \\
 k-2 & (k-2) (2 m-1) & 2 k (m-1)-4 m+3 & k-2 \\
 k-2 & -4 m+k (2 m-1)+1 & 2 (k-2) (m-1) & k-2 \\
 k-3 & -6 m+k (2 m-1)+2 & 2 (k-3) (m-1) & k-3 \\
\end{array}
\right) \ , \\
E &=
\left(
\begin{array}{cccc}
 2 m-\frac{2}{k-1} & \frac{-4 m (m+1)+k \left(2 m^2+m-1\right)+2}{-k+2 (k-2) m+1} & \frac{-4 m^2+k (m-1) (2 m+1)+3}{2 k (m-1)-4 m+3} & 1 \\
 2 m+\frac{1}{2-k} & \frac{k-3}{k-2}+m & \frac{-4 m^2+k (m-1) (2 m+1)+3}{2 k (m-1)-4 m+3} & 1 \\
 2 m & \frac{2 (k-2) m^2+(k-2) m-k}{-k+2 (k-2) m+1} & m+\frac{1}{2} & 1 \\
 2 m & \frac{2 (k-3) m^2+(k-3) m-k+1}{-k+2 (k-3) m+2} & m+\frac{1}{2} & 1 \\
\end{array}
\right) \ .
\end{align}
Finally, the central charge for the $\{4m{+}1,k\}$ (block) perfect Majorana dimer model for $m\geq 1, k \geq 4$ follows as
\begin{equation}
c_{\{4m+1,k\}}^\text{d} = \frac{6 \left(\frac{-3 k m+k+6 m+1}{-4 k m+k+8 m+2}+m\right) \ln 2}{\ln\frac{4 k m+\sqrt{(-4 k m+k+8 m)^2-4}-k-8 m}{2}} \ .
\end{equation}
In the large $k$ limit the central charge behaves as
\begin{equation}
c_{\{4m+1,k\}}^\text{d} = \frac{6\frac{ \left(4 m^2+2 m-1\right)}{4 m-1} \ln 2}{\ln\left( (4m-1) k - 8m \right)} + O\left( k^{-1} \right)  \ .
\end{equation}

\end{widetext}
\end{document}